\documentclass[aps,prb,twocolumn,superscriptaddress]{revtex4}
\usepackage{xtab,afterpage,longtable}
\usepackage{graphicx}% Include figure files
\usepackage{siunitx}
\usepackage{bm}% bold math
\usepackage[caption=false]{subfig}
\usepackage{mathtools}
\usepackage{multirow}
\usepackage{amssymb}
\usepackage{braket}
\usepackage{amsmath}
\usepackage{placeins}
\usepackage{cancel}
\usepackage{xcolor}
\usepackage{spreadtab}
\usepackage{wasysym}
\usepackage[colorlinks=true,
            linkcolor=blue,
	        citecolor=blue,
	         urlcolor=blue]{hyperref}
\usepackage[normalem]{ulem}
\begin{document}

\title{Long-range electrostatic contribution to the electron-phonon couplings and mobilities of two-dimensional and bulk materials}

\author{Samuel Ponc\'e}
\email{samuel.ponce@uclouvain.be}
\affiliation{%
Institute of Condensed Matter and Nanosciences, Universit\'e catholique de Louvain, Chemin des \'Etoiles 8, B-1348 Louvain-la-Neuve, Belgium
}%
\affiliation{%
Theory and Simulation of Materials (THEOS), \'Ecole Polytechnique F\'ed\'erale de Lausanne,
CH-1015 Lausanne, Switzerland
}%
\author{Miquel Royo}
\affiliation{%
Institut de Ci\`encia de Materials de Barcelona (ICMAB-CSIC), Campus UAB, 08193 Bellaterra, Spain
}%
\author{Massimiliano Stengel}
\affiliation{%
Institut de Ci\`encia de Materials de Barcelona (ICMAB-CSIC), Campus UAB, 08193 Bellaterra, Spain
}%
\affiliation{Instituci\'o Catalana de Recerca i Estudis Avançats (ICREA), Pg. Llu\'is Companys, 23, 08010 Barcelona, Spain}
\author{Nicola Marzari}
\affiliation{%
Theory and Simulation of Materials (THEOS), \'Ecole Polytechnique F\'ed\'erale de Lausanne,
CH-1015 Lausanne, Switzerland
}%
\affiliation{%
National Centre for Computational Design and Discovery of Novel Materials (MARVEL), \'Ecole Polytechnique F\'ed\'erale de Lausanne, CH-1015 Lausanne, Switzerland
}%
\author{Marco Gibertini}
\affiliation{%
Dipartimento di Scienze Fisiche, Informatiche e Matematiche, Universit\`a di Modena e Reggio Emilia, Via Campi 213/a I-41125 Modena, Italy
}%
\affiliation{%
Centro S3, Istituto Nanoscienze-CNR, Via Campi 213/a, I-41125 Modena, Italy
}%

\date{\today}

\begin{abstract}
Charge transport plays a crucial role in manifold potential applications of two-dimensional materials, including field effect transistors, solar cells, and transparent conductors.
At most operating temperatures, charge transport is hindered by scattering of carriers by lattice vibrations.
Assessing the intrinsic phonon-limited carrier mobility is thus of paramount importance to identify promising candidates for next-generation devices.
Here we provide a framework to efficiently compute the  drift and Hall carrier mobility of two-dimensional materials through the Boltzmann transport equation by relying on a Fourier-Wannier interpolation.
Building on a recent formulation of long-range contributions to dynamical matrices and phonon dispersions [\href{https://journals.aps.org/prx/abstract/10.1103/PhysRevX.11.041027}{Phys.\ Rev.\ X \textbf{11}, 041027 (2021)}], we extend the approach to electron-phonon coupling including the effect of dynamical dipoles and quadrupoles.
We identify an unprecedented contribution associated with the Berry connection that is crucial to preserve the Wannier-gauge covariance of the theory. 
This contribution is not specific to 2D crystals, but also concerns the 3D case, as we demonstrate via an application to bulk SrO. 
We showcase our method on a wide selection of relevant monolayers ranging from SnS2 to MoS2, graphene, BN, InSe, and phosphorene.
We also discover a non-trivial temperature evolution of the Hall hole mobility in InSe whereby the mobility increases with temperature above 150~K due to the mexican-hat electronic structure of the InSe valence bands.
Overall, we find that dynamical quadrupoles are essential and can impact the carrier mobility in excess of 75\%.
\end{abstract}

\maketitle

\section{Introduction}

Two-dimensional (2D) materials exhibit extraordinary properties that could lead to manifold technological applications ranging from heat dissipation and lubricants to electronic applications and energy storage~\cite{Wang2013,Fiori2014,Lv2016,Chhowalla2016,Kang2020}.
A family of 2D materials known as transition-metal dichalcogenides (TMDs) are in the scientific spotlight~\cite{Radisavljevic2011,Manzeli2017}.
TMD monolayers are atomically thin materials of the MX$_2$ family with a transition metal M and a chalcogen atom X.
They are believed to have potential technological impact, especially group-VI TMDs (M = Mo, W) that are semiconducting and exhibit a direct electronic bandgap, which makes them suitable for use in transistors, photodetectors, and emitters~\cite{Radisavljevic2011,Manzeli2017}.
Their lack of inversion symmetry in the 2H phase and strong SOC makes them promising candidates for applications in spintronics~\cite{Reyes-Retana2016,Ahn2020,Sierra2021} and valleytronics~\cite{Zeng2012,Schaibley2016}.
Overall, TMDs possess relatively high carrier mobility compared to other 2D semiconductors~\cite{Sohier2020,Cheng2020} and are seen as good candidate materials for electronic transport.

In an effort to understand the nanoscopic mechanisms underlying the transport properties of materials, density functional perturbation theory~\cite{Gonze1997a,Baroni2001,Giustino2017} provides a powerful tool by giving access to electron-phonon scattering rates fully from first principles, which has recently been extended to include 2D materials~\cite{Sohier2017}.
In particular, room-temperature resistive transport calculations based on such scattering rates are of key technological importance to orient experimental investigations and speed up materials discovery,  and have therefore been widely exploited both in bulk~\cite{Li2015,Zhou2016,Ma2018,Macheda2018,Ponce2018,Ponce2019b,Li2019,Ponce2019c,Macheda2020,Protik2020,Jalil2020,DSouza2020,Brunin2020,Brunin2020a,Ponce2020a,Jhalani2020,Park2020a,Ponce2021} and 2D materials~\cite{Kaasbjerg2012a,Li2013,Kaasbjerg2013,Zhang2014,Park2014,Sohier2014,Li2015,Gunst2016,Sohier2018,Li2019,Gaddemane2019,Guo2019,Cheng2020,Sohier2020,Sio2022}
 reaching remarkable accuracy with respect to experiments in prominent cases including graphene~\cite{Efetov2010,Park2014}.

In most cases, to obtain phonon-limited mobilities, first-principles electron-phonon scattering rates are injected in the Boltzmann transport equation (BTE), considered as an approximation to the Kadanoff-Baym equations of motion~\cite{Ponce2020}.
The linearized BTE describes the non-equilibrium steady-state situation whereby the forces driving the carriers (external electric and/or magnetic field) are equal to the resistive forces associated with electron-phonon scattering.
The resulting integro-differential equation must be solved self-consistently but various approximations exist to avoid it~\cite{Lundstrom2009}, including
the constant relaxation time approximation, the self-energy relaxation time approximation, and the momentum relaxation time approximation.
All have been found to be crude approximations with respect to the exact solution in many materials~\cite{Ganose2021,Ponce2021,Claes2022}, and should thus be avoided in favor of the latter, especially given the minimal computational overhead needed to solve iteratively the linearized BTE.

From a technical point of view, the convergence of the BTE solution requires dense momentum sampling of the electron-phonon interaction~\cite{Ponce2018}.
Although this is in principle achievable by performing direct calculations, especially when using non-uniform grids and in 2D owing to the reduced number of momentum directions to probe~\cite{Sohier2018}, a significant speed up can be obtained through Fourier interpolation~\cite{Giustino2007,Calandra2010,Ponce2016a,Gonze2020,Romero2020,Zhou2021} by using for example Wannier functions~\cite{Marzari2012}.
Importantly, this procedure can only be accurate if the quantities are smooth and localized in real-space to prevent Gibbs oscillations.
However, atomic motions in bulk and 2D semiconductors typically generate dynamical dipoles and quadrupoles~\cite{Vogl1976}, so that lattice vibrations are associated with long-range electrostatic potentials, thus preventing the desired real space localization.
To overcome this obstacle, the electron-phonon matrix elements can be decomposed into a problematic long-range part and a short-range part, which can be safely interpolated.
Indeed, if the long-range part can be expressed analytically in terms a few easy-to-compute macroscopic quantities~\cite{Sjakste2015,Verdi2015}, then it can be removed from the computed quantities, leaving only the short-range part to be interpolated, and then added back to evaluate the overall electron-phonon coupling at arbitrary transferred momentum.

Although this approach has proved to be very effective in 3D systems~\cite{Sjakste2015,Verdi2015,Brunin2020,Jhalani2020} and accurate calculations of the electron-phonon interactions in 2D materials can be performed in the correct electrostatic open-boundary conditions~\cite{Sohier2017}, the application to 2D materials has remained elusive, owing to the complexity of describing, in a unified way, both the in-plane and out-of-plane electrostatics.
It is the subject of the present manuscript.
Continuous efforts~\cite{Sohier2016,Sohier2017a,Deng2021} have delivered simplified formulations that neglect quadrupoles and approximate the out-of-plane dipoles.
Recently, Royo and Stengel~\cite{Royo2021} derived an exact 2D electrostatic framework using an image-charge decomposition for the long- and short-range contributions, and applied it to the interpolation of interatomic force constants (IFC).
In this manuscript, we extend this approach to the electron-phonon matrix elements of 2D materials.
In the process, we discover that including naively higher order terms (both in 2D and 3D) introduces a spurious dependence on the Wannier gauge that can be eliminated by including a contribution associated with the Berry connection (i.e. the position operator in the Wannier basis), restoring the gauge covariance of the method in the long-wavelength limit.
In Section~\ref{sec:theory}, we first recall the general framework and then apply it to the 2D electron-phonon coupling and present various
approximations to compute drift and Hall mobility in 2D materials.
In Section~\ref{sec:results}, we report the numerical parameters used in this study and we verify the quality of the Wannier interpolation and phonon dispersion.
We then assess the effect of quadrupoles and approximations on the accuracy of the deformation potential of SnS$_2$, MoS$_2$, graphene, hexagonal BN, InSe, and phosphorene.
We finish by calculating their drift and Hall carrier mobility and determining the dominant scattering mechanisms in these monolayers.

\section{Theory}\label{sec:theory}
The key quantity in studies of electron-phonon couplings is the matrix element
\begin{equation}\label{eq:ephm}
g_{mn\nu}(\mathbf{k,q})  \equiv \bra{\Psi_{m\mathbf{k+q}}} \Delta_{\mathbf{q}\nu}V \ket{\Psi_{n\mathbf{k}}},
\end{equation}
which provides information about the probability of an electron to be excited from
a quantum state described by the wave-function $\Psi_{n\mathbf{k}}$ to another state $\Psi_{m\mathbf{k+q}}$ via scattering with a
phonon of branch $\nu$, wave vector $\mathbf{q}$, and frequency $\omega_{\nu}(\mathbf{q})$.
In turn, $\Delta_{\mathbf{q}\nu}V$ is, within a density functional theory (DFT) framework, the first-order change in
the Kohn-Sham potential induced by the phonon.

In this section, we describe the formalism that we follow in order to efficiently and accurately obtain the matrix elements $g_{mn\nu}(\mathbf{k,q})$ over ultra-dense grids of $\mathbf{k}$ and $\mathbf{q}$ points.
To this end, first we write $\Delta_{\mathbf{q}\nu} V$ in terms of the potential induced by a simpler atomic displacement perturbation;
the latter being the quantity that is actually obtained via density functional perturbation theory (DFPT) calculations~\cite{Zein1984,Baroni1987,Gonze1997,Gonze1997a,Baroni2001}.
We then discuss the difficulties arising in the interpolation of the first-order potential for the case of semiconductors and insulators associated with the nonanalytical behaviour of the potential near the Brillouin-zone center.
To deal with this problem, we follow the formalism developed in Ref.~\onlinecite{Royo2021} to perform a range separation with the goal of concentrating all the nonanalyticities in a so-called \textit{long-range scattering potential}.
Practical formulas for both the long-range potential and long-range matrix elements, specific to quasi-2D systems, are subsequently derived in the long-wavelength limit and expressed in terms of few macroscopic coefficients readily available within existing public numerical implementations.
Finally, we describe the transport formalism that we use to obtain the carrier mobilities.

\subsection{Electron-phonon scattering potential}\label{sec:separaScat}

The first-order potential in Eq.~\eqref{eq:ephm} can be written in terms of a phase times a lattice-periodic part~\cite{Giustino2017,Brunin2020},
\begin{multline}
\label{eq:perpot}
\Delta_{\mathbf{q}\nu}V(\mathbf{r,r'}) = \Big[ \frac{\hbar}{2 \omega_{\nu}(\mathbf{q})}\Big]^{\frac{1}{2}} \\
 \times \sum_{\kappa\alpha p} \frac{e_{\kappa\alpha\nu}(\mathbf{q})}{\sqrt{M_{\kappa}}}
V_{\mathbf{q}\kappa\alpha}(\mathbf{r,r'}) e^{i\mathbf{q}\cdot (\mathbf{r}'+\mathbf{R}_p)},
\end{multline}
 where $e_{\kappa\alpha\nu}$ is the phonon eigenvector describing the displacement along the Cartesian direction $\alpha$ of the
atom $\kappa$ in the unit cell, and $M_{\kappa}$ the atomic mass.
The lattice-periodic part $V_{\mathbf{q}\kappa\alpha}$ can be exactly obtained via DFPT by solving the following self-consistent Sternheimer equation~\cite{Gonze1997,Baroni2001}
\begin{equation}\label{eq:stern}
 \hat{P}_{\mathbf{k+q}}^{\rm c} \big[ \hat{H}_{\mathbf{k+q}} - \varepsilon_{m\mathbf{k}} \big] \hat{P}_{\mathbf{k+q}}^{\textrm{c}}
|u_{m\mathbf{k}\mathbf{q}}^{\tau_{\kappa\alpha}} \rangle = - \hat{P}_{\mathbf{k+q}}^{\textrm{c}} V_{\mathbf{q}\kappa\alpha}  |u_{m\mathbf{k}} \rangle,
\end{equation}
where $P^{\rm c}$ is the projector on the conduction band manifold, $\hat{H}$, $\varepsilon$ and $u$ are the ground-state Hamiltonian, Kohn-Sham energies and lattice-periodic part of the Bloch functions, respectively.
The $u^{\tau_{\kappa\alpha}}$ are the first-order response functions and the first-order potential $V_{\mathbf{q}\kappa\alpha}$ is
\begin{multline}\label{eq:atdis_pot}
  V_{\mathbf{q}\kappa\alpha}(\mathbf{r,r'})=  V^{\textrm{nl},\tau_{\kappa\alpha}}_{\mathbf{q}}(\mathbf{r,r'}) + \\
  \delta(\mathbf{r-r'}) \big[ V^{\textrm{loc},\tau_{\kappa\alpha}}_{\mathbf{q}}(\mathbf{r}) + V_{\mathbf{q}}^{\textrm{Hxc},\tau_{\kappa\alpha}}(\mathbf{r}) \big],
\end{multline}
which includes both local and nonlocal pseudopotential terms as well as the induced self-consistent field (SCF) potential,
\begin{equation} \label{eq:atdis_scpot}
V_{\mathbf{q}}^{\textrm{Hxc},\tau_{\kappa\alpha}}(\mathbf{r}) =
\int_{\textrm{uc}} d^3r' K_\mathbf{q}(\mathbf{r, r'}) n^{\tau_{\kappa\alpha}}_\mathbf{q}(\mathbf{r'}),
\end{equation}
where $n^{\tau_{\kappa\alpha}}_\mathbf{q}(\mathbf{r})=e^{-i\mathbf{q\cdot r}} n^{\tau_{\kappa\alpha}}(\mathbf{r})$ is the cell-periodic part of the electron density response and the integral is performed over the unit-cell volume.
The kernel in Eq.~\eqref{eq:atdis_scpot} includes both Coulomb and exchange-correlation interactions,
\begin{equation}\label{eq:sckernel}
K(\mathbf{r, r'})= e^2 v(\mathbf{r,r'}) + f^{\mathrm{xc}}(\mathbf{r,r'}),
\end{equation}
where $e$ is the electronic charge and the cell-periodic $K_\mathbf{q}(\mathbf{r, r'})$ is obtained from the above \emph{all-space} kernel by means of the Fourier transforms described in Appendix~\ref{app:representations}.
In the context of DFT, Eq.~\eqref{eq:atdis_pot} is the first-order perturbation from the Kohn-Sham potential
\begin{multline}
 V^{\textrm{KS}}(\mathbf{r,r'}) = V^{\textrm{nl}}(\mathbf{r,r'}) + \\
  \delta(\mathbf{r-r'}) \big[ V^{\textrm{loc}}(\mathbf{r}) + V^{\textrm{Hxc}}(\mathbf{r}) \big].
\end{multline}

In principle, Eq.~\eqref{eq:stern} can be solved at any value of $\mathbf{q}$.
However, such process is time consuming and the exact DFPT potentials are, in practice, only obtained for a set of symmetry-compliant $\mathbf{q}$ points building a coarse grid which is subsequently used to Fourier interpolate the potentials, or the ensuing matrix elements, over much denser grids.
The \emph{bare} Fourier interpolation procedure consists in, first, calculating the real-space representation of the potential using the transformations
shown in Appendix~\ref{app:representations} for the local- and nonlocal-like contributions.
Then, if the real-space potential turns out to decay fast enough so that it vanishes at the boundaries of the supercell, one can safely convert it back at any arbitrary $\mathbf{q}$ point.

\subsection{Range separation}
\label{sec:range_sep}

In the case of semiconductors and insulators, the direct Fourier interpolation of $V_{\mathbf{q}\kappa\alpha}$ is thwarted in the long wavelength limit by the presence of electrostatic fields that decay slowly with the distance.
These long-range fields arise from the nonanalytic behaviour of the Coulomb potential for $\mathbf{q}\rightarrow 0$ and need to be treated separately.
To this end, it is common to carry out a range separation of the Coulomb kernel into a short- and long-range part
\begin{equation}\label{eq:bareCoul}
v_\mathbf{q}(\mathbf{r,r'})=v_{\mathbf{q}}^{\mathcal{S}}(\mathbf{r,r'})+v_{\mathbf{q}}^{\mathcal{L}}(\mathbf{r,r'}),
\end{equation}
where  the short-range Coulomb kernel is analytic in $\mathbf{q}$ and describes the so-called \textit{local fields}, whereas the long-range part ideally acts on a smaller space and includes the nonanalyticities of the total kernel.
Correspondingly, the scattering potential can be separated into short- and long-range contributions stemming from the underlying splitting in the Coulomb kernel,
\begin{equation}\label{eq:vka_sep}
V_{\mathbf{q}\kappa\alpha}(\mathbf{r,r'}) = V^{\mathcal{S}}_{\mathbf{q}\kappa\alpha}(\mathbf{r,r'}) + \delta(\mathbf{r-r'})V^{\mathcal{L}}_{\mathbf{q}\kappa\alpha}(\mathbf{r}),
\end{equation}
where the local dependence on the spatial coordinate adopted for the long-range potential is based on the short-range character of the nonlocal pseudopotential term~\cite{Gonze1997a}, see Eq.~\eqref{eq:atdis_pot}.

In a DFPT framework, $V^{\mathcal{S}}_{\mathbf{q}\kappa\alpha}$ can be obtained by solving the Sternheimer Eq.~\eqref{eq:stern} with a short-range self-consistent field kernel.
The latter has the same structure as Eq.~\eqref{eq:sckernel} but the total Coulomb interaction is replaced with the short-range one $v_\mathbf{q}^{\mathcal{S}}$.
Since the exchange-correlation interaction is a smooth function of $\mathbf{q}$ in both the local and semilocal flavours of DFT, the resulting SCF kernel is short-ranged.
The short-range character of $V^{\mathcal{S}}_{\mathbf{q}\kappa\alpha}$ ensures its correct Fourier interpolation.

The interpolation of the long-range part of the potential $V^{\mathcal{L}}_{\mathbf{q}\kappa\alpha}$ is the main challenge.
To deal with it, we follow Ref.~\onlinecite{Royo2021} and assume that the long-range part of the cell-periodic Coulomb kernel can be written in a separable form as follows,
\begin{equation}\label{eq:separable}
v_\mathbf{q}^\mathcal{L}(\mathbf{r,r'}) = \sideset{}{'}\sum_{NN'} \varphi_{\mathbf{q}N} (\mathbf{r}) \tilde v^{\mathcal{L}}_{\mathbf{q}} (N,N') \left(\varphi_{\mathbf{q}N'}(\mathbf{r'})\right)^\dagger,
\end{equation}
where $N=(\mathbf{G},l)$ is a combined index consisting in a reciprocal-space Bravais lattice vector $\mathbf{G}$ and another index $l$ that characterizes the basis functions along non-periodic directions in finite systems, and where the basis functions $\varphi_{\mathbf{q}N}(\mathbf{r})$ are macroscopic.
This means that they are smooth over the primitive cell volume, analytical in $\mathbf{q}$, and that they span a reduced space, indicated with $'$ over the sum, that we call \emph{small space}~\cite{Martin2016,Royo2021}.
We use a tilde to identify small-space quantities.
Using this small-space representation, the cell-periodic part of the long-range scattering potential can be expressed as~\cite{Royo2021}
\begin{equation}\label{eq:delta2Deq}
V_{\mathbf{q}\kappa\alpha}^{\mathcal{L}}(\mathbf{r}) =  -e \sideset{}{'}\sum_{NN'} \varphi_{\mathbf{q}N}^{\mathcal{S}}(\mathbf{r})  \tilde{W}_{\mathbf{q}}^{\mathcal{L}}(N,N')\tilde{\rho}_{\mathbf{q}\kappa\alpha}^{\mathcal{S}}(N'),
\end{equation}
where we have explicitly indicated the electron charge $-e$ to emphasize the nature of $V_{\mathbf{q}\kappa\alpha}^{\mathcal{L}}(\mathbf{r})$ as an electron potential energy, $\varphi_{\mathbf{q}N}^{\mathcal{S}}(\mathbf{r})$ is the dressed basis function, $\tilde{\rho}_{\mathbf{q}\kappa\alpha}^{\mathcal{S}}(N')$ the charge-density response to an atomic displacement, and $\tilde{W}_{\mathbf{q}}^{\mathcal{L}}(N,N')$ the long-range screened Coulomb interaction.
The latter can be written in terms of the long-range Coulomb operator and the short-range polarizability $\tilde{\chi}_{\mathbf{q}}^{\mathcal{S}}(N,N')$ as follows
\begin{equation}\label{eq:tildew}
 \tilde{W}_{\mathbf{q}}^{\mathcal{L}}= \left(I-e^2\tilde{v}^{\mathcal{L}}_\mathbf{q}\tilde{\chi}^{\mathcal{S}}_\mathbf{q} \right)^{-1} \tilde{v}^{\mathcal{L}}_\mathbf{q},
\end{equation}
where $I$ is the identity matrix and the elements of $\tilde{\epsilon}^{\mathcal{L}}_\mathbf{q}=I- e^2\tilde{v}^{\mathcal{L}}_\mathbf{q} \tilde{\chi}^{\mathcal{S}}_\mathbf{q}$ constitute a long-range dielectric matrix.
The derivation of Eq.~\eqref{eq:delta2Deq} can be found in Eq.~(20) of Ref.~\onlinecite{Royo2021}, while the derivation of Eq.~\eqref{eq:tildew} is given in Eqs.~(8) and (19) as well as in Eqs.~(A1) and (A2) of Ref.~\onlinecite{Royo2021}.

From Eqs.~\eqref{eq:delta2Deq} and~\eqref{eq:tildew}, we conclude that the long-range scattering potential is constructed from a mathematical object, the long-range Coulomb operator, and three short-range and material-specific quantities.
In order to obtain them, it is useful to define the  perturbation
\begin{equation}\label{eq:scalpot}
 \Delta_{\mathbf{q}N} V^{\mathrm{ext}}(\mathbf{r}) = V_{\mathbf{q}N} \varphi_{\mathbf{q}N}(\mathbf{r}) e^{i\mathbf{q \cdot r}},
\end{equation}
which is due to a scalar function from the basis set $\varphi_{\mathbf{q}N}(\mathbf{r})$ modulated at some wave vector $\mathbf{q}$, with $V_{\mathbf{q}N}$ being the perturbation parameter.
Then, the relevant response functions can be obtained via DFPT, while employing the aforementioned short-range SCF kernel, as follows.
On the one hand, the charge-density response to an atomic displacement and the polarizability are obtained as second-order derivatives of the total energy,
\begin{align}\label{respv}
\tilde{\rho}^\mathcal{S}_{\mathbf{q}\kappa \alpha}(N) &= -e \frac{\partial^2 E}{\partial [V_{\mathbf{q}N}]^*  \partial \tau_{\mathbf{q}\kappa\alpha}}, \\
 \tilde{\chi}_{\mathbf{q}}^{\mathcal{S}}(N,N')        &= \frac{\partial^2 E}{\partial [V_{\mathbf{q}N}]^*  \partial V_{\mathbf{q}N'}},
\end{align}
where $\tau_{\mathbf{q}\kappa\alpha}$ in an atomic displacement in the unit cell modulated by the wavevector \textbf{q}.
On the other hand, the dressed basis function can be obtained by adding  the first-order potential response to the perturbation,
\begin{equation}\label{eq:srbf}
 \varphi_{\mathbf{q}N}^{\mathcal{S}}(\mathbf{r})=  \varphi_{\mathbf{q}N}(\mathbf{r}) + e^{-i\mathbf{q\cdot r}} \frac{\partial V^{\mathrm{Hxc}}(\mathbf{r})}{\partial V_{\mathbf{q}N}}.
\end{equation}

At this point, it is important to observe that all the objects entering the definition of the long-range potential in Eq.~\eqref{eq:delta2Deq}, except the long-range Coulomb potential $\tilde{v}^{\mathcal{L}}_{\mathbf{q}}$, are analytic functions of $\mathbf{q}$.
This means that such quantities, once obtained via the aforementioned DFPT scheme on the coarse grid of $\mathbf{q}$ points, can be efficiently interpolated over the whole Brillouin zone.
This fact can be therefore exploited to obtain the \emph{exact} $V_{\mathbf{q}\kappa\alpha}^{\mathcal{L}}(\mathbf{r})$ at any
arbitrary value of $\mathbf{q}$. %once a valid formulation of $\tilde{v}^{\mathcal{L}}_{\mathbf{q}}(N,N')$ is attained.
Despite the apparent appeal of such strategy, in the context of the present work we use the \emph{approximate} method described in Section~\ref{sec:lw_v} since its implementation requires less modifications to the existing codes.
Yet, this \emph{exact} approach might represent a secure fallback to use in systems where the interpolation of the long-range electrostatic fields becomes problematic~\cite{Royo2020}.

\subsection{Long-range Coulomb in two dimensions}
\label{sec:2dcoul}

As stated in the previous section, the core of our formalism rests on a proper separation into short- and long-range Coulomb operators.
This separation is nonunique, but needs to satisfy two main conditions: (i) the long-range kernel $v^\mathcal{L}$ must reproduce the entire nonanalytic behavior of the full kernel, thereby yielding a strictly short ranged $v^\mathcal{S}$; and (ii), $v^\mathcal{L}$ must be smooth in real space, consistent with its macroscopic nature.
In the following paragraphs we discuss conditions (i--ii) in a 2D context, thereby providing an alternative justification to the image-charge construction of Ref.~\onlinecite{Royo2021}.
We first use the \textit{all-space} representation in our derivations, and switch to the cell-periodic convention in Sec.~\ref{sec:lw_v} -- see Appendix~\ref{app:representations} for details about the notation.

In 2D, the bare Coulomb kernel reads
\begin{align}
v(\mathbf{K}z, \mathbf{K'}z') \equiv & \bar{v}(\mathbf{K},z-z') \delta^{(2)}(\mathbf{K-K'}), \\
             \bar{v}(\mathbf{K},z) = & \frac{2\pi}{K} e^{-K|z|}.
\end{align}
Here $\delta^{(2)}$ is the two-dimensional Dirac delta function, $\mathbf{K}$ and $\mathbf{K'}$ are in-plane wave 
vectors in reciprocal space, $z$ denotes the out-of-plane direction in real space, and $K=|\mathbf{K}|$.
Our goal in the following consists in separating $\bar{v}(\mathbf{K},z)$ into two parts: one that is nonanalytic with respect to the parameter 
$\mathbf{K}$ in the vicinity of $\mathbf{K}=\mathbf{0}$, and a remainder that is strictly analytic in ${\bf K}$. 
We note that the Taylor expansion of $\bar{v}(\mathbf{K},z)$ at small $\mathbf{K}$ contains a nonanalytic leading $2\pi / K$ divergence and an infinite number of terms involving odd powers of $K$, which are also nonanalytic.
Based on this observation, one would be tempted to separate the kernel as follows,
\begin{equation}\label{eq:exp_idtty2}
\bar{v}(\mathbf{K},z)  = \frac{2\pi}{K} \big[ \cosh({Kz}) - \sinh({K|z|}) \big].
\end{equation}
Here $\cosh(Kz)/K$ contains all the $K$-odd, and hence nonanalytic contributions to $\bar{v}(\mathbf{K},z)$, while the remainder $\sinh{(K|z|)}/K$ 
is analytic in $K$. 
The obvious nonanaliticity in $z$ of the latter is irrelevant to our present purposes.
While the long-range kernel resulting from Eq.~\eqref{eq:exp_idtty2} complies with condition (i) above, it clearly violates (ii): 
$\cosh({Kz})/K$ diverges exponentially as a function of $K$, which implies that its Fourier transform to real space is not smooth.

To move forward, we shall consider a more general separation by allowing an arbitrary analytic piece to be transferred between the
first and second terms on the right-hand side of Eq.~\eqref{eq:exp_idtty2}.
More specifically, we define
\begin{align}\label{eq:longrangdef}
\bar{v}^{\mathcal{L}}({\bf K},z) &= \frac{2\pi}{K}  \cosh({Kz}) + \Delta \bar{v}({\bf K},z),
\end{align}
where $\Delta \bar{v}(\mathbf{K},z)$ is an arbitrary analytic function of $\mathbf{K}$.
Such form of $\bar{v}^{\mathcal{L}}$ still satisfies (i) as it reproduces the nonanalyticities of the full kernel by construction.
However, the freedom in the additional $\Delta \bar{v}(\mathbf{K},z)$ term can now be exploited to take care of (ii).
We find it convenient at this stage to introduce a \emph{range-separation function} $f(K)$, and 
use it to write the long-range kernel as
\begin{equation}
 \bar{v}(\mathbf{K},z) = \frac{2\pi}{K}   f(K)  \cosh({Kz}),
\end{equation}
which corresponds to setting $\Delta \bar{v}(\mathbf{K},z) = {2\pi}  \left[ f(K) - 1\right]  \cosh({Kz})/K$ in Eq.~\eqref{eq:longrangdef}.
The two requirements (i--ii) on $\bar{v}^{\mathcal{L}}$ can now be both satisfied provided that $f(K)$ vanishes exponentially fast for large $K$, and it linearly approaches unity for small $K$. 
The latter property is essential to ensure that $[f(K) - 1]/K$, and hence $\Delta \bar{v}(\mathbf{K},z)$, is analytic.
This implies that
\begin{align}
\bar{v}^{\mathcal{S}}(\mathbf{K},z) &= \frac{2\pi}{K}  \sinh({K|z|}) - \Delta \bar{v}({\bf K},z) \\
&= \bar{v}(\mathbf{K},z) - \bar{v}^{\mathcal{L}}(\mathbf{K},z), \notag
\end{align}
is strictly short-ranged after a Fourier transformation to real space.

The nonunique separation into a short- and long-range contributions to the Coulomb kernel is thus reflected in the arbitrariness of the range separation function $f(K)$. 
In Ref.~\onlinecite{Royo2021}, a convenient form has been obtained using an image-charge construction 
that leads to the following expression:
\begin{equation}\label{eq:ffunction}
f(K) = 1 - \tanh\Big[ \frac{K L}{2} \Big],
\end{equation}
with $L$ being the parameter that defines the length scale of the range separation: 
the long-range kernel is restricted to only those $\mathbf{K}$ vectors with magnitude sufficiently smaller than $2/L$. 
It is important to stress that a real space representation of the long-range kernel is restricted to $|z|<L/2$, for which Eq.~\eqref{eq:longrangdef} decays exponentially with $K$ and the Fourier transform can be performed. 
Hereafter we use the same expression for $f(K)$, which has been shown to perform well in the practical interpolation of long-range IFCs~\cite{Royo2021}.

\subsection{Long-wavelength approximation of the two-dimensional long-range potential}
\label{sec:lw_v}

In Section~\ref{sec:range_sep}, we have demonstrated that the response functions building the long-range scattering potential, Eq.~\eqref{eq:delta2Deq}, are analytic in the wave vector $\mathbf{q}$.
This fact implies that one can describe them near the Brillouin-zone center via a longwave expansion.
It is therefore possible to reach an approximate expression for the potential written in terms of few macroscopic properties of the system (typically dielectric constants and Born effective charges) as long as a truly macroscopic small space is used to expand the long-range Coulomb operator.
In the present work we describe and use this approach for the specific case of quasi-2D materials since its application to bulk 3D systems has been carried out in earlier studies~\cite{Sjakste2015,Verdi2015,Brunin2020,Brunin2020a,Jhalani2020,Ponce2021}.

We start by introducing the small-space representation of the long-range Coulomb operator~\cite{Royo2021} using the choice motivated in Sec.~\ref{sec:2dcoul} and here expressed in a cell-periodic form:
\begin{multline}\label{eq:vbare2D}
\tilde{v}_{\mathbf{q}}^{\mathcal{L}}(N = \mathbf{G}l,N'=\mathbf{G}'l') \\
\equiv \delta_{\mathbf{G,G}'}\delta_{ll'}(-1)^{l+1}\frac{2\pi f(|\mathbf{G+q}|)}{|\mathbf{G+q}|},
\end{multline}
where $\mathbf{G}$ and $\mathbf{q}$ are assumed to be in plane and where $l=1,2$ is the index for the basis function in the out-of-plane direction.
Accordingly, the basis functions are chosen as:
\begin{equation}
\varphi_{\mathbf{q}N}(\mathbf{r}) \equiv \frac{1}{\sqrt{S}} e^{i \mathbf{G} \cdot \mathbf{r}} \phi_{\mathbf{q}N}(z),
\label{eq:2dbasis}
\end{equation}
where the 2D plane-waves ($S$ is the unit-cell area) form a complete orthonormal set on the Hilbert space of the cell-periodic 
functions, and $\phi_{\mathbf{q}N}(z)$ introduce an explicit dependence on the out-of-plane coordinate $z$ via two hyperbolic functions:
\begin{align}\label{eq:basisfunction}
\phi_{\mathbf{q}N}(z) &=
\begin{cases}
\cosh\left(  |\mathbf{G+q}|  z \right) \quad l = 1\\
\sinh\left(  |\mathbf{G+q}|  z \right) \quad l = 2\\
\end{cases}.
\end{align}
As discussed in Ref.~\onlinecite{Royo2021}, the even/odd symmetry of the cosh/sinh functions with respect to an out-of-plane reflection implies that the cosh and sinh respectively mediate in-plane and out-of-plane electrostatic interactions.

We shall now assume a large enough value of the range-separation parameter $L$ such that $f(|\mathbf{G+q}|)$ vanishes except for $\mathbf{G}=0$.
This leaves us with a single $\mathbf{q}$-dependence for the quantities entering the long-range scattering potential in Eq.~\eqref{eq:delta2Deq} and a small space of dimension 2 spanned by the two $\mathbf{G}=0$ components of the basis functions given in Eqs.~\eqref{eq:2dbasis} and \eqref{eq:basisfunction}.
In such a regime, one can proceed to write the small-space response functions in terms of few macroscopic coefficients by expanding them in a long-wave series as~\cite{Royo2021}:
\begin{align}\label{eq:polarizability2}
\lim_{\mathbf{q} \to \mathbf{0}}
e^2 \tilde{\chi}_{\mathbf{q}}^{\mathcal{S}}(1,1) &= -\mathbf{q} \cdot \boldsymbol{\alpha}^{\parallel} \cdot \mathbf{q} + \mathcal{O}(\mathbf{q}^4)\\
\lim_{\mathbf{q} \to \mathbf{0}}
e^2 \tilde{\chi}_{\mathbf{q}}^{\mathcal{S}}(2,2) &= |\mathbf{q}|^2 \alpha^{\perp} +  \mathcal{O}(\mathbf{q}^4)
\label{eq:polarizability3}
\end{align}
where the round-bracketed indexes refer to values of $l$ and $l'$ in which we neglect the cross terms $(1,2)$ and where the in-plane and out-of-plane macroscopic polarizabilities are given by~\cite{Royo2021}:
\begin{align}
\boldsymbol{\alpha}^{\parallel}=&(\breve{\varepsilon}_{\alpha\beta} - \delta_{\alpha\beta})\frac{c}{4\pi} \\
\alpha^{\perp} =& (1-\breve{\varepsilon}_{zz}^{-1} )\frac{c}{4\pi},\label{eq:outofplanepola}
\end{align}
where $\breve{\varepsilon}_{\alpha\beta}$ and $\breve{\varepsilon}_{zz}$ are the macroscopic in-plane and out-of-plane dielectric constants computed over a unit cell with size $c$ along the out-of-plane direction.
The breve indicates that the dielectric constants depend on the vacuum size dimension, i.e.\ on $c$, while the polarizabilities do not.
We note that when assuming the use of the 2D Coulomb truncation scheme from Ref.~\onlinecite{Sohier2017}, which effectively multiplies the out-of-plane macroscopic polarizability by $\breve{\varepsilon}_{zz}$, then Eq.~\eqref{eq:outofplanepola} becomes $\alpha^{\perp} = (\breve{\varepsilon}_{zz}-1 )\frac{c}{4\pi}$.

The dressed charge-response functions, in turn, are expanded as~\cite{Royo2021}:
\begin{align}
\lim_{\mathbf{q} \to \mathbf{0}} \tilde{\rho}_{\mathbf{q}\kappa\alpha}^{\mathcal{S}}(1)  &= -\sum_{\beta} \frac{iq_{\beta}}{\sqrt{S}} e^{-i\mathbf{q}\cdot \boldsymbol{\tau}_\kappa} \boldsymbol{\mathcal{Z}}_{\kappa\alpha}^{\parallel}(\mathbf{q}), \label{eq:rho1} \\
\lim_{\mathbf{q} \to \mathbf{0}} \tilde{\rho}_{\mathbf{q}\kappa\alpha}^{\mathcal{S}}(2) &= \frac{|\mathbf{q}|}{\sqrt{S}} e^{-i\mathbf{q}\cdot \boldsymbol{\tau}_\kappa}
\mathcal{Z}_{\kappa\alpha}^{\perp}(\mathbf{q}), \label{eq:rho2}
\end{align}
where $\tau_{\kappa\alpha}$ is the atomic displacement of atom $\kappa$ in the Cartesian direction $\alpha$ and $\boldsymbol{\mathcal{Z}}$ are the dipolar expansion defined as:
\begin{align}
\!\! \boldsymbol{\mathcal{Z}}_{\kappa\alpha}^{\parallel}\!(\mathbf{q}) &\equiv Z_{\kappa\alpha\beta} - i \sum_{\gamma}\frac{q_\gamma}{2}(Q_{\kappa\alpha\beta\gamma} \!-\! \delta_{\beta\gamma} Q_{\kappa\alpha zz}) + \!\cdots\!  \\
\!\! \mathcal{Z}_{\kappa\alpha}^{\perp}(\mathbf{q}) &\equiv Z_{\kappa\alpha z} \!-\! i \sum_{\beta}  q_\beta Q_{\kappa\alpha z \beta}  + \cdots,
\end{align}
where $Z_{\kappa\alpha\beta}$ is the dynamical in-plane Born effective charge tensor corresponding to the polarization response along the direction $\beta$ to an atomic displacement of atom $\kappa$ in the Cartesian direction $\alpha$; $Z_{\kappa\alpha z}$ is the dynamical out-of-plane Born effective charge equivalent.
The $Q_{\kappa\alpha\beta\gamma}$ are the dynamical quadrupoles, which describe the polarization response along the direction $\beta$ to a gradient in the Cartesian direction $\gamma$ of an atomic displacement of atom $\kappa$ in the Cartesian direction $\alpha$.
Both dipoles and quadrupoles are here expressed in open-circuit electrical boundary conditions as done for example in Appendix~B of Ref.~\onlinecite{Royo2021}.
We note that in some first-principles software the momentum $\mathbf{q}$ are expressed in unit of $2\pi/a$.
In such cases, one needs to be careful as Eqs.~\eqref{eq:rho1} and \eqref{eq:rho2} assume the $\mathbf{q}$ wave vector to be in inverse length units.

Next, we consider the expansion in $\mathbf{q}$ of the dressed basis functions, $\varphi_{\mathbf{q}l}^{\mathcal{S}}(\mathbf{r})$,
which specifically enter the long-range scattering potential and were not elaborated in Ref.~\onlinecite{Royo2021}.
To this end, we start by performing the longwave expansion of the hyperbolic functions of Eq.~\eqref{eq:basisfunction} as $\cosh(qz) = 1+q^2z^2/2+\cdots$ and $\sinh(qz) = q z+ q^3z^3 / 6  + \cdots$, while retaining the first-order term only.
Then, recalling Eq.~\eqref{eq:srbf} and considering the macroscopic limit ($\mathbf{G}=0$), the perturbation from the basis function reduces to a scalar- ($l=1$) or electric field-like ($l=2$) potential as
\begin{align}
\lim_{\mathbf{q} \to \mathbf{0}}
\varphi_{\mathbf{q}1}^{\mathcal{S}}(\mathbf{r}) =& 
\frac{1}{\sqrt{S}} \big[ 1 + i \sum_\alpha q_\alpha \frac{V^{\textrm{Hxc},\mathcal{E}_\alpha}(\mathbf{r})}{e} + \mathcal{O}(\mathbf{q}^2) \big], \label{eq:basis2Dinplane} \\
\lim_{\mathbf{q} \to \mathbf{0}}
\varphi_{\mathbf{q}2}^{\mathcal{S}}(\mathbf{r}) = & 
\frac{|\mathbf{q}|}{\sqrt{S}} \big[ z + \frac{V^{\textrm{Hxc},\mathcal{E}_z}(\mathbf{r})}{e} +  \mathcal{O}(\mathbf{q}^1) \big],\label{eq:basis2Dout}
\end{align}
where one can notice the unusual units due to the normalization factors in Eq.~\eqref{eq:2dbasis}, and where $V^{\textrm{Hxc},\mathcal{E}_{\alpha / z}}(\mathbf{r})$ is the self-consistent potential (including the Hartree and exchange-correlation terms) induced by an in-plane ($\mathcal{E_\alpha}$) or an out-of-plane ($\mathcal{E}_z$) uniform electric-field perturbation interacting at the level of a short-range kernel.
Notice that $V^{\textrm{Hxc},\mathcal{E}_{z}}(\mathbf{r})$ has to be calculated in open-circuit electrical boundary conditions along the out-of-plane direction or, in other words, as the response to an electric displacement field ($\mathcal{D}_z$).
Eqs.~\eqref{eq:basis2Dinplane} and~\eqref{eq:basis2Dout} can be derived by noting that, to lowest order in $\mathbf{q}$, Eq.~\eqref{eq:scalpot} reduces to $ \Delta_{\mathbf{q}1} V^{\mathrm{ext}}(\mathbf{r}) = V_{\mathbf{q}1} [1 + i\mathbf{q \cdot r}]/\sqrt{S}$ and $ \Delta_{\mathbf{q}2} V^{\mathrm{ext}}(\mathbf{r}) = V_{\mathbf{q}2} |\mathbf{q}| z/\sqrt{S}$.
Since there is no response to a uniform scalar potential, we can further simplify  $ \Delta_{\mathbf{q}1} V^{\mathrm{ext}}(\mathbf{r}) = V_{\mathbf{q}1} i\mathbf{q \cdot r}/\sqrt{S}$, so that both cases can be related to uniform electric field perturbations $\Delta V_{\boldsymbol{\mathcal{E}}}^{\mathrm{ext}}(\mathbf{r}) = e \boldsymbol{\mathcal{E}}\cdot\mathbf{r}$ and the derivatives in Eq.~\eqref{eq:srbf} can be written as $\partial/\partial V_{\mathbf{q}1} = i q_\alpha/(e\sqrt{S}) \partial/\partial \mathcal{E}_\alpha$ and $\partial/\partial V_{\mathbf{q}2} = |\mathbf{ q}|/(e\sqrt{S}) \partial/\partial \mathcal{E}_z$.

Finally, by plugging the above expansions into Eq.~\eqref{eq:delta2Deq} one obtains a formula for the long-range
scattering potential which is valid at any order in $\mathbf{q}$.
In the context of the calculations reported in the present work, we truncate the expansion of the potential at order $\mathcal{O}(\mathbf{q})$, which yields the following practical formula
\begin{multline}\label{eq:vlr_lw}
V_{\mathbf{q}\kappa\alpha}^{\mathcal{L}}(\mathbf{r}) = \frac{\pi e}{S} \frac{f(|\mathbf{q}|)}{|\mathbf{q}|} e^{-i \mathbf{q} \cdot \boldsymbol{\tau}_\kappa} \bigg[ \frac{1}{\tilde{\epsilon}^\parallel(\mathbf{q})} \Big\{ 2 i\mathbf{q} \cdot \mathbf{Z}_{\kappa\alpha}   \\
+ \mathbf{q}\cdot \mathbf{q} \cdot \mathbf{Q}_{\kappa\alpha}  - |\mathbf{q}|^2Q_{\kappa\alpha zz} - 2 \mathbf{q} \cdot \mathbf{Z}_{\kappa\alpha} \mathbf{q} \cdot V^{\textrm{Hxc},\boldsymbol{\mathcal{E}}}(\mathbf{r})/e \Big \} \\
 + \frac{1}{\tilde{\epsilon}^\perp(\mathbf{q})} \Big\{ 2|\mathbf{q}|^2 Z_{\kappa\alpha z} \big[ z + V^{{\rm Hxc},\mathcal{E}_z} (\mathbf{r})/e \big]\Big\} \bigg],
\end{multline}
where the dielectric functions appearing at the denominators are
\begin{align}\label{dielectric}
\tilde{\epsilon}^\parallel(\mathbf{q}) &= 1 + \frac{2\pi f(|\mathbf{q}|)}{|\mathbf{q}|} \mathbf{q} \cdot \boldsymbol{\alpha}^\parallel \cdot \mathbf{q}, \\
\tilde{\epsilon}^\perp    (\mathbf{q}) &= 1 - 2\pi |\mathbf{q}| f(|\mathbf{q}|)  \, \alpha^\perp.
\end{align}

Similarly to the long-range IFC of Ref.~\onlinecite{Royo2021}, which are reproduced in Appendix~\ref{app:longrangeIFC} for completeness, here we end up with two differential contributions, one mirror-even ($\parallel$) and a another mirror-odd ($\perp$), which respectively describe the in-plane and out-of-plane interactions.
However, each one of these two contributions is here shown to incorporate, apart from a macroscopic constant term similar to the ones observed in the IFC formula, an additional local-fields  term with the self-consistent potential induced by an electric field.
These quadrupolar terms, which enter as second order in $\mathbf{q}$ at the numerators, are the 2D generalization of equivalent contributions previously elaborated for the 3D case in Ref.~\onlinecite{Vogl1976} and Refs.~\onlinecite{Brunin2020,Brunin2020a} via alternative approaches.
The latter are recovered with the present formalism based on Eq.~\eqref{eq:delta2Deq} when applied to a 3D crystal, which is demonstrated in Appendix~\ref{app:longrange3D}.

\subsection{Comparison with existing formalism}\label{sub:sohierdip}
In this section, we shall formally compare our Eq.~\eqref{eq:vlr_lw} with the long-range potential equation extracted from Refs.~\onlinecite{Sohier2016, Sohier2017a}.
The 2D long-range scattering potential was there developed up to the dipole level while neglecting both the out-of-plane (mirror-odd) electrostatic fields and the local-fields potentials.
The ensuing long-range scattering potential can be written as
\begin{equation}\label{eq:lr_sohier}
V_{\mathbf{q}\kappa\alpha}^{\mathcal{L},\mathrm{eDS}} (\mathbf{r}) = \frac{2\pi e}{S} \frac{g(|\mathbf{q}|)}{|\mathbf{q}|} \frac{i\mathbf{q} \cdot \mathbf{Z}_{\kappa \alpha}}
 {\varepsilon^{\text{2D}}(|\mathbf{q}|)} e^{-i\mathbf{q} \cdot \boldsymbol{\tau}_\kappa},
\end{equation}
where $g(|\mathbf{q}|)=e^{-{|\mathbf{q}|^2}/{4\Lambda^2}}$ plays the role of the range-separation function where $\Lambda$ is chosen to be large enough for the long-range potential to be truly macroscopic.
The corresponding 2D screening is given, to linear order in $\mathbf{q}$, by~\cite{Sohier2017a}
\begin{equation}
\varepsilon^{\text{2D}}(\mathbf{q}) = \varepsilon^{\text{ext}} + \frac{ \mathbf{q} \cdot \mathbf{r}^{\text{eff}}\cdot \mathbf{q} }{|\mathbf{q}|^2}|\mathbf{q}|,
\label{eq:diel_sohier}
\end{equation}
where $\varepsilon^{\text{ext}}$ is the external dielectric constant which in the case of an isolated monolayer in vacuum is $\varepsilon^{\text{ext}}=1$.
Using classical electrostatics in the limit of vanishing monolayer thickness, the effective screening length is given by
\begin{equation}
r_{\alpha\beta}^{\text{eff}} = (\breve{\varepsilon}_{\alpha\beta}^\infty - \delta_{\alpha\beta})\frac{c}{2} = 
 2\pi{\alpha^\parallel_{\alpha\beta}},
\end{equation}
where $\breve{\varepsilon}_{\alpha\beta}$ is the in-plane high-frequency dielectric constants computed over a unit cell with size $c$ along the out-of-plane direction.
The breve indicates that the dielectric constant depends on the vacuum size dimension, i.e.\ on $c$.

Eq.~\eqref{eq:lr_sohier} can be directly compared with our mirror-even term in Eq.~\eqref{eq:vlr_lw}.
Apart from the lack of any quadrupolar contribution, the main difference comes from the distinct range-separation function.
Both $g(|\mathbf{q}|)$ and $f(|\mathbf{q}|)$ tend to unity in the $\mathbf{q} \rightarrow \mathbf{0}$ limit.
However, the Gaussian function of Sohier \emph{et al.} is reminiscent of the Ewald summation approach in 3D~\cite{Gonze1994} and does so quadratically.
Conversely, our $f(|\mathbf{q}|)$ directly emerging from the analytical derivations of $v^\mathcal{L}$ with a linear behavior at small $\mathbf{q}$. 
The latter property is crucial to correctly reproduce the nonanalytic behaviour of the Coulomb kernel, as we demonstrated in Section~\ref{sec:2dcoul}.
Another difference lies in the fact that, within our formalism, the range separation function also enters the definition of the dielectric functions given in Eq.~\eqref{dielectric}.
The advantages of this improved description of the in-plane screening was already demonstrated in the context of phonon frequencies interpolation~\cite{Royo2021}.
In Section~\ref{sec:results}, we assess its role for the specific case of interpolating electron-phonon scattering quantities.

Two alternative formalism have been more recently reported.
The first one from Deng \textit{et al.}~\cite{Deng2021} introduces an out-of-plane dipolar interaction, which has been understood~\cite{Royo2021} as a dynamical BEC contribution to the $Q_{\kappa\alpha zz}$ quadrupole, and  impacts the in-plane fields in Eq.~\eqref{eq:rho1}.
The second one, from Sio and Giustino~\cite{Sio2022}, proposes a unified long-range description that allows to go smoothly from bulk, to multilayers, to monolayers.
However such formalism treats LO modes, neglects quadrupoles, and recovers the approximate 2D limit of Refs.~\onlinecite{Sohier2016,Sohier2017a}.
These are nonetheless approximate approaches that do not fully capture all electrostatic effects discussed here.

\subsection{Two-dimensional long-range electron-phonon matrix elements}\label{sub:glong}

Although we have shown that the problematic long-range fields are deeply rooted at the level of the scattering potential $V_{{\bf q},\kappa\alpha}^{\mathcal{L}}$, our implementation is based on the direct interpolation of the electron-phonon matrix elements by exploiting the wannierization of the electronic wave functions~\cite{Marzari2012}.
To this end, we make use of an equivalent range separation for the  matrix elements via the incorporation of Eq.~\eqref{eq:vka_sep} into Eqs.~\eqref{eq:perpot}, and~\eqref{eq:ephm}.
It is convenient to operate a rotation to the Wannier gauge to guarantee a smooth behavior as a function of $\mathbf{q}$ by writing the cell-periodic part of the Bloch eigenstates $\ket{u_{n\mathbf{k}}} = e^{-i\mathbf{k}\cdot\mathbf{r}} \ket{\Psi_{n\mathbf{k}}} =  \sum_p U_{np\mathbf{k}}^{*} \ket{u^{\rm W}_{p\mathbf{k}}}$, so that:
\begin{equation}\label{eq:glong-final2D}
g^\mathcal{L}_{mn,\kappa\alpha}(\mathbf{k,q}) =
\sum_{sp} U_{ms\mathbf{k+q}} \langle u_{s\mathbf{k+q}}^{\rm W}| V_{\mathbf{q}\kappa\alpha}^{\mathcal{L}} | u_{p\mathbf{k}}^{\rm W} \rangle U_{pn\mathbf{k}}^{\dagger},
\end{equation}
where $U_{ms\mathbf{k}}$ are the Wannier rotation matrices and $V_{\mathbf{q}\kappa\alpha}^{\mathcal{L}}$ is given by Eq.~\eqref{eq:vlr_lw}.

While the  $U_{mn\mathbf{k}}$ matrices can be obtained at arbitrary wave vectors by diagonalizing the Hamiltonian in the Wannier basis~\cite{Marzari2012}, the problem of interpolating the matrix element $g^\mathcal{L}_{mn,\kappa\alpha}(\mathbf{k,q})$ thus relies on obtaining the correct long-wavelength non-analytic behavior of $\langle u_{s\mathbf{k+q}}^{\rm W}| V_{\mathbf{q}\kappa\alpha}^{\mathcal{L}} | u_{p\mathbf{k}}^{\rm W} \rangle $, which we obtain using Eqs.~\eqref{eq:basis2Dinplane} and \eqref{eq:basis2Dout} as
\begin{multline}
\langle u_{s\mathbf{k+q}}^{\rm W}| V_{\mathbf{q}\kappa\alpha}^{\mathcal{L}} | u_{p\mathbf{k}}^{\rm W} \rangle = \frac{ 2 \pi e f(\textbf{q})}{S |\textbf{q}|}  e^{-i\mathbf{q}\cdot \boldsymbol{\tau}_{\kappa}} \bigg[  \frac{ i\mathbf{q}\cdot \boldsymbol{\mathcal{Z}}_{\kappa\alpha}^{\parallel}(\mathbf{q}) }{\tilde{\varepsilon}^{\parallel}(\mathbf{q})} \\
\!\!\! \times \! \langle u_{s\mathbf{k+q}}^{\rm W}| \varphi_{\mathbf{q}1}^{\mathcal{S}}(\mathbf{r})| u_{p\mathbf{k}}^{\rm W} \rangle 
\!+\! \frac{q \mathcal{Z}_{\kappa\alpha}^{\perp} }{\tilde{\varepsilon}^{\perp}(\mathbf{q})}\langle u_{s\mathbf{k+q}}^{\rm W}| \varphi_{\mathbf{q}2}^{\mathcal{S}}(\mathbf{r})| u_{p\mathbf{k}}^{\rm W} \rangle  \bigg].
\end{multline}

In addition to the $\mathbf{q}\to0$ limit of the potential in Eq.~\eqref{eq:vlr_lw}, we also need the expansion to first order
\begin{equation}\label{eq:uexp}
\langle u_{s\mathbf{k+q}}^{\rm W}| = \langle u_{s\mathbf{k}}^{\rm W}| + \sum_\alpha q_\alpha \left\langle \frac{\partial u_{s\mathbf{k}}^{\rm W}}{\partial k_\alpha}\right| + \cdots,
\end{equation}
where we have exploited that the Wannier gauge is smooth everywhere in the Brillouin zone.
By introducing the Berry connection $A_{sp\mathbf{k},\alpha}^{\text{W}} \equiv -i\left\langle \frac{\partial u_{s\mathbf{k}}^{\rm W}}{\partial k_\alpha}\middle| u_{p\mathbf{k}}^{\rm W} \right\rangle$, we obtain
\begin{align}\label{eq:long-range-new}
 \langle u_{s\mathbf{k+q}}^{\rm W}| \varphi_{\mathbf{q}1}^{\mathcal{S}}(\mathbf{r})| u_{p\mathbf{k}}^{\rm W} \rangle =& \delta_{sp} \nonumber \\ 
 + i\mathbf{q} \cdot  \Big[  \mathbf{A}_{sp\mathbf{k}}^{\text{W}} & + \langle u_{s\mathbf{k}}^{\rm W}| \frac{V^{\textrm{Hxc},\boldsymbol{\mathcal{E}}}(\mathbf{r})}{e} | u_{p\mathbf{k}}^{\rm W} \rangle  \Big] \\
 \langle u_{s\mathbf{k+q}}^{\rm W}| \varphi_{\mathbf{q}2}^{\mathcal{S}}(\mathbf{r})| u_{p\mathbf{k}}^{\rm W} \rangle =& q A_{sp\mathbf{k},z}^{\text{W}}  \nonumber \\
 & + q \langle u_{s\mathbf{k}}^{\rm W}| \frac{V^{\textrm{Hxc},\mathcal{E}_z}(\mathbf{r})}{e} | u_{p\mathbf{k}}^{\rm W} \rangle. \label{eq:long-range-new2}
\end{align}
We note that the current state-of-the art is to take the first-order only:
\begin{align}
 \langle u_{s\mathbf{k+q}}^{\rm W}| \varphi_{\mathbf{q}1}^{\mathcal{S}}(\mathbf{r})| u_{p\mathbf{k}}^{\rm W} \rangle =& \delta_{sp} \\
 \langle u_{s\mathbf{k+q}}^{\rm W}| \varphi_{\mathbf{q}2}^{\mathcal{S}}(\mathbf{r})| u_{p\mathbf{k}}^{\rm W} \rangle =& 0.  \label{eq:long-range-new4}
\end{align}
In practice, $\mathbf{A}_{sp\mathbf{k}}^{\text{W}}$  can be obtained via the Fourier transform of the position operator $\mathbf{r}_{sp,\mathbf{R}}$ in the Wannier basis as:
\begin{equation}\label{eq:positionoperator}
\mathbf{A}_{sp\mathbf{k}}^{\text{W}} = \sum_{\mathbf{R}} e^{i\mathbf{k}\cdot \mathbf{R}} \mathbf{r}_{sp,\mathbf{R}}.
\end{equation}
Here it is  crucial that the position operator be Hermitian, i.e. $\mathbf{r}_{sp,\mathbf{R}} = \mathbf{r}_{ps,\mathbf{-R}}^*$, and translationally invariant.
These conditions are typically not met in discretized forms on the coarse grid~\cite{Marzari1997}, but alternative invariant formulations exist such as the one recently implemented by Lihm in the \textsc{WannierBerri} software~\cite{Tsirkin2021} starting from the expression for Wannier centers by Stengel and Spaldin~\cite{Stengel2006}:
\begin{widetext}
\begin{align}\label{eq:jaemo}
\mathbf{r}_{sp,\mathbf{R}} =
\begin{cases} -\sum_{\mathbf{b}} w_{\mathbf{b}} \mathbf{b} \sum_{\mathbf{k}}\frac{e^{-i\mathbf{k}\cdot \mathbf{R}}}{N_k} \Im \ln \Big[ \sum_{nm} U_{sn\mathbf{k}}^{\dagger} M_{nm}(\mathbf{k,b})U_{mp\mathbf{k+b}} \Big] = \mathbf{r}_{s} &\text{ for } s = p \text{ and } \mathbf{R=0} \\
 i\sum_{\mathbf{b}} w_{\mathbf{b}} \mathbf{b} e^{i\mathbf{b} \cdot (\frac{\mathbf{r}_s + \mathbf{r}_p - \mathbf{R}}{2})}  \sum_{\mathbf{k}}\frac{e^{-i\mathbf{k}\cdot \mathbf{R}}}{N_k} \sum_{nm} U_{sn\mathbf{k}}^{\dagger} M_{nm}(\mathbf{k,b})U_{mp\mathbf{k+b}} &\text{ otherwise } 
 \end{cases}
\end{align}
\end{widetext}
where $\mathbf{r}_s$ and $\mathbf{r}_p$ are the Wannier centers and $M_{nm}(\mathbf{k,b})=\langle u_{n\mathbf{k}} | u_{m\mathbf{k+b}} \rangle $ is the overlap matrix between the cell-periodic Bloch eigenstates at neighboring points \textbf{k} and \textbf{k}+\textbf{b}.

Interestingly, the inclusion of the second term in Eq.~\eqref{eq:uexp} for the $\parallel$ part allows to write both in-plane ($\alpha=x,y$) and out-of-plane ($\alpha=z$) components in Eq.~\eqref{eq:long-range-new} in a similar way in terms of the matrix element between Wannier functions $\ket{p \mathbf{R}}$ of the total (bare+induced) electric field perturbation 
\begin{multline}
e A_{sp\mathbf{k},\alpha}^{\text{W}} + \langle u_{s\mathbf{k}}^{\rm W}| V^{\textrm{Hxc},\mathcal{E}_\alpha}(\mathbf{r}) | u_{p\mathbf{k}}^{\rm W} \rangle = \\
 \sum_{\mathbf{R}} e^{i\mathbf{k}\cdot \mathbf{R}}  \bra{s \mathbf{0}} \left [e r_\alpha +  V^{\textrm{Hxc},\mathcal{E}_\alpha}(\mathbf{r}) \right] \ket{p\mathbf{R}} .
\end{multline} 

\begin{figure}[t]
  \centering
  \includegraphics[width=0.95\linewidth]{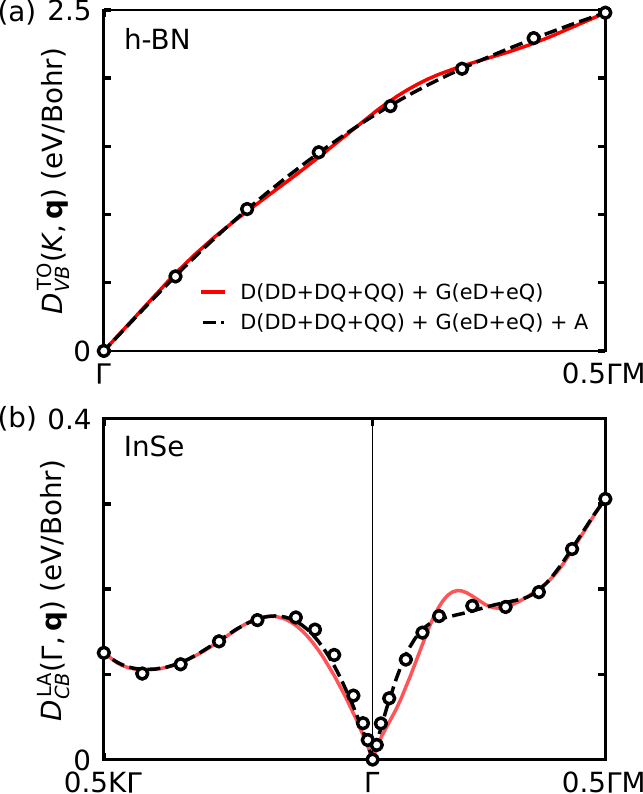} %Ponce_2d_elph/deformation-A1.py
  \caption{\label{fig:BN-test}
Deformation potential, Eq.~\eqref{eq:deformation}, for a specific state and phonon mode as a function of \textbf{q} along high symmetry directions of (a) hexagonal BN monolayer and (b) InSe monolayer.
Results computed using DFPT with electrostatic open-boundary conditions~\cite{Sohier2017} (black empty circles) are compared with Fourier interpolation where the long-range part of the dynamical matrix (D) and electron-phonon matrix elements (G) includes dipole-dipole (DD), dipole-quadrupole (DQ), quadrupole-quadrupole (QQ), monopole-dipole (eD), and monopole-quadrupole (eQ). 
Adding the new Berry connection term $\mathbf{A}_{sp\mathbf{k}}^{W}$ (dashed black line) improves the interpolation quality compared to the result without (red line). 
The potential change $V^{\textrm{Hxc},\boldsymbol{\mathcal{E}}}$ has been added as well (dashed black line) but has a negligible effect (not visible). 
} 
\end{figure}

The corrections associated with $\mathbf{A}_{sp\mathbf{k}}$ and $V^{\textrm{Hxc},\boldsymbol{\mathcal{E}}}$ in Eq.~\eqref{eq:long-range-new}  are necessary to capture the full nonanalytic behavior of $g^\mathcal{L}_{mn,\kappa\alpha}$ up to (and including) the first order in $\mathbf{q}$, going beyond the current approaches based on dipoles only~\cite{Verdi2015,Sjakste2015}.
Even more compelling, by setting $\mathbf{A}_{sp\mathbf{k}}=0$, as typically done also in 3D~\cite{Verdi2015}, the long range contribution \eqref{eq:glong-final2D} depends on the specific Wannier gauge adopted in the calculation through the $U_{mn\mathbf{k}}$  rotation matrices. 
Indeed, even fixing the gauge of the Bloch eigenstates -- and thus of $g^\mathcal{L}_{mn,\kappa\alpha}(\mathbf{k,q})$ -- and restricting to maximally localized Wannier functions, there are multiple Wannier gauge choices available, associated with  rotation matrices $\tilde U_{np\mathbf{k}}$ that differ by a right multiplication by a smooth $\mathbf{k}$-dependent unitary matrix $ W_{lp\mathbf{k}}$ that relates the two Wannier gauges, i.e.\  $\tilde U_{np\mathbf{k}} =  U_{nl\mathbf{k}} W_{lp\mathbf{k}}$ with $ \ket{\tilde u^{\rm W}_{p\mathbf{k}}}  =  \sum_p W_{lp\mathbf{k}} \ket{u^{\rm W}_{l\mathbf{k}}}$. 
In current implementations, this leads to an explicit dependence on $W_{lp\mathbf{k}}$ and thus inconsistently different long range expressions depending on the specific Wannier gauge. 
The terms associated $\mathbf{A}_{sp\mathbf{k}}$ instead restore a Wannier gauge independence to lowest order in $\mathbf{q}$.

In addition, the $\mathbf{A}_{sp\mathbf{k}}$ term also improves interpolation quality of the deformation potential at quadrupolar order.  
We show in Fig.~\ref{fig:BN-test} two such examples in the hexagonal BN and InSe monolayers cases.
Note however that the oscillations in the case of BN (Fig.~\ref{fig:BN-test}(a)) are mild and, as it will be shown in Section~\ref{sec:results}, the quality of the interpolation is excellent in most cases without taking the $\mathbf{A}_{sp\mathbf{k}}$ and $V^{\textrm{Hxc},\boldsymbol{\mathcal{E}}}$ terms into account.
If not explicitly stated in the following, we neglect gauge consistency and $V^{\textrm{Hxc},\boldsymbol{\mathcal{E}}}$.

\subsection{Carrier mobility}\label{sub:mobility}

Finally, once the matrix elements have been obtained on ultra-dense grids,
we can compute the hollow transverse low-field phonon-limited carrier mobility in the presence of a small finite magnetic field \textbf{B}~\cite{Ponce2020}
\begin{equation}
\!\! \mu_{\alpha\beta}^{\rm T}(B_\gamma) = \frac{-1}{S n^{\rm c}}\!\sum_{\mathbf{k}n} w_{\mathbf{k}} v_{n\mathbf{k}\alpha} \big[ \partial_{E_\beta}  f_{n\mathbf{k}}(B_\gamma) - \partial_{E_\beta}  f_{n\mathbf{k}} \big],
\end{equation}
where $w_{\mathbf{k}}$ is the $\textbf{k}$-point weight in the first Brillouin zone surface, $v_{n\mathbf{k}\alpha}$ the band velocity for the eigenstate $\varepsilon_{n\mathbf{k}}$, and $n^{\rm c}$ the carrier concentration.
The linear variation of the electronic occupation $\partial_{E_\beta} f_{n\mathbf{k}}(B_\gamma)$ due to an electric field $\textbf{E}$ and in the presence of a magnetic field $\textbf{B}$ can be obtained by solving the Boltzmann transport equation (BTE)~\cite{Macheda2018,Ponce2020,Ponce2021}:
\begin{multline}\label{eq:iterwithbimpl}
 \Big[ 1 - \frac{e}{\hbar}\tau_{n\mathbf{k}} ({\bf v}_{n\mathbf{k}} \times {\bf B}) \cdot \nabla_{\bf k}
\Big]\partial_{E_{\beta}} f_{n\mathbf{k}}(\mathbf{B}) = e v_{n\mathbf{k}\beta} \frac{\partial f_{n\mathbf{k}}^0}{\partial \varepsilon_{n\mathbf{k}}} \tau_{n\mathbf{k}} \\
+ \frac{2\pi\tau_{n\mathbf{k}}}{\hbar}
  \sum_{\mathbf{q} m\nu} w_{\mathbf{q}} \partial_{E_{\beta}} f_{m\mathbf{k}+\mathbf{q}}(\mathbf{B})   | g_{mn\nu}(\mathbf{k},\mathbf{q})|^2 \\
 \times  \Big[(n_{\mathbf{q}\nu}+1-f_{n\mathbf{k}}^0)\delta(\varepsilon_{n\mathbf{k}} - \varepsilon_{m\mathbf{k+q}}  + \hbar \omega_{\mathbf{q}\nu} )   \\
  +  (n_{\mathbf{q} \nu} + f_{n\mathbf{k}}^0)\delta(\varepsilon_{n\mathbf{k}} - \varepsilon_{m\mathbf{k+q}}  - \hbar \omega_{\mathbf{q}\nu}) \Big] ,
\end{multline}
with $\tau_{n\mathbf{k}}$ being the total scattering lifetime and the inverse $\tau_{n\mathbf{k}}^{-1}$  is the scattering rate given by:
\begin{align}\label{eq:scattering_rate}
  \tau_{n\mathbf{k}}^{-1} =& \frac{2\pi}{\hbar} \sum_{\mathbf{q}m\nu} w_{\mathbf{q}} | g_{mn\nu}(\mathbf{k,q})|^2 \nonumber \\
  \times & \big[ (n_{\mathbf{q}\nu} +1 - f_{m\mathbf{k+q}}^0) \delta( \varepsilon_{n\mathbf{k}} - \varepsilon_{m\mathbf{k+q}}   -  \hbar \omega_{\mathbf{q}\nu}) \nonumber \\
   + &  (n_{\mathbf{q}\nu}  +   f_{m\mathbf{k+q}}^0 )\delta(\varepsilon_{n\mathbf{k}} - \varepsilon_{m\mathbf{k+q}}  +  \hbar \omega_{\mathbf{q}\nu}) \big],
\end{align}
where $f_{n\mathbf{k}}^0$ is the Fermi-Dirac occupation function at equilibrium (in the absence of fields) and $n_{\mathbf{q}\nu}$ is the Bose-Einstein equilibrium distribution function.
Finally we obtain the drift mobility as:
\begin{equation}\label{eq:mobilitydrift}
\mu_{\alpha\beta}^{\rm drift} = \frac{-1}{S n^{\rm c}}\sum_{\mathbf{k}n} w_{\mathbf{k}} v_{n\mathbf{k}\alpha} \partial_{E_\beta}  f_{n\mathbf{k}},
\end{equation}
where $\partial_{E_\beta}  f_{n\mathbf{k}}$ is obtained by solving Eq.~\eqref{eq:iterwithbimpl} without \textbf{B}.

From this we can define the dimensionless Hall tensor, which is
defined as the ratio between the mobility with and without magnetic field as~\cite{Reggiani1983,Popovic1991}
\begin{equation}\label{eq:hallfactor}
  r_{\alpha\beta}(\hat{\mathbf{B}}) \equiv  \lim_{\mathbf{B} \rightarrow 0} \sum_{\delta\epsilon} \frac{[\mu_{\alpha\delta}^{\rm drift}]^{-1} \, \mu_{\delta\epsilon}^{\rm T}(\mathbf{B}) \, [\mu_{\epsilon\beta}^{\rm drift}]^{-1}}{|\mathbf{B}|},
\end{equation}
where $\hat{\mathbf{B}}$ is the direction of the magnetic field and Eq.~\eqref{eq:hallfactor} is the tensorial generalization of Ref.~\onlinecite{Reggiani1983}.
The Hall mobility is computed as:
\begin{equation}\label{eq:mobilityhall}
\mu_{\alpha\beta}^{\rm Hall}(\hat{\mathbf{B}}) = \sum_{\gamma} \mu_{\alpha\gamma}^{\rm drift} r_{\gamma\beta}(\hat{\mathbf{B}}).
\end{equation}

A common approximation to Eq.~\eqref{eq:mobilitydrift} is called the self-energy relaxation time approximation (SERTA)~\cite{Ponce2018} and consists in neglecting the second term in the right-hand side of Eq.~\eqref{eq:iterwithbimpl} which therefore does not need to be solved iteratively and yields:
\begin{equation}\label{eq:mobilityserta}
\mu_{\alpha\beta}^{\rm SERTA} = \frac{-e}{S n^{\rm c}}\sum_{\mathbf{k}n} w_{\mathbf{k}} \frac{\partial f_{n\mathbf{k}}^0}{\partial \varepsilon_{n\mathbf{k}}} v_{n\mathbf{k}\alpha} v_{n\mathbf{k}\beta}  \tau_{n\mathbf{k}}.
\end{equation}

\section{Results}\label{sec:results}

In this work, we have decided to study 6 monolayers (SnS$_2$, BN, MoS$_2$, InSe, phosphorene, and graphene) that represent various cases, from polar to non-polar materials, semiconductors and semimetals, to highlight the accuracy
of the electron-phonon matrix element interpolation presented in the theory Section~\ref{sec:theory}.
More specifically, we study SnS$_2$ and hexagonal BN monolayers, for which phonon dispersions including the effect of quadrupoles have already been studied in Ref.~\onlinecite{Royo2021}, as validation of our theory and implementation.
We then choose MoS$_2$, InSe, graphene, and phosphorene for their technological relevance and richness, and their extensive investigations available in the literature.
Finally, we study bulk SrO where the quadrupole tensor is null by symmetry to highlight the importance of the Berry connection term.
We note that dynamical quadrupoles are zero in centrosymmetric crystals where all the atoms are placed in centrosymmetric sites, which are fairly common~\cite{Claes2022}.

\subsection{Computational details}

In the practical interpolation of $g_{mn\nu}^{\mathcal{L}}$, we shall follow the common practice~\cite{Verdi2015, Brunin2020} of extending Eqs.~\eqref{eq:vlr_lw} and ~\eqref{eq:glong-final2D} to finite $\mathbf{G}$ vectors by incorporating a sum over $\mathbf{G}$ and replacing every occurrence of $\mathbf{q}$ by $\mathbf{G+q}$.
This replacement emulates the periodic nature of the potential and should be seen as a model which numerically improves the quality of the interpolation of the matrix elements beyond the long-wavelength limit.
The Cartesian short-range matrix elements in real-space $g_{mn\kappa\alpha}^{\mathcal{S}}(\mathbf{R}_p,\mathbf{R}_{p'})$ are Fourier interpolated in Bloch space and the long-range contribution is added:
\begin{multline}\label{eq:glong}
g_{mn,\kappa\alpha}(\mathbf{k},\mathbf{q}) =  g_{mn,\kappa\alpha}^{\mathcal{S}}(\mathbf{k},\mathbf{q}) + \sum_{\mathbf{G} \neq -\mathbf{q}}  \sum_{sp}\\
\times  U_{ms\mathbf{k+q+G}}   \langle u_{s\mathbf{k+q+G}}^{\rm W}| V_{\mathbf{q+G}\kappa\alpha}^{\mathcal{L}} | u_{p\mathbf{k}}^{\rm W} \rangle U_{pn\mathbf{k}}^{\dagger}.
\end{multline}

Finally, the interpolated electron-phonon matrix element is rotated in the mode basis to recover Eq.~\eqref{eq:ephm}:
\begin{equation}\label{eq:final_formula}
g_{mn\nu}(\mathbf{k},\mathbf{q}) = \\
\Big[ \frac{\hbar}{2 \omega_{\nu}(\mathbf{q})}\Big]^{\frac{1}{2}}  \sum_{\kappa\alpha}  \frac{e_{\kappa\alpha\nu}(\mathbf{q})}{\sqrt{M_{\kappa}}} g_{mn,\kappa\alpha}(\mathbf{k},\mathbf{q}).
\end{equation}

We further obtain the long-range matrix elements in the mode basis from the rotation of Eq.~\eqref{eq:glong-final2D} as:
\begin{equation}\label{eq:final_formula2}
g_{mn\nu}^{\mathcal{L}}(\mathbf{k},\mathbf{q}) = \\
\Big[ \frac{\hbar}{2 \omega_{\nu}(\mathbf{q})}\Big]^{\frac{1}{2}}  \sum_{\kappa\alpha}  \frac{e_{\kappa\alpha\nu}(\mathbf{q})}{\sqrt{M_{\kappa}}} g_{mn,\kappa\alpha}^{\mathcal{L}}(\mathbf{k},\mathbf{q}),
\end{equation}
where we define the monopole-dipole contribution $g_{mn\nu}^{\mathcal{L}, \rm eD}$ by setting $\mathbf{Q} = 0$ in Eq.~\eqref{eq:vlr_lw} and the monopole-quadrupole one $g_{mn\nu}^{\mathcal{L}, \rm eQ}$ by setting $\mathbf{Z} = 0$ in Eq.~\eqref{eq:vlr_lw}.

\begin{figure}[t]
  \centering
  \includegraphics[width=0.85\linewidth]{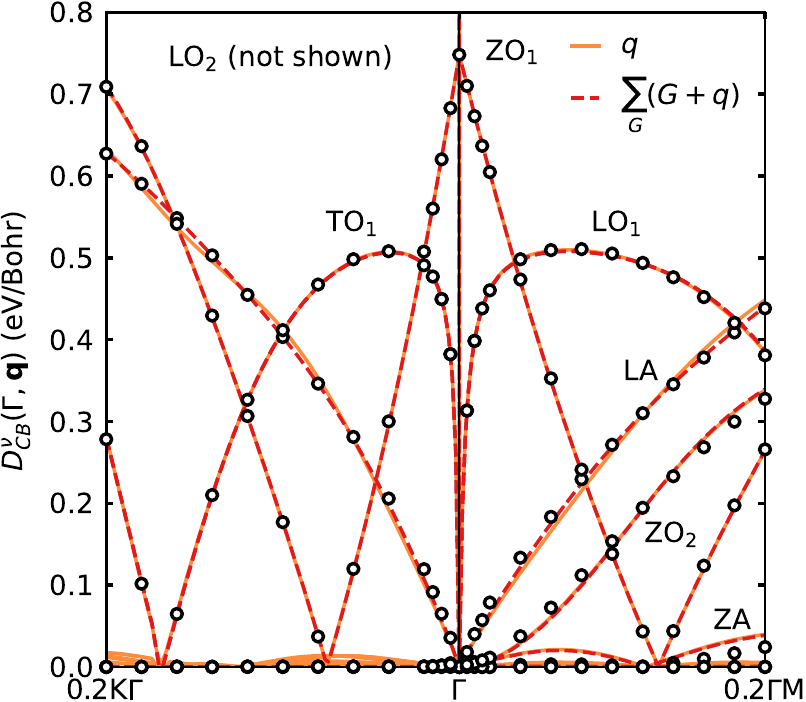}
  \caption{\label{fig:Sns2-defo-test}
Comparison between the deformation potential Eq.~\eqref{eq:deformation} of the conduction band of SnS$_2$ where the initial state is $\mathbf{k}=\boldsymbol{\Gamma}$ along high-symmetry lines
interpolated using Eq.~\eqref{eq:glong} with (dashed red line) and without (orange line) the sum over $\mathbf{G}$-vector.
We used a 16$\times$16$\times$1 coarse $\mathbf{k}$-point and $\mathbf{q}$-point grids.
The interpolation of the LA mode is slightly improved with the sum over $\mathbf{G}$-vector.
The LO$_2$ mode occurs at higher deformation potential and is not shown.
}
\end{figure}

\begin{figure}[ht]
  \centering
  \includegraphics[width=0.8\linewidth]{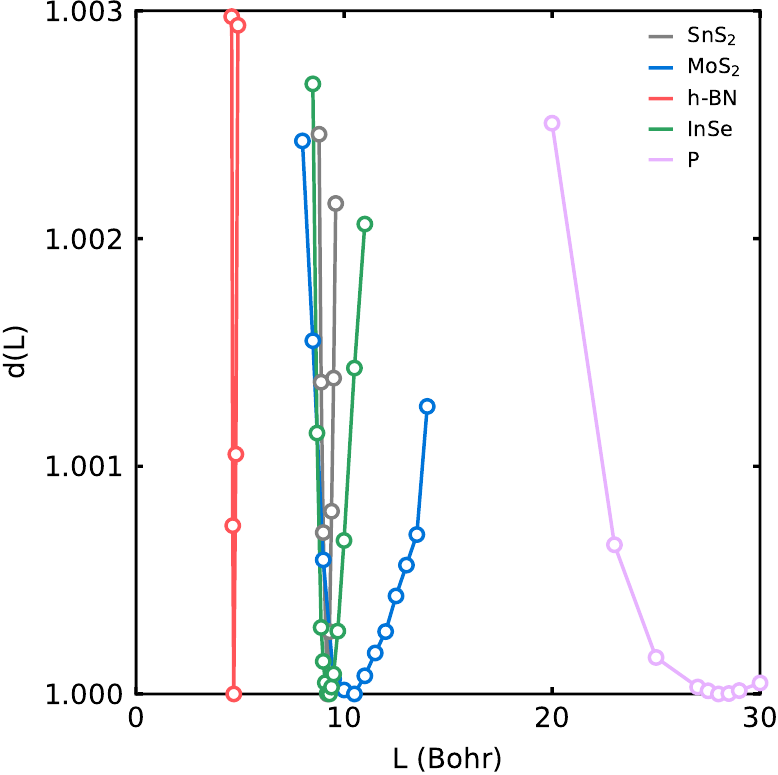}
  \caption{\label{fig:lpara}
 Average and normalized short-range interatomic force constant as a function of the $L$ parameter using Eq.~\eqref{eq:findingL}.
 They have been divided by their value at the minimum to be able to compare them.
  }
\end{figure}

We illustrate the effect of summing $g^{\mathcal{L}}$ over the grid of \textbf{G} vectors in Fig.~\ref{fig:Sns2-defo-test} for the case of SnS$_2$ which shows a small improvement.
Therefore, instead of using the purposely large value of $L$ assumed in deriving Eq.~\eqref{eq:vlr_lw}, in our calculations we shall choose an optimal value of $L$ such that not all the $\mathbf{G}\ne \mathbf{0}$ contributions are filtered out.
To this end, we follow the approach successfully employed in Ref.~\onlinecite{Royo2021} to interpolate the IFC, which consists in
taking the value of $L$ that minimizes the sum of the real-space short-range IFC ($\Phi^{\mathcal{S}}$):
\begin{equation}\label{eq:findingL}
d(L) = \frac{1}{N}\sum_{\kappa\kappa'l}^{*}\sum_{\alpha\beta} |\Phi^{\mathcal{S}}_{\kappa\alpha,\kappa'\beta}(0,l)|,
\end{equation}
where $^{*}$ indicates that the $\kappa = \kappa'$ terms are excluded in the reference unit cell ($l=0$) and $N$ is the number of cells in the real-space supercell.
The optimal values of the $L$ parameter using Eq.~\eqref{eq:findingL} are show in Fig.~\ref{fig:lpara} for the materials studied in this manuscript.
We also proceed following Ref.~\onlinecite{Royo2021} to interpolate the phonon frequencies and eigenvectors with the formulas reported in Appendix~\ref{app:longrangeIFC}.

\begin{table}[ht]
  \caption{\label{table1}
Cartesian components of the dynamical dipole $Z (e)$ and quadrupole $Q (e \text{bohr})$, separation length $L (\text{bohr})$, dielectric and polarizability tensors in atomic units for SnS$_2$, MoS$_2$, BN, InSe, and P monolayer.
Only independent components are shown.
}
\begin{footnotesize}
  \begin{tabular}{ r r r | r r | r r | r r | r}
  \toprule\\
                    & Sn    & S$_1$  &     Mo &  S$_1$ &    B   &   N    & In$_1$  & Se$_2$ & P \\
\hline
$Z_{\kappa xx}$ & 4.813 & -2.407 & -0.988 &  0.494 &  2.685 & -2.685 &  2.444 & -2.444  &    0.000 \\
$Z_{\kappa yz}$ &   -    &   -     &   -     &   -     &    -    &  -      &    -    &     -    &    0.011 \\
$Z_{\kappa zy}$ &   -    &   -     &   -     &  -      &     -   &   -     &    -    &    -     &    0.351 \\
$Z_{\kappa zz}$ & 0.343 & -0.172 & -0.070 &  0.035 &  0.246 & -0.246 &  0.169 & -0.169  &    0.000 \\
$Q_{\kappa xxy}$ &   -    &  3.699 & -5.533 & -0.391 &  4.261 &  0.384 & -7.158 & -1.547  &  -1.732 \\
$Q_{\kappa xxz}$ &   -    &   -     &  -      &    -    &  -      &    -    & -1.042 &  0.450  &   0.206 \\
$Q_{\kappa yxx}$ &   -    &  3.699 & -5.533 & -0.391 &  4.261 &  0.384 & -7.158 & -1.547  & -15.68 \\
$Q_{\kappa yyy}$ &   -    & -3.699 &  5.533 &  0.391 & -4.261 & -0.384 &  7.158 &  1.547  &  -2.256 \\
$Q_{\kappa yzz}$ &   -    &   -     &  -      &    -    &   -     &   -     &     -   &   -      &   0.238 \\
$Q_{\kappa yyz}$ &   -    & -0.298 &   -     & -0.174 &     -   &     -   & -1.042 &  0.450  &   0.213 \\
$Q_{\kappa zxx}$ &   -    & -2.932 &   -     &  7.858 &     -   &     -   & -7.116 & -1.167  &  -8.970 \\
$Q_{\kappa zyy}$ &   -    &   -     &  -      &    -    &   -     &   -     & -7.116 & -1.167  &   1.448 \\
$Q_{\kappa zzz}$ &   -    &  0.230 &    -    & -0.297 &     -   &     -   & -0.369 &  0.395  &   0.341 \\
$Q_{\kappa zyz}$ &   -    &   -     &   -     &    -    &    -    &   -     &    -    &    -     &   0.429 \\
$\varepsilon_{xx}^{\parallel}$ & \multicolumn{2}{c}{3.079} & \multicolumn{2}{c}{6.105} & \multicolumn{2}{c}{1.591} & \multicolumn{2}{c}{3.777} & 3.838\\
$\varepsilon_{yy}^{\parallel}$ & \multicolumn{2}{c}{3.079} & \multicolumn{2}{c}{6.105} & \multicolumn{2}{c}{1.591} & \multicolumn{2}{c}{3.777} & 4.898 \\
$\varepsilon^{\perp}$     & \multicolumn{2}{c}{1.226} & \multicolumn{2}{c}{1.299} & \multicolumn{2}{c}{1.098} & \multicolumn{2}{c}{1.342} & 1.215 \\
$\alpha_{xx}^{\parallel}$      & \multicolumn{2}{c}{6.632} & \multicolumn{2}{c}{13.050}& \multicolumn{2}{c}{1.881} & \multicolumn{2}{c}{9.043} &  9.036 \\
$\alpha_{yy}^{\parallel}$      & \multicolumn{2}{c}{6.632} & \multicolumn{2}{c}{13.050}& \multicolumn{2}{c}{1.881} & \multicolumn{2}{c}{9.043} & 12.40 \\
$\alpha^{\perp}$          & \multicolumn{2}{c}{0.720} & \multicolumn{2}{c}{0.765} & \multicolumn{2}{c}{0.310} & \multicolumn{2}{c}{1.088} & 0.683 \\
$L$  & \multicolumn{2}{c}{9.0} & \multicolumn{2}{c}{10.5} &  \multicolumn{2}{c}{5.1} &  \multicolumn{2}{c}{9.3} &  28.0 \\
  \botrule
  \end{tabular}
\end{footnotesize}
\end{table}

\begin{figure*}[t]
  \centering
  \includegraphics[width=0.85\linewidth]{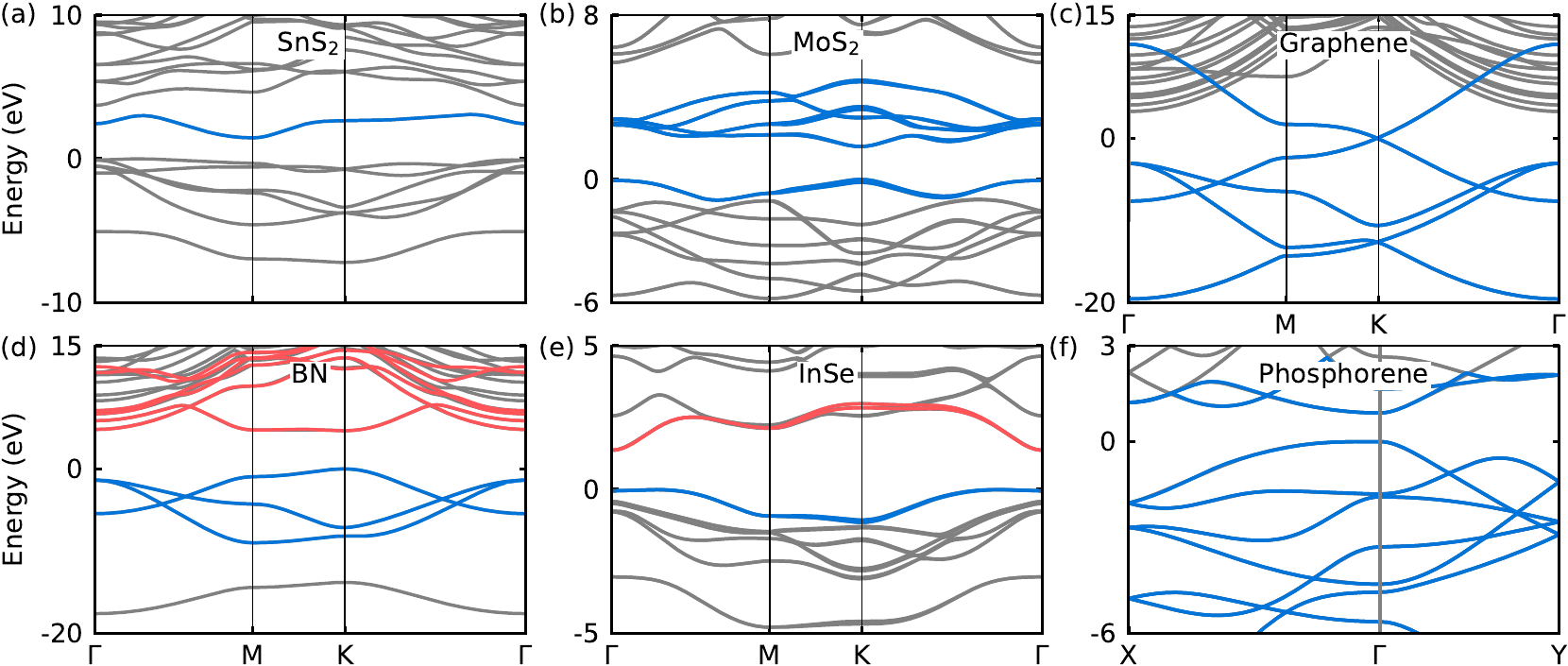}
  \caption{\label{fig:BS}
Comparison between the electronic band structure of SnS$_2$, MoS$_2$, graphene, hexagonal BN, InSe, and phosphorene
between a direct density functional theory calculation (gray lines) and the Wannier interpolated bands along high-symmetry directions.
In some case, we use a different set of Wannier functions for the valence (blue) and conduction (red) bands.
The zero in the energy axis is aligned to be at the valence band maximum.
  }
\end{figure*}

We perform all calculations within density functional theory (DFT) and density functional perturbation theory (DFPT)~\cite{Gonze1997a,Baroni2001} using the \textsc{Quantum Espresso} (\textsc{QE})~\cite{Giannozzi2009,Giannozzi2017}  and \textsc{Abinit}~\cite{Gonze2016,Royo2019,Romero2020,Gonze2020} suites of codes, with plane-wave basis sets and pseudopotentials to include the effects of core electrons.
In QE we introduce a cutoff on Coulomb interactions~\cite{Sohier2017} to prevent spurious interactions with artificial periodic replicas of the monolayers in the vertical direction.
The macroscopic dielectric tensor and dynamical Born effective charges are computed within \textsc{QE}, while the dynamical quadrupoles are computed using the linear response implementation in \textsc{Abinit}.
In the latter case, the same computational parameters as in \textsc{QE} are considered, including the same pseudopotentials, although without non-linear core corrections and without spin-orbit coupling.
All resulting materials parameters are summarized in Table~\ref{table1}.
First principles results for electronic eigenvalues, phonon frequencies, and electron-phonon matrix elements are interpolated on ultra dense Brillouin-zone grids using a generalized Wannier-Fourier approach~\cite{Giustino2007,Calandra2010} using  \textsc{EPW}~\cite{Ponce2016a} and \textsc{Wannier90}~\cite{Pizzi2020}.
Details about the Wannier functions used in this study for interpolation as well as interpolated electronic band structures are presented in Sec.~\ref{sec:wannier}, while here we provide a brief description of the systems investigated and the corresponding parameters adopted in the \textsc{QE} simulations.

SnS$_2$ crystallizes in the trigonal P$\bar{3}$m1 [164] space group with point group $\bar{3}$m.
For better comparison with Ref.~\onlinecite{Royo2021}, we use the same lattice parameter of 6.837~bohr with 40~bohr of vacuum and the two sulphur atoms positioned 2.774~bohr away from the Sn layer.
We also use the same norm-conserving pseudopotential~\cite{Hamann2013} from \textsc{PseudoDojo}~\cite{Setten2018} within the local density approximation (LDA)~\cite{Perdew1992}.
A plane-wave cutoff of 160~Ry and a 16$\times$16$\times$1 \textbf{k}-point grid is adopted.
DFPT calculations are performed on a 16$\times$16$\times$1 \textbf{q}-point with a tight 10$^{-24}$ threshold on the perturbed wavefunction.
The resulting in-plane and out-of-plane dielectric tensor and Born effective charges are given in Table~\ref{table1}.
As in the original publication\cite{Royo2021}, we neglect spin-orbit coupling (SOC) in SnS$_2$.

\begin{figure*}[ht]
  \centering
  \includegraphics[width=0.75\linewidth]{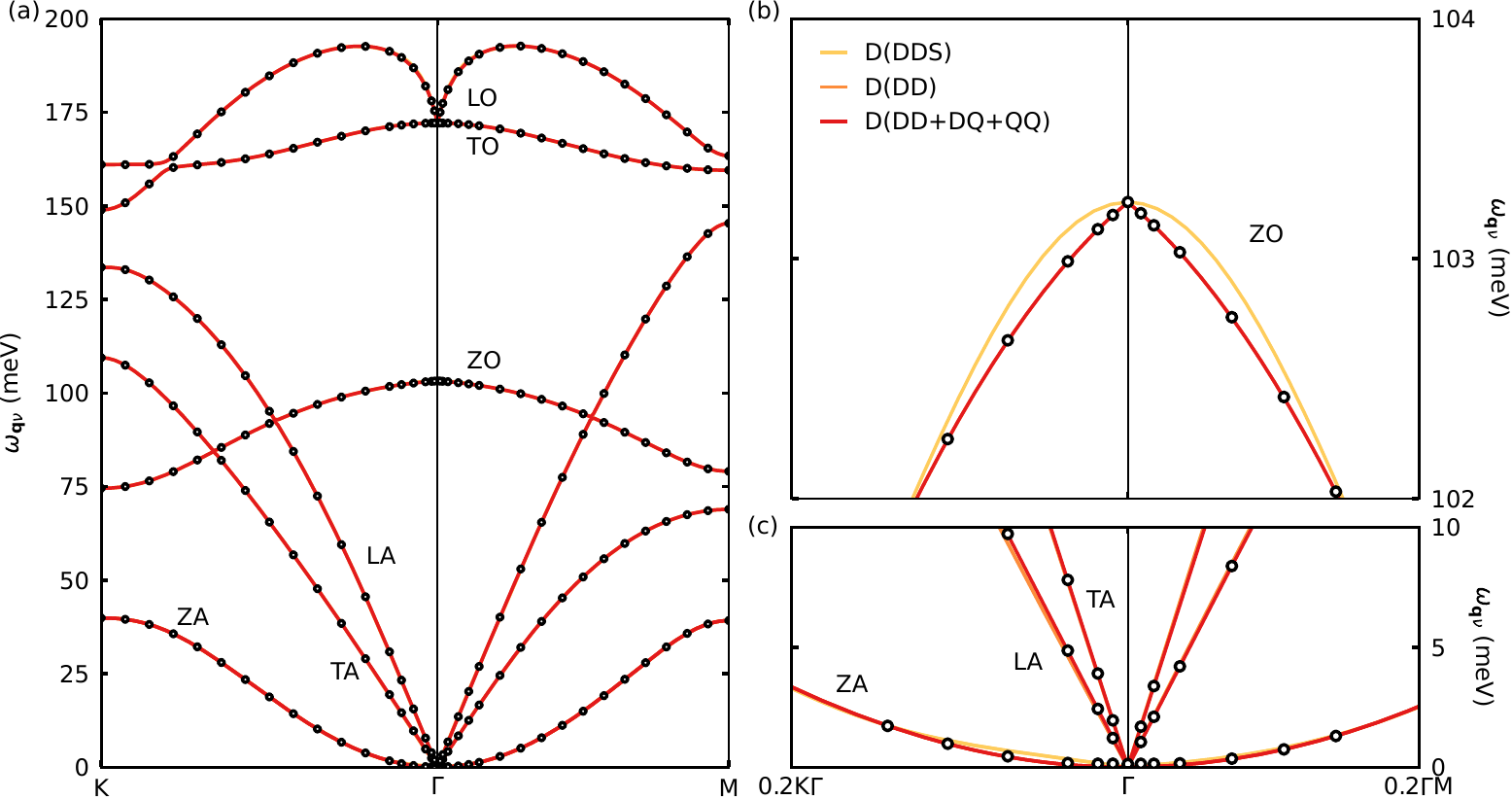} %BN/phonon5.pdf
  \caption{\label{fig:BN-phonon}
 Phonon dispersion of BN monolayer calculated with direct density functional perturbation theory calculations (black empty circles) compared with Fourier interpolation where the long-range part of the dynamical matrix (D) includes dipole-dipole with the scheme of Ref.~\onlinecite{Sohier2016,Sohier2017a} (DDS, , Eq.~\eqref{eq:dynmatlong2D-sohier}), dipole-dipole (DD, Eq.~\eqref{eq:dynmatlong-dip}), dipole-quadrupole (DQ, Eq.~\eqref{eq:dynmatlong-dipquad}) and quadrupole-quadrupole (QQ, Eq.~\eqref{eq:dynmatlong-quadquad}). The subfigures (b) and (c) are energy and momentum magnification.
}
\end{figure*}

We consider also  monolayer hexagonal BN (h-BN), both to compare our results with the phonon frequencies of Ref.~\onlinecite{Royo2021} but
also because h-BN has attracted much attention in recent years due to its high dielectric constant, wide bandgap, chemical intertness, flexibility and good mechanical strength~\cite{Liu2019}.
It is also seen as one of the best dielectric interface material for novel electronics.
We use the same lattice parameter as Ref.~\onlinecite{Royo2021} of 4.689~bohr with 40~bohr vacuum and the same scalar relativistic norm conserving LDA pseudopotential without SOC.
We choose a 160~Ry plane-wave energy cutoff with a 16$\times$16$\times$1 \textbf{k}-point and \textbf{q}-point grids and a tight 10$^{-20}$ threshold on the perturbed wavefunction.

Next, we study the prototypical TMD monolayer MoS$_2$ for its technological relevance and because many theoretical and experimental data exist for this material.
MoS$_2$ is a piezoelectric material with an experimental relaxed-ion piezoelectric coefficients of 2.9~10$^{-10}$~C/m~\cite{Zhu2015a}.
The primitive cell contains three atoms with broken inversion symmetry and space group P$\bar{6}$m2.
In crystal coordinates, the Mo atom occupies the [1/3,2/3,0] position while the two S atoms have [2/3,1/3,$\pm z$] coordinate.
We use fully relativistic norm-conserving Perdew-Burke-Ernzerhof (PBE)~\cite{Perdew1996} pseudopotentials that allow to introduce  SOC effects self-consistently in the calculations and that have been generated using the \textsc{ONCVPSP} code~\cite{Hamann2013} and optimized via the \textsc{PseudoDojo} initiative~\cite{Setten2018}, taking the $4s^2$, $4p^6$, $4d^5$, $5s^1$ as valence states for Mo and $3s^2$, $3p^4$ as valence states for S.
The electron wave functions are expanded in a plane-wave basis set with kinetic energy cutoff of 140~Ry.
We perform response calculations using DFPT on a 18$\times$18$\times$1 electron and 18$\times$18$\times$1 phonon grids to ensure good convergence of the dielectric properties.
After structural relaxation, we obtain a lattice parameter of 6.020~bohr with a 5.907~bohr atomic distance between the two sulfur atoms in the out-of plane direction with a direct DFT bandgap at $K$ of 1.60~eV and large spin-orbit splitting of 148~meV of the valence band maximum, while the conduction band minimum has a much smaller 3~meV splitting.
The momentum-averaged electron and hole effective masses at the band edges are 0.42~$m_e$ and 0.52~$m_e$, respectively.
These values are in agreement with previous theoretical and experimental works~\cite{Zhang2014,Molina-Sanchez2015}.

We study also monolayer  InSe, which is a piezoelectric material with a calculated piezoelectric coefficient of 0.57~10$^{-10}$~C/m~\cite{Li2015a}.
Due to its high carrier mobility~\cite{Sucharitakul2015,Bandurin2017,Ho2017,Li2018}, it is considered a good candidate for post-silicon electronics.
Also in this case, we use fully relativistic norm-conserving PBE pseudopotentials~\cite{Setten2018}, which include  $4d^{10}$, $5s^2$, and $5p^1$ as valence states for In and $3d^{10}$, $4s^2$, and $4p^4$ as valence states for Se.
The electron wave functions are expanded in a plane-wave basis set with kinetic energy cutoff of 160~Ry.
For the response calculations, we use a 16$\times$16$\times$1 electron and 16$\times$16$\times$1 phonon grids to ensure good convergence of the dielectric properties.
After structural optimization, we obtain a lattice parameter of 7.721~bohr, in close agreement with prior studies that found values ranging from 7.46 to 7.728~bohr~\cite{Sun2016,Hu2017,Peng2017a,Wang2019a,Gopalan2019}.
We also find an In-In bond length of 5.333~bohr and an In-Se bond length of 4.978~bohr.

We also examine graphene, being the seminal non-polar 2D material~\cite{Geim2007}, whose transport properties are well studied~\cite{Peres2010}.
It therefore serves as a reference test-case to ensure our scheme works also in (semi) metals, where screening plays a relevant role but the presence of long range contributions cannot be discarded a priori.
In contrast to Ref.~\onlinecite{Macheda2020} that focuses on doped graphene, here we compute the \textit{intrinsic} carrier mobility, i.e. assuming the Fermi level to be at the Dirac point, with electrons and holes being purely generated by thermal effects and not by doping.
We use a relativistic norm-conserving PBE pseudopotential with a 100~Ry plane-wave energy cutoff, a cold smearing~\cite{Marzari1999} of 7.5~mRy, and a dense $96\times96\times1$ $\mathbf{k}$-point grid coupled with a $18\times18\times1$ \textbf{q}-point grid for the response calculations.
The computed relaxed lattice parameter is 4.661~bohr, close to the experimental one of 4.648~bohr~\cite{Yang2018}.

The last material that we investigate in this work is phosphorene, an elemental group-V 2D material, which displays a buckled orthorhombic structure with  4 atoms per unit cell and space group Pmna, a direct band gap promising for optoelectronic applications~\cite{Xia2014}, and large mobility and on/off ratios in field effect transistors~\cite{Liu2014,Li2014,Long2016,Long2016b}.
We adopt a relativistic norm-conserving PBE pseudopotential with a 160~Ry plane-wave energy cutoff and a $16\times16\times1$ $\mathbf{k}$-point grid coupled with a $16\times16\times1$ \textbf{q}-point grid for the response calculations.
A denser $32 \times 24 \times 2$ \textbf{k}-point grid is considered for the \textsc{Abinit} dynamical quadrupole calculation.
The calculated lattice parameters of phosphorene are $a = 6.242$~bohr and $b = 8.741$~bohr and an out-of-plane buckling distance of 3.986~bohr.

Finally, we report for each material the choice of initial projections for the Wannier functions in Appendix~\ref{sec:wannier}
and show in Fig.~\ref{fig:BS} that interpolated electronic band structures reproduces perfectly the direct DFT calculations.

\begin{figure*}[ht]
  \centering
  \includegraphics[width=0.99\linewidth]{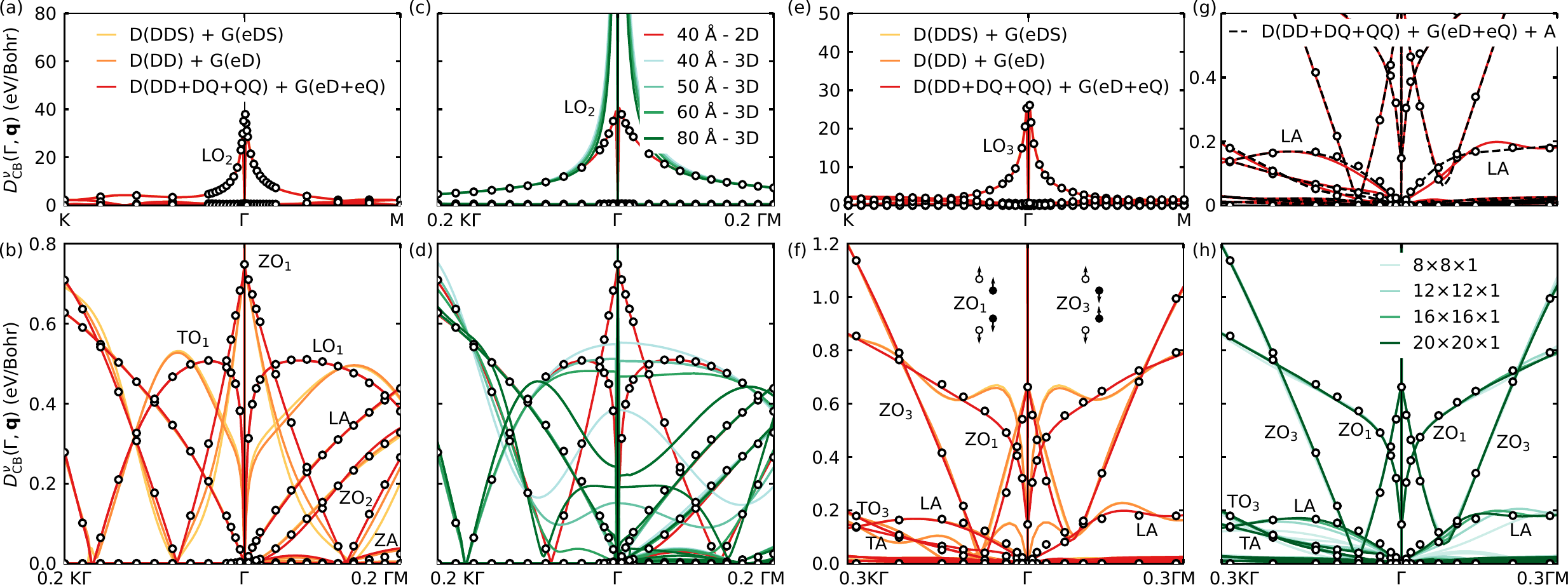}
  \caption{\label{fig:Sns2-defo}
(a) Deformation potential showing all phonon modes along high-symmetry lines and computed at $\mathbf{k}=\Gamma$ for the conduction band of SnS$_2$ monolayer calculated with direct density functional perturbation theory (black empty circles) compared with Fourier interpolation where the long-range part of the dynamical matrix (D)
and electron-phonon matrix elements (G) includes dipole-dipole (DDS) and monopole-dipole (eDS) with the scheme of Ref.~\onlinecite{Sohier2016}, dipole-dipole (DD), dipole-quadrupole (DQ), quadrupole-quadrupole (QQ), monopole-dipole (eD, Eqs.~\eqref{eq:final_formula2},~\eqref{eq:glong-final2D} and \eqref{eq:vlr_lw} with $\mathbf{Q}=0$), and monopole-quadrupole (eQ,  Eqs.~\eqref{eq:final_formula2},~\eqref{eq:glong-final2D} and \eqref{eq:vlr_lw} with $\mathbf{Z}=0$), all with a 40~\AA~vacuum.
(b) Momentum and deformation potential range zoom of (a).
(c) Comparison of the deformation potential between a 3D long-range scheme with quadrupoles for vacuum sizes ranging
from 40 to 80~\AA~with the deformation potential with the 2D long-range scheme and dynamical quadrupoles.
(d) Deformation potential range zoom of (c).
The subfigures (e-h) show the deformation potential of InSe where
(f) is a momentum and deformation potential range zoom of (e); 
(g) is a deformation potential range zoom of (f) where the results with quadrupoles are compared with and without the Berry connection term $A$; and
(h) is a comparison of the deformation potential in the quadrupole case for different coarse \textbf{k}-point and \textbf{q}-point grids with quadrupoles but without the $A$ term.
}
\end{figure*}

\subsection{Phonon dispersion}

We now analyze the effect of the out-of-plane electrostatics, dipoles and quadrupoles on the phonon dispersion of all the materials along high-symmetry lines of the Brillouin zone.
We compare in Fig.~\ref{fig:BN-phonon} the phonon dispersion of BN monolayer calculated with direct DFPT (black empty circles) with a Fourier interpolation where the long range dynamical matrix includes either dipoles only D(DD), Eq.~\eqref{eq:dynmatlong-dip}, or dipole-dipole, dipole-quadrupole, and quadrupole-quadrupole contributions D(DD+DQ+QQ), Eqs.~\eqref{eq:dynmatlong-dip}-\eqref{eq:dynmatlong-quadquad}.
We also show the results obtained using the 2D approach for dipole-dipole interactions of Ref.~\onlinecite{Sohier2017a}, D(DDS) using Eq.~\eqref{eq:dynmatlong2D-sohier}.
In the case of BN, the phonon frequencies with or without dynamical quadrupoles are numerically the same.
This was also observed in the case of many bulk materials~\cite{Ponce2021} but we stress that even in these cases, the eigenvectors associated with these phonon modes are not the same, yielding for example different deformation potentials.

However this finding is not general.
For example in the case of the TO$_1$ and TO$_2$ modes of SnS$_2$, a clear difference is observed between the phonon frequencies with and without quadrupoles around the zone center, as shown in Fig.~\ref{fig:Sns2-phonon} in  App.~\ref{app:phonon}.
A similar observation can be made for InSe in Fig.~\ref{fig:Sns2-phonon}, where the TA, LA, and LO$_3$ modes show small differences.
This has been observed before~\cite{Royo2021} and confirmed here.
We also mention that in piezoelectric bulk materials, the quality of the phonon interpolation can be improved by using quadrupoles~\cite{Royo2020}.

Coming back to the case of BN, we observe an important difference between the DDS approximation and the DD case for the ZO branch interpolation close to $\Gamma$.
While the DDS solution approaches the zone center quadratically, the DFPT results display linearity and a derivative discontinuity at $\Gamma$.
This behaviour of the DDS approach is due to the neglect of the out-of-plane electrostatics given by the second term in Eq.~\eqref{eq:dynmatlong-dip}.
Interestingly, in the case of MoS$_2$ and phosphorene shown in Figs.~\ref{fig:Sns2-phonon} and \ref{fig:p-phonon} there are no visible differences in the phonon dispersion between the three approaches.
We nonetheless mention that a 3D long-range scheme would instead fail~\cite{Sohier2017a}.

Finally, the case of the phonon dispersion of graphene is shown in Fig.~\ref{fig:p-phonon} of  App.~\ref{app:phonon} where we do not use any long-range treatment due to the semi-metallic nature of graphene.
The agreement with direct DFPT is excellent and the parabolicity of the flexural ZA mode is recovered.

Overall, we can thus conclude that, although some differences can be observed between DFPT results and interpolated phonon dispersions without quadrupole effects, they are always quite small.
This is however not the case for the scattering potential where much bigger discrepancies are found and the effect of dynamical quadrupoles becomes crucial.

\begin{figure*}[ht]
  \centering
  \includegraphics[width=0.90\linewidth]{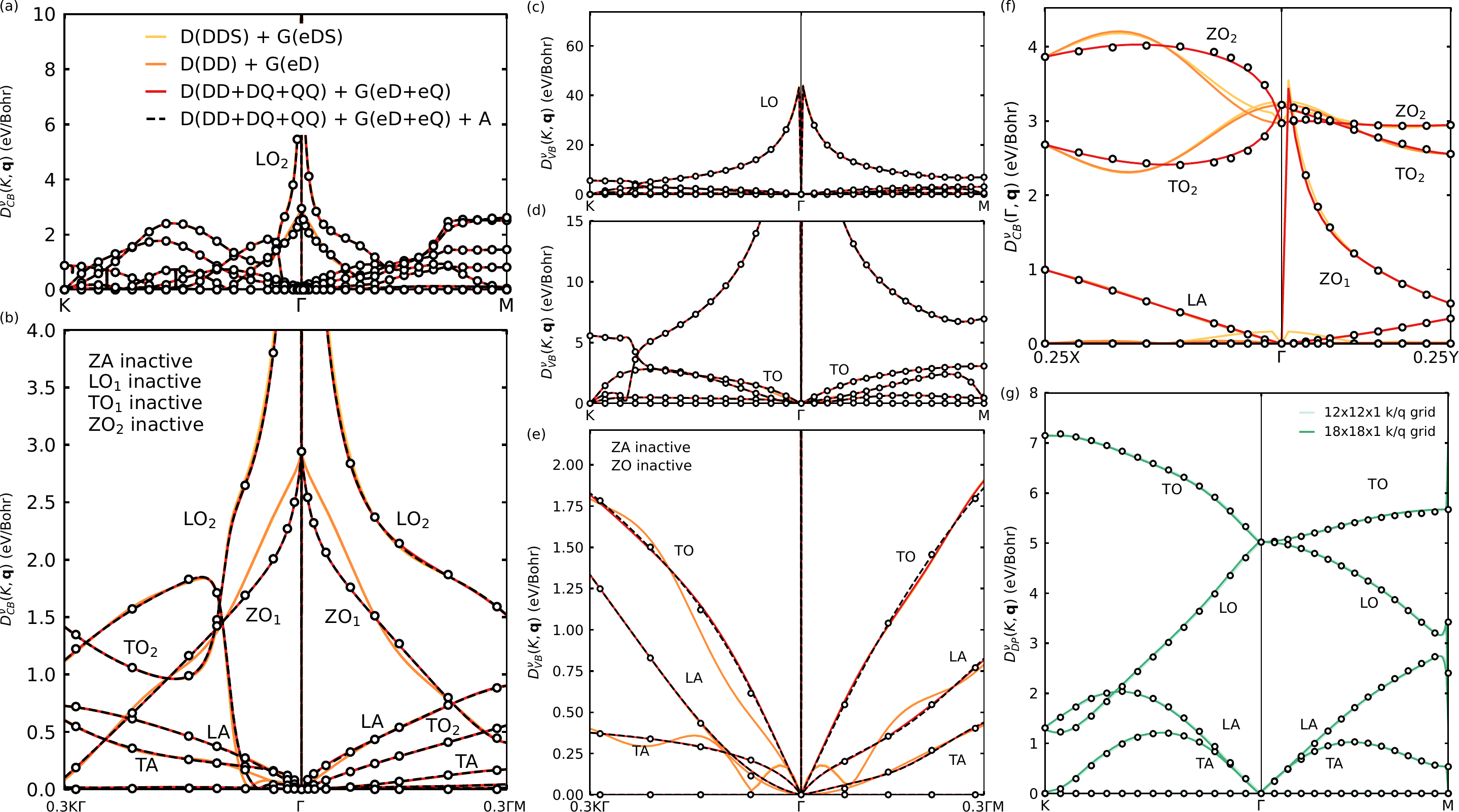}
  \caption{\label{fig:def-SI}
(a,b) Comparison of the deformation potential of the conduction band of MoS$_2$ monolayer at the \textbf{k} = \textbf{K} point with the three schemes to treat the long-range interaction during interpolation, as well as with the Berry connection term (dashed black line), compared with direct DFPT calculation (black empty circles).
(c,d,e) Deformation potential of the valence bands of h-BN at the \textbf{k} = \textbf{K} point for the same comparison as in (a,b).
(f) Deformation potential of the conduction bands of phosphorene at the \textbf{k} = $\boldsymbol{\Gamma}$ point.
(g) Deformation potential of graphene at the Dirac point (DP) with no long-range treatment for two coarse grids compared with DFPT references (black empty circles).
}
\end{figure*}

\subsection{Deformation potential}\label{sec:deform}

We now turn to the assessment of the accuracy of the various interpolation schemes for the scattering potential presented in this work.
For an easier comparison of the electron-phonon coupling, we compute the total deformation potential~\cite{Zollner1990,Sjakste2015}:
\begin{equation}\label{eq:deformation}
\! D^{\nu}(\mathbf{k}, \mathbf{q}) =  \frac{1}{\hbar N_{\rm w}}\bigg[ 2 \rho S  \hbar\omega_{\nu}(\mathbf{q}) \! \sum_{mn}  |g_{mn\nu}(\mathbf{k},\mathbf{q})|^2  \bigg]^{1/2}\!\!\!,
\end{equation}
where the sum over bands is carried over the $N_{\rm w}$ states of the Wannier manifold (or a subset), and $\rho$ is the mass density of the crystal.
Eq.~\eqref{eq:deformation} has the advantage to factor out the contribution from the phonon frequency $\omega_{\nu}(\mathbf{q})$ and to sum multiple electronic bands at once.

\begin{figure*}[t]
  \centering
  \includegraphics[width=0.9\linewidth]{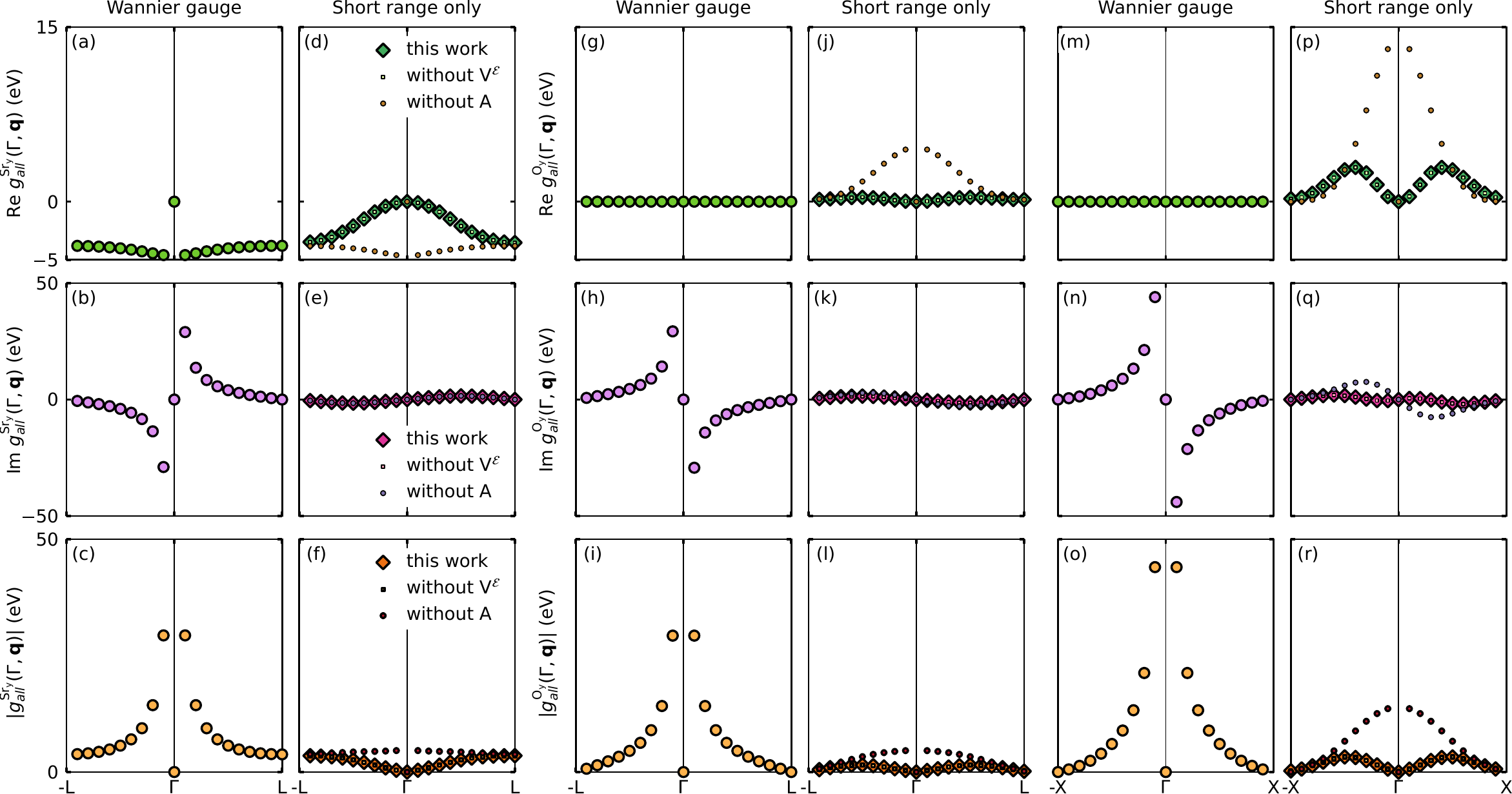}
  \caption{\label{fig:SrO}
Real, imaginary and absolute value of the electron-phonon matrix element $g$ as a function of phonon momentum $\mathbf{q}$ of bulk SrO 
where we perform a sum over all Wannier functions in the calculation which describe the valence band at the $\mathbf{k}$=$\Gamma$ point and for the displacement of the Sr atom (a-f), O atom (g-l) in the $y$ Cartesian direction, O atom (m-r) in the direction of the $X$ point, respectively.
The other displacements are null along the $\Gamma L$ line.
(a-c, g-i) Direct DFPT results in the smooth Wannier gauge as well as the short range component (d-f, j-l) with
the new gauge restoring term $A$ and the local fields response to an electric field $V^{\mathcal{E}}$ as well as without each term separately.
}
\end{figure*}

We start by presenting in Fig.~\ref{fig:Sns2-defo} the deformation potential of the conduction band of SnS$_2$ where the initial state is located at $\mathbf{k}=\Gamma$ and the final state spans high-symmetry directions.
The first striking feature is that, differently to what typically happens in 3D systems, the coupling to the longitudinal optical LO$_2$ mode, also called Fr\"ohlich mode, goes to a finite value with a cusp in the long wavelength ($\mathbf{q}\rightarrow \mathbf{0}$) limit~\cite{Sohier2016}.
The finite value  can be computed analytically from the knowledge of the Born effective charges and macroscopic dielectric function only by considering the $\mathbf{q}\rightarrow \mathbf{0}$ limit of Eq.~\eqref{eq:vlr_lw} without quadrupole.
We verify that this limit is the same as the dipole approximation of Eq.~\eqref{eq:lr_sohier}.
This is a general feature of all 2D polar materials.

In contrast, in the case of bulk materials the polar Fr\"ohlich interaction diverges as $1/|\mathbf{q}|$ in the long wavelength limit as can be seen with the dipole term of Eq.~\eqref{eq:glong-final3D}.
It has been suggested in the past~\cite{Li2019} that a good approximation to the 2D deformation potential could be obtained by
using the 3D long-range formulation and increasing the vacuum space. %or densifying the out-of-plane $\mathbf{k}$-point grid.
However, as seen in Fig.~\ref{fig:Sns2-defo}(c) and (d), the deformation potential converges slowly with the vacuum distance making it a difficult approach in practice.

If we now focus on the low region of the deformation potential close to the zone center of SnS$_2$, Fig.~\ref{fig:Sns2-defo}(b), we can see that the D(DDS)+G(eDS) in-plane dipole of Eqs.~\eqref{eq:dynmatlong2D-sohier} and \eqref{eq:lr_sohier} is
a good approximation to the D(DD)+G(eD) dipole approach of Eqs.~\eqref{eq:dynmatlong-dip} and \eqref{eq:glong} with $\mathbf{Q}=0$.
Still, both approaches fail to reproduce correctly the DFTP results, especially for the TO$_1$ and ZO$_1$ modes, both quantitatively and qualitatively.  Adding the contribution of quadrupoles through Eqs.~\eqref{eq:dynmatlong-dipquad}, \eqref{eq:dynmatlong-quadquad}, and  ~\eqref{eq:glong} allows to recover the first principles results, showing that including dynamical quadrupoles is therefore crucial to accurately describe the scattering potential in SnS$_2$ monolayer.

Another case in which the quadrupoles have a significant impact is InSe monolayer, shown in Fig.~\ref{fig:Sns2-defo}(e-h).
In InSe, there are two mirror-even A$^\prime$ out-of-plane modes, the ZO$_1$ and ZO$_3$ modes, which are depicted in a schematic way in Fig.~\ref{fig:Sns2-defo}(f).
Both contribute to the deformation potential and are strongly affected by quadrupole corrections.
In contrast the ZO$_2$ mode is mirror-odd with respect to the $z=0$ plane and of symmetry A$^{\prime\prime}$  which means that its contribution to the deformation potential is forbidden in this case.
However, even with quadrupoles the LA mode still oscillates significantly.
In Fig.~\ref{fig:Sns2-defo}(g) we show that including the new Berry connection term $\mathbf{A}_{sp\mathbf{k}}^{\rm W}$ yields a much better interpolation of the LA mode.
and that correct interpolation cannot be achieved without it by brute force coarse grid convergence as shown in Fig.~\ref{fig:Sns2-defo}(h).
Also here, we verified that the $V^{\text{Hxc},\mathcal{E}}(\mathbf{r})$ term in Eq.~\eqref{eq:glong-final2D} was negligible.

The deformation potentials for MoS$_2$, h-BN, phosphorene, and graphene are instead reported in Fig.~\ref{fig:def-SI}.
In the case of MoS$_2$ shown in Fig.~\ref{fig:def-SI}(a,b), only the ZO$_1$ and LO$_2$ modes strongly couple to the electrons in the long-wavelength limit.
For the ZO$_1$ mode, the DFPT coupling approaches a constant value at $\Gamma$ linearly in \textbf{q} with a significant quadratic component that becomes dominant already at small wave vectors. When only dipole contributions are retained, the interpolated deformation potential is purely linear for $\mathbf{q}\to0$, and the inclusion of quadrupoles is essential to recover the quadratic correction.
We also notice a small improvement in the LA mode when quadrupoles are added.
Interestingly, in the case of h-BN in Fig.~\ref{fig:def-SI}(c-e), the inclusion of 2D quadrupoles has little effect because the out-of-plane mode is forbidden by symmetry.
As for the case of InSe, we observe a slow convergence of the LA mode with coarse grids density in BN.
Moreover, in Fig.~\ref{fig:def-SI}(f) we show the deformation potential of phosphorene.
This is an interesting case because it has no polar Fr\"ohlich component since the in-plane Born effective charges vanish, although  the out-of-plane Born effective charge is 0.351 as reported in Table~\ref{table1}.
Remarkably, phosphorene has instead very large quadrupoles that strongly impact the scattering potential, especially for the ZO$_1$, ZO$_2$, and TO$_2$ modes that are finite in the long-wavelength limit. We note that the coupling to the ZO$_1$ mode is suppressed by symmetry along the $\Gamma-$X high-symmetry direction.
We report the case of graphene in Fig.~\ref{fig:def-SI}(g), which is semi-metallic and calculations do not include any long-range treatment.
We find that the interpolation of the deformation potential is excellent without long-range treatment and therefore proceeds as such.

\subsection{The 3D case}\label{sec:deform3D}

%%%%%%%%%%%%%%%%%%%%%%%%%%%%%%%%%%%%%%%%%%%%%%%
% SrO

\begin{figure}[t]
  \centering
  \includegraphics[width=0.99\linewidth]{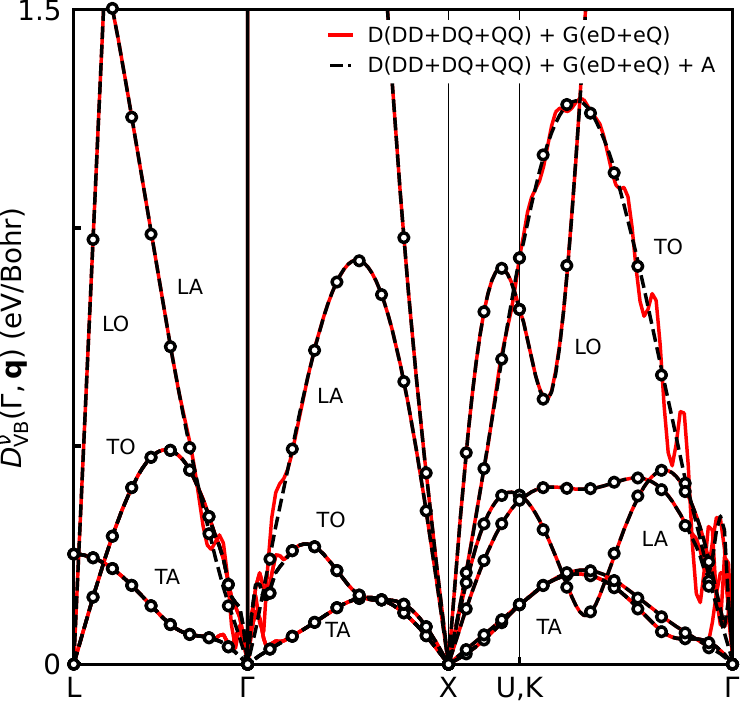} %SrO/deformation-noSOC4.py
  \caption{\label{fig:SrOdefo}
Comparison of the deformation potential of the valence band of SrO at the \textbf{k} = $\boldsymbol{\Gamma}$ point with the long-range interaction including or not the Berry connection term  $\mathbf{A}_{sp\mathbf{k}}^{\rm W}$ during interpolation, and compared with direct DFPT calculation (black empty circles).
}
\end{figure}

Finally, we decide to revisit the case of bulk SrO that we studied in Ref.~\onlinecite{Ponce2021}.
Indeed, the symmetry in SrO yields a null dynamical quadrupole tensor such that the poor interpolation quality in Ref.~\onlinecite{Ponce2021} 
could not be improved by including multipolar terms beyond dipoles and should therefore be an ideal platform to test the importance of the new Berry connection term.
We performed the same calculation as in Ref.~\onlinecite{Ponce2021} but without SOC and with a 20$\times$20$\times$20 coarse \textbf{k}-point and \textbf{q}-point grids.
We here only consider the Wannierziation of the highest three valence bands with three Wannier functions of $p$ character and centred around the oxygen atom. 
In Fig.~\ref{fig:SrO}(a-c,g-i,m-o) we present the direct DFPT calculation of all the zone-centered non-zero electron-phonon matrix elements along a high-symmetry line, rotated in the smooth Wannier gauge and summed over all the Wannier functions.
The real part is discontinuous at \textbf{q}=$\Gamma$ and the imaginary part diverges, preventing accurate interpolation. 
To overcome this problem, we remove the long-range part using Eq.~\eqref{eq:glong-final2D} and compare the resulting short-range solution with and without the Berry connection $\mathbf{A}_{sp\mathbf{k}}^{\rm W}$ and  the $V^{\text{Hxc}, \boldsymbol{\mathcal{E}}}$ terms.
We find the latter to be negligible while the Berry connection term is crucial for a smooth real part of the matrix elements. 
Importantly, and in analogy with the 2D case, we tune the range separation parameter $L^2$ in Eq.~\eqref{eq:bareCoulomb3DGaussian} and we determine that a value of $L^2 = 4$~bohr$^2$ is optimal. 
In Fig.~\ref{fig:SrO} we can see that the absolute value of the matrix elements are the same for strontium and oxygen displacements while their imaginary parts are almost opposite. 
Interestingly, we find that neglecting the Berry connection term has a larger impact on the imaginary part of the short-range term in the case of oxygen, Fig.~\ref{fig:SrO}(k), than strontium, Fig.~\ref{fig:SrO}(e).
We also find that along the $\Gamma$-$X$ high symmetry direction, Fig.~\ref{fig:SrO}(m-r), the short range matrix elements are compressed closer to the zone center due to the finite mesh.
We therefore expect a lower interpolation quality along that direction that can be improved with a denser coarse grid or a gauge restoration to higher order in \textbf{q}.

\begin{figure}[t]
  \centering
  \includegraphics[width=0.99\linewidth]{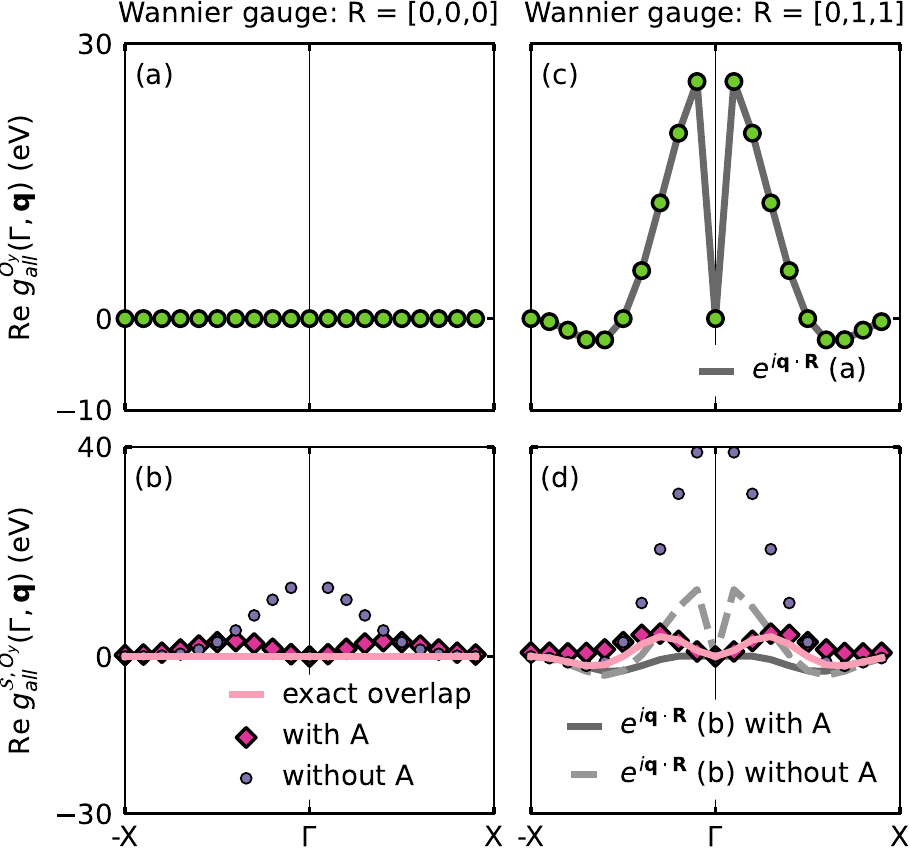} 
  \caption{\label{fig:SrOgauge}
Real part of the electron-phonon matrix element $g$ as a function of phonon momentum $\mathbf{q}$ of bulk SrO 
where we have performed a sum over all Wannier functions in the calculation which describe the valence band at the $\mathbf{k}$=$\Gamma$ point and for the displacement of the oxygen in the $y$ Cartesian direction.
We show in (a) the direct DFPT results in the smooth Wannier gauge where the Wannier centers are located in the primitive cell as well as (b) the
short range component $g^{\mathcal{S}}$  with and without the new Berry connection term $A$ as well as an exact overlap reference using Eq.~\eqref{eq:exact}.
In (c-d), we show the same results but where all the Wannier centers are located one lattice site away in the $\mathbf{X}$ direction.
In both (c) and (d) we show gauge covariance by multiplying (a) and (b) with the corresponding gauge transformation.  
}
\end{figure} 

Importantly, the Berry connection term has two effects: (i) it improves the interpolation quality by removing long-range effects at quadrupolar order, see Eqs.~\eqref{eq:long-range-new}-\eqref{eq:long-range-new2}, and (ii) it restores gauge independence to lowest order in \textbf{q}.

We first show point (i) in Fig.~\ref{fig:SrOdefo} where the improvement in interpolation quality resulting from the use of the Berry connection term is striking. 
The spurious oscillations are removed and the interpolation quality systematically improves and is excellent for all modes except for the TO modes close to the zone center where a smooth overestimation is observed.
We attribute this overestimation to the sharply varying short range shown in Fig.~\ref{fig:SrO}(m-r) in the $X$ and $K$ directions. 
Regardless, the improvement is clear and showcases the importance and applicability of our findings about the Berry connection term in 2D and 3D materials.

To highlight the point (ii), we show in Fig.~\ref{fig:SrOgauge} the real part of the electron-phonon matrix element $g$ along the $X$ $\mathbf{q}$-point direction where 
we perform a sum over all Wannier functions in the calculation which describes the valence band at the $\mathbf{k}$=$\Gamma$ point and for the displacement of the oxygen atom in the $y$ Cartesian direction.  
Explicitly, the green dots in Figs.~\ref{fig:SrOgauge}(a,c) are obtained via direct DFPT calculations of the real part of the electron-phonon matrix elements in the Wannier basis while the dots and diamonds in Figs.~\ref{fig:SrOgauge}(b,d) are the corresponding short-range components with and without the new Berry connection term $A$, see Eqs.~(1) and (2) of the supplemental information of our companion manuscript for more details. 
We show in Fig.~\ref{fig:SrOgauge} that the Berry connection term ensures smoothness of the real part of the electron-phonon matrix element which is also preserved if we shift all the Wannier centers by a lattice vector. 
Moreover it preserves, to lowest order in $\mathbf{q}$, the gauge covariance when the Wannier centers are translated. 
The gauge covariance is demonstrated by multiplying the short range obtained with and without $\mathbf{A}_{sp\mathbf{k}}^{\rm W}$  by the gauge transformation
$W_{sp\mathbf{k}} = e^{-i\mathbf{k}\cdot \mathbf{R}}$ and shown in Fig.~\ref{fig:SrOgauge}(d) with dashed and plain gray lines, respectively. 
As can be seen, only the case with the Berry connection recovers the results obtained by the gauge transformation, validating gauge covariance for $\mathbf{q}\to0$. 
Even stronger, we compare our results to lowest order in $\mathbf{q}$ with the exact overlap solution by directly computing the wavefunction overlap instead of using Eq.~\eqref{eq:long-range-new}, which gives in the bulk case:
\begin{multline}\label{eq:exact}
\langle u_{s\mathbf{k+q}}^{\rm W}| V_{\mathbf{q}\kappa\alpha}^{\mathcal{L}} | u_{p\mathbf{k}}^{\rm W} \rangle = \frac{ 4 \pi e f(\textbf{q})}{\Omega |\textbf{q}|^2 \tilde{\varepsilon}(\mathbf{q}) }  e^{-i\mathbf{q}\cdot \boldsymbol{\tau}_{\kappa}}   i\mathbf{q}\cdot \boldsymbol{\mathcal{Z}}_{\kappa\alpha}^{\parallel}(\mathbf{q}) \\
 \times \langle u_{s\mathbf{k+q}}^{\rm W}| u_{p\mathbf{k}}^{\rm W} \rangle.
\end{multline}

As seen with a pink line in  Fig.~\ref{fig:SrOgauge}(d), the short range matrix element recovers the exact overlap results in a large momentum region close to the zone center. 
However, in contrast to the position operator for the Berry connection, the exact overlap cannot be easily interpolated but we see in Fig.~\ref{fig:SrOgauge}(b,d) that the Berry connection term makes the short-range matrix element close to the exact overlap solution.

\subsection{Carrier mobility}

\begin{figure*}[ht]
  \centering
  \includegraphics[width=0.8\linewidth]{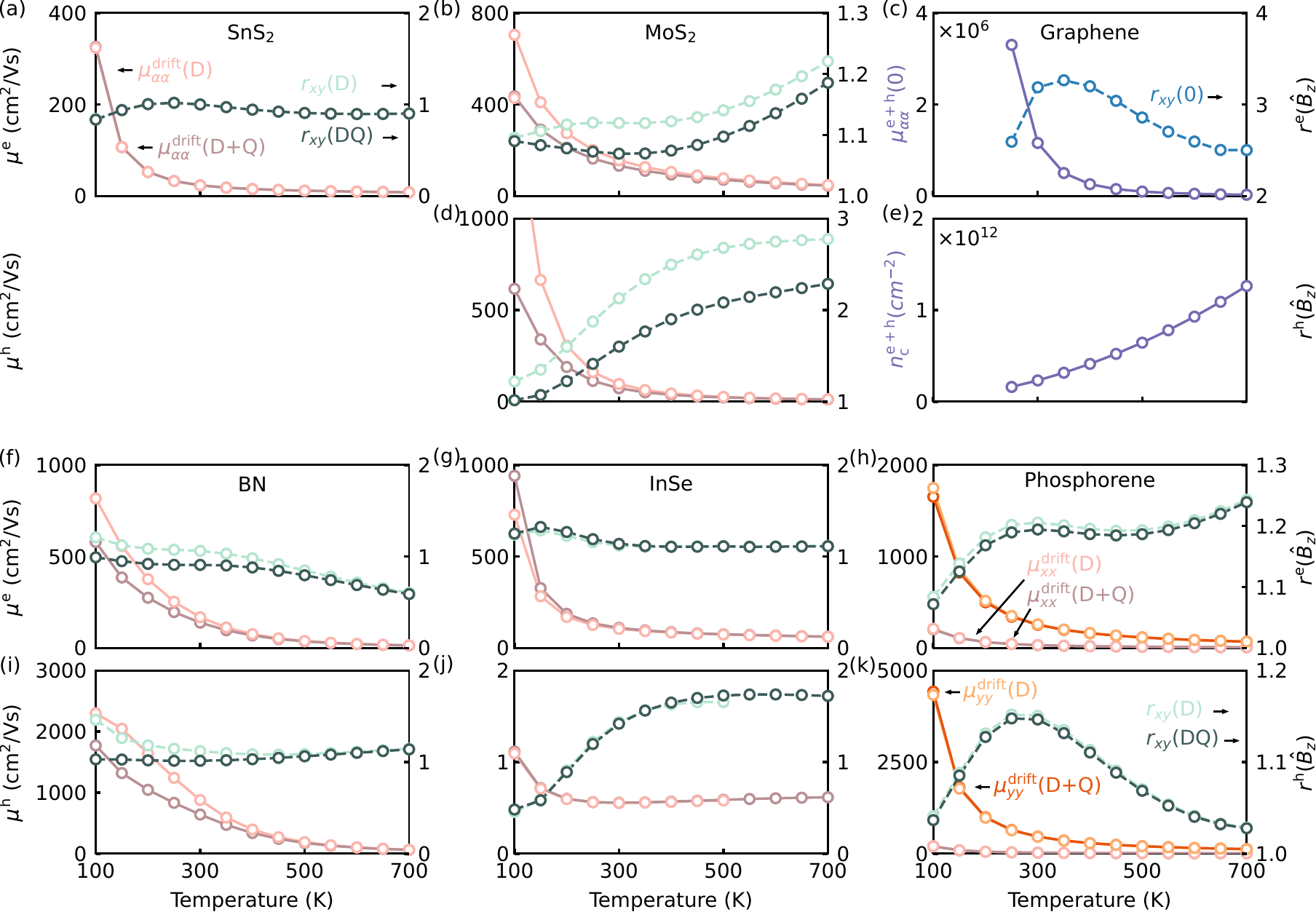}
  \caption{\label{fig:mobility}
Electron and hole carrier drift BTE mobility including dipole-dipole (DD) in the long-range part of the dynamical matrix (D) as well as the monopole-dipole (eD) in the electron-phonon matrix elements (G) (plain salmon line), compared with also adding the dipole-quadrupole (DQ) and quadrupole-quadrupole (QQ) term in D as well as monopole-quadrupoles (eQ) in the G (plain mauve line).
The dashed light green line are the Hall factor including dipoles and the dashed dark green line the same with also quadrupole contributions.
}
\end{figure*}

Now that we have validated and assessed the quality of the 2D deformation potentials in Section~\ref{sec:deform}, we proceed to study
the intrinsic drift and Hall carrier mobility of the six monolayers considered here using Eqs.~\eqref{eq:mobilitydrift} and \eqref{eq:mobilityhall}.

To be clear, in all calculations the intrinsic carrier mobility is obtained by placing the Fermi level in the band gap and then determining the position of the Fermi level such that the tail of the Fermi-Dirac distribution, for a given temperature, gives a fixed carrier concentration.
Here we choose a carrier concentration of 10$^{10}$~cm$^{-2}$ and we verified that the mobilities are independent of that value.
In the case of graphene, the Fermi level is placed at the Dirac point and the carrier concentration is computed accordingly and reported.
For each material, a convergence study is performed to find the smallest energy window required to compute the mobility and Hall factors.
The values of the resulting energy windows are reported for each material in Table~\ref{table2}.
In all mobility calculations, if not otherwise stated, we include the effect of dynamical quadrupoles for the interpolation of electron-phonon matrix elements and dynamical matrices.
We use an adaptive smearing in all calculations as described in Ref.~\onlinecite{Ponce2021}.
The used coarse \textbf{k}-point and \textbf{q}-point grids are also reported in Table~\ref{table2} for each material.
Finally, band velocities are obtained by direct evaluation of the non-local part of the pseudopotential (see Eq.~(24) of Ref.~\onlinecite{Ponce2021} for example).

\begin{figure*}[ht]
  \centering
  \includegraphics[width=0.8\linewidth]{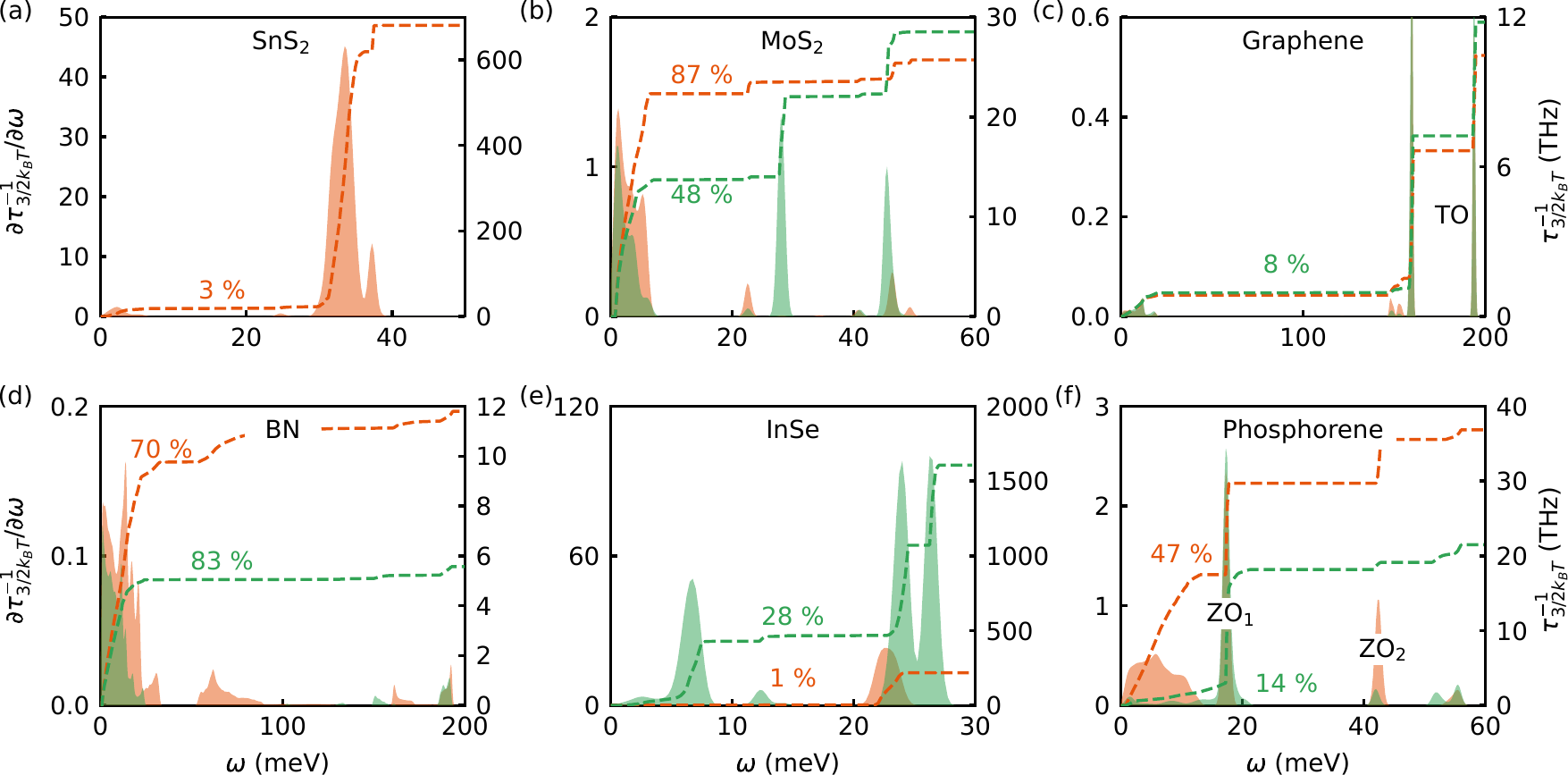}
  \caption{\label{fig:spectralFCT}
Spectral decomposition of the electron (orange) and hole (green) scattering rates as a function of phonon energy
at 300 K.
The rates are calculated as angular averages for carriers at an energy of 3 $k_{\rm B}T /2 = 39$~meV away from the band edges.
The dashed lines represent the cumulative integrals of the calculated rates, and add up to the carrier scattering rate $\tau_{3/2 k_{B\rm}T}^{-1}$.
The percentage indicates the contributions from acoustic modes.
  }
\end{figure*}

We start by looking at the electron mobility of SnS$_2$ as a function of fine momentum grids.
As explained in Ref.~\onlinecite{Ponce2021}, the carrier mobility and Hall factor converge linearly as a function of inverse fine grid density.
One can therefore extrapolate the results to the theoretical infinite grid density as shown in Table~\ref{table2} for all materials.
In the case of SnS$_2$, we obtain a room temperature electron drift mobility in the SERTA of 16.5~cm$^2$/Vs, which increases to 23.4~cm$^2$/Vs
if the BTE is solved self-consistently.
If we account for the Hall factor, the electron Hall mobility is 23.9~cm$^2$/Vs.
However, as it can be seen in Table~\ref{table2}, the extrapolated values are usually quite close to their high-density grid values.
For simplicity we therefore report the carrier mobility as a function of temperature and spectral decomposition results at finite (although very fine) grid density.
In the case of SnS$_2$, the temperature dependence of electron mobility and Hall factor computed with a 600$\times$600$\times$1 fine \textbf{k}- and \textbf{q}-point grids is shown in Fig.~\ref{fig:mobility}(a)
where one can see, despite differences in deformation potential, that there is almost no visible effect of including quadrupoles on the mobility and Hall factor.
As seen in Table~\ref{table3}, the only important effect is when we neglect long-range treatments entirely, which yields 31~cm$^2$/Vs.
In all cases, the room-temperature values that we report here are in stark contrast with the only prior computed electron mobility of 756.6~cm$^2$/Vs using deformation potential theory which only accounts for acoustic scattering~\cite{Shafique2017}.
This difference can be explained by looking at the spectral decomposition of the electron scattering rate of SnS$_2$ shown in Fig.~\ref{fig:spectralFCT}(a) where most of the scattering comes from the optical modes at 35~meV associated with the LO$_2$ and ZO$_1$ phonons.
Experimentally, a field-effect mobility of 0.04~cm$^2$/Vs was reported for SnS$_2$ obtained with exfoliation and characterized via a field-effect transistor with a high-$\kappa$ dielectric screening~\cite{Zschieschang2014},
18~cm$^2$/Vs for SnS$_2$ bulk crystals~\cite{Shibata1991},
50~cm$^2$/Vs for SnS$_2$ monolayer field-effect transistors grown with chemical vapor deposition~\cite{Song2013},
230~cm$^2$/Vs for a thin SnS$_2$  field effect transistor screened by a high-$\kappa$ dielectric consisting of deionized water~\cite{Huang2014}, and
330~cm$^2$/Vs was reported for vertical SnS$_2$ nanoflakes~\cite{Giri2019}.
Given the range of experimental measurement available, the precise experimental intrinsic electron mobility of SnS$_2$ monolayer remains an open question.

Next, we look at the mobility of MoS$_2$ monolayer which has a rich history.
The first reports of high mobility MoS$_2$ monolayer date from 2011 using a HfO$_2$ gate dielectric and achieving about 200~cm$^2$/Vs~\cite{Radisavljevic2011},
quickly followed by early first-principles predictions using Monte Carlo simulations and reporting a mobility of 130~cm$^2$/Vs~\cite{Li2013}.
Computationally, the mobility was reported almost exclusively for electrons with the exception of Ref.~\onlinecite{Guo2019} which reported a value of 26~cm$^2$/Vs for the hole mobility of MoS$_2$, neglecting SOC -- a number that we confirm in table~\ref{table3} with a value of 23~cm$^2$/Vs in our case.
Previous theoretical values ranged from 320~cm$^2$/Vs to 410~cm$^2$/Vs using LDA in the SERTA~\cite{Kaasbjerg2012a,Kaasbjerg2013,Zhang2014,Gunst2016}, which
reduce to 127~cm$^2$/Vs by solving the BTE iteratively~\cite{Gaddemane2019}.
In addition, theoretical results using the PBE exchange-correlation functional and the BTE range from 97 to 150~cm$^2$/Vs~\cite{Li2015,Sohier2018,Gaddemane2019,Guo2019}.
Comparison with experimental mobilities must be performed with care as many factors influence the measurements, from the actual thickness of the system (not necessarily a single layer) to the carrier density of the material.
In particular, exfoliated samples seem to outperform the ones grown by chemical vapor deposition (CVD) with values ranging from 23 to 217~cm$^2$/Vs for the exfoliated samples~\cite{Radisavljevic2011,Yu2014,Liu2015c,Yu2016} while the CVD ones range from 24 to 60~cm$^2$/Vs~\cite{Sanne2015,Kang2015,Cui2015,Huo2018} on various substrate and encapsulation.
One notable experiment is the hole mobility obtained for CVD sample deposited on an SiO$_2$ substrate and measured with Ag contacts and four probes which gave a mobility of 76~cm$^2$/Vs~\cite{Momose2018}.

We also find that MoS$_2$ is a complex material with a room temperature electron drift and Hall mobility of 132~cm$^2$/Vs and 142~cm$^2$/Vs, respectively.
In agreement with experimental reports, we find that the hole drift and Hall mobility are smaller with values of 74~cm$^2$/Vs and 121~cm$^2$/Vs, respectively.
The temperature dependence of the mobility and Hall factor are given in Fig.~\ref{fig:mobility}(b,d) and display a significant variation with temperature.
We see that the Hall factor increases with temperature both for  electrons and  holes and that the hole Hall factor is significantly larger than unity, demonstrating the importance of accounting for it when comparing to Hall mobility measurements.

What is particularly remarkable with MoS$_2$, is that the neglect of quadrupolar interaction during interpolation yields an overestimation of the electron and hole room-temperature Hall mobility by 23\% and 76\%, respectively.
The neglect of quadrupoles therefore leads to a situation where the hole mobility is actually larger than the electron one, a remarkable qualitative difference.
Another important aspect is that the neglect of SOC  strongly suppresses the Hall hole mobility to 23~cm$^2$/Vs by enhancing intervalley scattering~\cite{Sohier2019}, while leaving the electron mobility almost unaffected (see table~\ref{table3}).
We also find in table~\ref{table3} that the most crucial aspect is to include quadrupoles corrections at the level of the electron-phonon matrix elements. 
These results can be compared with the state-of-the art one from Deng \textit{et al.}~\cite{Deng2021} who included a partial quadrupolar contribution to the in-plane
fields and obtained a room-temperature drift electron mobility in MoS$_2$ of 176.6~cm$^2$/Vs.
Since these results are close to our values with only 2D dipoles, we conclude that the scheme of Deng \textit{et al.}~\cite{Deng2021} misses most of the quadrupolar effects, in agreement with the authors claim of a dipolar effect.
Overall we find that including SOC and dynamical quadrupoles in MoS$_2$ is also crucial to reproduce the temperature scaling with a T$^{-1.08}$ and T$^{-1.46}$ dependence for  electrons and holes, respectively,  in agreement with experiments~\cite{Yu2016,Cui2015,Liu2015c,Momose2018}.

\begin{figure}[t]
  \centering
  \includegraphics[width=0.85\linewidth]{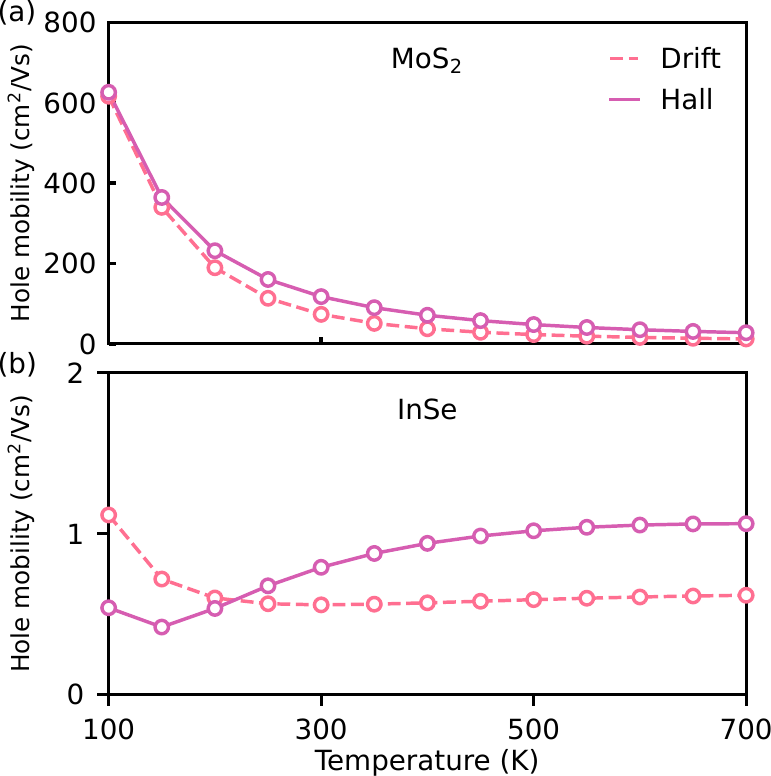} %mobilityInSe1.py
  \caption{\label{fig:InSe-mob}
Hole drift and Hall mobility of (a) MoS$_2$ and (b) InSe including quadrupole contributions.
In both cases the Hall factor strongly increases with temperature, see Fig.~\ref{fig:mobility}(d) and (j) but only in the case of InSe does that yield
an increase of mobility with temperature, highlighting the remarkable nature of InSe.
}
\end{figure}

Next, we turn to  h-BN monolayers which is the second material in our list to display differences upon quadrupoles inclusion, as seen in Fig.~\ref{fig:mobility}(f,i).
h-BN is mostly used as a support or encapsulation layer for graphene or other 2D materials and therefore its intrinsic mobility is seldom investigated.
In the literature, we found  high temperature hole mobility reports of 18~cm$^2$/Vs using the Van der Pauw–Hall method~\cite{Doan2014} as well as 2~cm$^2$/Vs
with h-BN doped through Mg implantation~\cite{Grenadier2021}.
Interestingly the hole mobility was also obtained by time of flight measurement and gave 35.5~cm$^2$/Vs for holes and 34.2~cm$^2$/Vs for electrons~\cite{Grenadier2019}.
For the electron mobility, typical doping include silicon or carbon and gives high temperature Hall mobility value of 48~cm$^2$/Vs~\cite{Grenadier2021}.
In addition, in the same experiment they also measured the Hall hole mobility to about 70~cm$^2$/Vs~\cite{Grenadier2021} at high temperature.
Therefore even if not definitive, it seems the hole mobility could be larger than the electron one in h-BN.

We confirm this numerically and obtain a room temperature Hall electron and hole mobility of 124.61~cm$^2$/Vs and 637.15~cm$^2$/Vs, respectively.
The convergence of SERTA and drift mobilities are reported in Table~\ref{table2}.
As for the case of MoS$_2$, the neglect of quadrupoles increases the Hall electron and hole mobility to 179~cm$^2$/Vs and 985~cm$^2$/Vs, respectively.
This is again a significant overestimation caused by the neglect of quadrupoles which should be accounted for.
We can rationalize this by noting in Fig.~\ref{fig:spectralFCT} that the materials for which the effect of quadrupoles is predominant are the materials with strong acoustic scattering since their piezoelectric constants are directly related to their dynamical dipoles and quadrupoles~\cite{Martin1972}.
Such findings make sense as small corrections in the low energy region of the deformation potential close to the zone center, such as the LA mode of h-BN shown in Fig.~\ref{fig:def-SI}(e), will have a strong contribution to the acoustic scattering and hence noticeably reduce the mobility.
Note that here we have used the LDA scalar relativistic pseudopotential without non-linear core correction (NLCC) for a direct comparison with Ref.~\onlinecite{Royo2021}.

To assess the effect of exchange-correlation functional and the effect of NLCC, we recomputed the mobility of h-BN using a PBE fully relativistic norm-conserving pseudopotential with NLCC from the \textsc{PseudoDojo} table~\cite{Setten2018}. 
We use the same quadrupole tensor, lattice and convergence parameters as for the LDA case above. 
The only difference being that we used two Wannier functions located on the boron atom and of initial $p_z$ and $s$ characters, instead of 6, for the conduction band manifold. 
The detailed room-temperature mobilities are reported in table~\ref{table3} showing an increased Hall mobility to 235~cm$^2$/Vs and 847~cm$^2$/Vs for electron and hole, respectively.
Interestingly, both the electron and hole mobilities are unaffected by the inclusion of SOC since the direct bandgap is located at the $K$ point and composed of a single band whose degeneracy is not lifted by SOC.

We now look at the results for InSe.
There is quite a large variability in the experimental results with values for the electron mobility ranging from 10 to 1200~cm$^2$/Vs~\cite{Feng2014,Sucharitakul2015,Yang2017,Bandurin2017,Ho2017,Chang2018,Li2018,Jiang2019}.
From the theoretical side, a study of the mobility of bulk, few layers and monolayer InSe~\cite{Li2019} reports a 120~cm$^2$/Vs room temperature electron and 0.5~cm$^2$/Vs hole mobility for the monolayer, respectively.
However, the system was treated as bulk with \textbf{k} and \textbf{q}-points along the vacuum directions such that it is worth to revisit this material.
By including the correct electrostatic boundary condition, the electron mobility was found to be as large as 500~cm$^2$/Vs~\cite{Sohier2020} (among the largest in 2D materials), an increase that might also arise from the explicit inclusion of a large carrier density ($10^{13}$~cm$^{-2}$) with the ensuing screening of the Fr\"ohlich interaction and with a larger carrier velocity.
The GW effective mass of InSe was studied in Ref.~\onlinecite{Li2020} and after including many-body renormalization effects, the calculated electron effective masses of InSe are 0.12 and 0.09 in the in-plane and out-of-plane directions, respectively.
S.~Gopalan~\textit{et al.}~\cite{Gopalan2019} studied the BTE mobility of InSe monolayer using Monte Carlo and found a low-field electron mobility 110~cm$^2$/Vs at room temperature.
Finally, L.-B. Shi~\textit{et al.}~\cite{Shi2019} obtained a room temperature electron mobility of about 300~cm$^2$/Vs and also discuss its variation with strain.
In general, the lack of horizontal mirror symmetry in materials such as silicene and germanene yields a strong ZA coupling, and correspondingly low mobilities.
However in the materials studied here such as InSe, MoS$_2$, and BN, the scattering potential linked with the flexural displacement is odd under the mirror symmetry making these flexural modes forbidden to first order and thus increasing mobility~\cite{Fischetti2016}.
Interestingly, we find that SnS$_2$ is an exception to this rule~\cite{Fischetti2016} as it does not have a horizontal mirror plane but still enjoys ZA mode suppression as the conduction band of SnS$_2$ is dominated by a spherically symmetric $s$-character orbitals around the Sn atom, making that electron-phonon coupling inactive by symmetry.

\begin{figure}[t]
  \centering
  \includegraphics[width=0.99\linewidth]{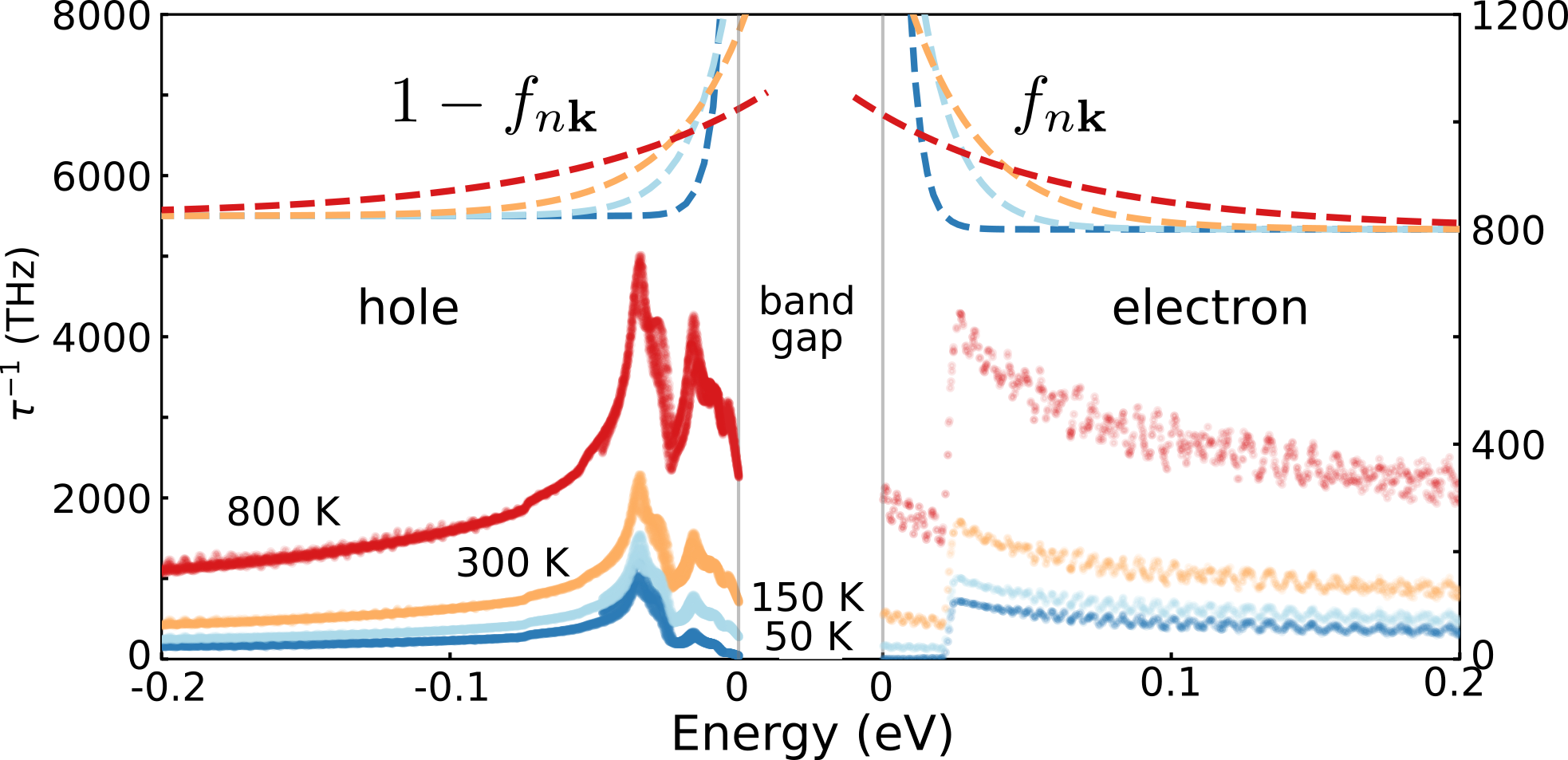} %InSe/epw16-quad-h/scattering2.pdf
  \caption{\label{fig:InSe-elph-zoom}
Scattering rate of InSe for 50~K, 150~K, 300~K, and 800~K for the top of the valence band (hole) and bottom of the conduction band (electron).
On the top of the image, we are showing the tail of the Fermi-Dirac distribution which are arbitrarily scaled in the same way and can therefore be compared with each other.}
\end{figure}

Our computed Hall electron mobility is 122.6~cm$^2$/Vs while the hole mobility is heavily suppressed to 0.78~cm$^2$/Vs due to the Mexican-hat shape of the valence band, in agreement with prior published values.
The good agreement is due to the fact that quadrupoles have very little impact in InSe as seen in Fig.~\ref{fig:mobility}(g,j).
We also look at the effect of neglecting SOC, with results summarized in Table~\ref{table3}.
We find that also here the electron mobility is weakly affected by SOC whereas the impact is larger for the hole mobility and in particular when the dipole of Ref.~\onlinecite{Sohier2016} is used.

What is remarkable is the temperature dependence of the hole mobility in InSe as seen in Fig~\ref{fig:InSe-mob}(b).
Because the drift mobility in Fig.~\ref{fig:mobility}(j) is almost flat above 200~K and the Hall factor increases with temperature,
we have a situation where the resulting Hall hole mobility, Fig.~\ref{fig:InSe-mob}, shows a minimum at 150~K and then \textit{increases} with temperature until reaching a plateau above 500~K.
To the authors' knowledge, this is the first time such non-monotonic behavior of the mobility calculated with the BTE is ever reported.
This behavior can be understood by looking at the scattering rate as a function of energy from the valence band maximum shown in Fig.~\ref{fig:InSe-elph-zoom}.
The scattering shows an unconventional double peak structure, due to a ``Mexican-hat" shape valence band, which gets accessed progressively as the temperature increases.
In particular, the first scattering peak is accessed at around 150~K yielding the mobility minimum while the dip between the two peaks is accessed around 300~K.
We note that the Mexican-hat structure in InSe has been theoretically predicted~\cite{Rybkovskiy2014,Lugovskoi2019,Li2019} and confirmed by angle resolved photoemission spectroscopy~\cite{Kibirev2018}.
Since similar valence bands structures have been predicted in other materials~\cite{Zolyomi2014}, we do not expect this unconventional
temperature dependence of the Hall hole mobility to be exclusive to InSe.

We also looked at graphene which is the seminal and most studied semi-metal.
For suspended graphene, carrier mobilities over 200,000~cm$^2$/Vs~\cite{Du2008,Bolotin2008,Castro2010} have been measured.
As it is not a semiconductor, the mobility is computed using a slightly different approach.
Instead of fixing a low carrier concentration and computing the concentration-independent mobility, we  place
the Fermi level at the Dirac point and compute the corresponding carrier concentration which is reported in Fig.~\ref{fig:mobility}(e).
We compute the electron and hole mobility separately but do include all interband and intraband transitions within a $\pm$0.5~eV energy window around the Dirac point.
The added electron and hole mobility as well as Hall factor reported it in Fig.~\ref{fig:mobility}(c) correspond to the intrinsic values (only thermal carriers) and result from that temperature-dependent concentration.
We recall that we do not include long-range contributions in non-polar graphene.
Moreover we do not consider temperatures below 250~K
due to the difficulty in sampling accurately the Dirac point at low temperatures even though we are using ultra-dense momentum grids of 2400$\times$2400$\times$1 \textbf{k}-/\textbf{q}-point grids, as reported in Table~\ref{table2}.

We finish our study with the investigation of phosphorene.
This material has four atoms per unit cell with a nonplanar distorted honeycomb lattice with structural ridges.
As a result, it is the only material in our list which has in-plane anisotropy with high mobility in the $y$ (armchair) direction and low mobility in the $x$ (zigzag) one.
Our computed room-temperature Hall electron and hole mobility in the $y$ direction is 304~cm$^2$/Vs and 520~cm$^2$/Vs, respectively, while
the Hall electron and hole mobility in the $x$ direction is 39~cm$^2$/Vs and 31~cm$^2$/Vs, respectively.
Although there are no available experimental values for the monolayer, few-layer samples have been studied with mobilities decreasing for thinner crystals~\cite{Liu2014,Li2014,Long2016,Long2016b}, an effect mainly attributed to sample degradation. In agreement with our findings, the electron mobility is typically lower than the hole one, with values up to 6000~cm$^2$/Vs and 8400~cm$^2$/Vs at cryogenic temperatures~\cite{Long2016b}, respectively. The current was also found to be strongly anisotropic~\cite{Liu2014}.
For what concerns theoretical results, previous calculations for the electron mobility range between 20 and 738~cm$^2$/Vs in the armchair direction and 5 to 114~cm$^2$/Vs in the zigzag direction~\cite{Liao2015a,Rudenko2016,Jin2016,Trushkov2017,Gaddemane2018,Sohier2018}.
For the hole mobility the calculations range from 19 to 460~cm$^2$/Vs in the armchair direction and 2 to 157~cm$^2$/Vs in the zigzag one~\cite{Liao2015a,Rudenko2016,Jin2016,Gaddemane2018,Sohier2018}.
The values reported here are in close agreement with Ref.~\onlinecite{Sohier2018}, despite the different doping considered in that study.
Although several modes have relatively large electron-phonon coupling~\cite{Sohier2018}, most of them give rise to ``side" scattering while only few--in particular LA and ZO--to back scattering~\cite{Sohier2018}, thus explaining the high mobility~\cite{Sohier2020}.
As shown in Fig.~\ref{fig:spectralFCT}(f), the ZO$_1$ peak alone accounts for 33\% of electron scattering and 70\% of hole scattering and
the ZO$_2$ peak to an additional 16\% for electrons.
This assignment seems consistent with the identification of back-scattering modes in Ref.~\onlinecite{Sohier2018}.
The differences observed in Fig.~\ref{fig:def-SI}(f) for the ZO$_2$ and TO$_2$ modes by including the effects of quadrupoles have therefore a limited impact on the mobility and Hall factor of phosphorene, consistent with Fig.~\ref{fig:mobility}(h,k), apart from a marginal correction for electrons associated with the contribution of ZO$_2$ phonons.

\section{Conclusions}

We have derived a theory for the long-range scattering potential and corresponding electron-phonon matrix elements in two-dimensional
materials.
Equations~\eqref{eq:glong-final2D}-\eqref{eq:long-range-new2} are the  most general formula to quadrupolar order.
Together, they form the main theoretical contribution from this work.
Importantly, the derivation enables the calculation of the long range part in terms of a small set of physical parameters, including dynamical Born effective charges, dynamical quadrupoles, and the macroscopic dielectric tensor, which are all accessible via first-principles density functional perturbation theory.
The ability to compute the long-range part is crucial to perform realistic calculations of electron-phonon coupling because it allows for accurate interpolation of the deformation potential or the electron-phonon matrix elements.
We have discovered that state-of-the art approaches, when expressing these matrix elements in a Wannier-based framework, introduce a gauge dependence that appears at quadrupolar order. 
Remarkably, we have identified a previously overlooked Berry connection term that restores gauge covariance at least in the long-wavelength limit.

In this manuscript, we have validated our theory and implementation by studying six representative monolayer materials including a semi-metal, non-polar and polar, as well as acoustic scattering dominated and optical scattering dominated ones.
We included spin-orbit coupling, drift and Hall mobility, electron and hole carriers and showed the importance of including dynamical quadrupoles in MoS$_2$, BN, and to a smaller extent in phosphorene.
The Berry connection term is found to be important for the interpolation quality of hexagonal BN and InSe monolayers as well as bulk SrO, demonstrating its relevance.

\appendix

\section{cell-periodic and all-space representations}
\label{app:representations}

Although we mainly work with the cell-periodic part of scalar potentials or two-body operators, in the course of our formal derivations we occasionally resort to the alternative all-space representation.
Both frameworks are characterized by the basis sets used to represent our mathematical objects.
For a bulk crystal, the all-space representation is spanned by the basis functions
\begin{equation}
\langle \mathbf{r}| \mathbf{K} \rangle = \frac{1}{\sqrt{(2\pi)^3}} e^{i\mathbf{K \cdot r}},
\end{equation}
where both $\mathbf{r}$ and $\mathbf{K}$ are continuous vectors of, respectively, the  real and reciprocal $\mathbb{R}^3$ spaces.
In turn, for the cell-periodic representation we use
\begin{equation}
\langle \mathbf{r}| \mathbf{G} \rangle = \frac{1}{\sqrt{\Omega}} e^{i\mathbf{G \cdot r}},
\end{equation}
where, instead, $\mathbf{r}$ run over the primitive cell and $\mathbf{G}$ are discrete vectors spanning the reciprocal-space Bravais lattice of the crystal.

The cell-periodic response functions (denoted with a $\mathbf{q}$ subscript), such as the first-order densities or potentials  due to an atomic displacement perturbation, are then related to their all-space counterparts as follows,
\begin{align}
\mathcal{F}^{\tau_{\kappa \alpha}}(\mathbf{r-R}_l) = &\frac{\Omega}{(2\pi)^3}\int_{\rm BZ} d^3 q \, \mathcal{F}^{\tau_{\kappa \alpha}}_{\mathbf{q}}(\mathbf{r})
  e^{i \mathbf{q}\cdot (\mathbf{r-R}_l)}, \label{fka1} \\
\mathcal{F}^{\tau_{\kappa \alpha}}_{\mathbf{q}}(\mathbf{r}) =& \sum_{m} \mathcal{F}^{\tau_{\kappa \alpha}}(\mathbf{r-R}_m) e^{-i \mathbf{q}\cdot (\mathbf{r-R}_m)},\label{fka2}
\end{align}
with $\mathbf{R}_l$ being the real-space lattice vector of cell $l$ in the so-called Born-von Karman supercell.
Note that the Brillouin-zone integral is related with its discretized version as
\begin{equation}
 \frac{\Omega}{(2\pi)^3} \int_{\rm BZ} d^3 q \simeq \frac{1}{N_{\mathbf{q}}} \sum_{\mathbf{q}},
\end{equation}
where $N_{\mathbf{q}}$ is the total number of wave vectors in the finite grid used in the calculation.

We can similarly write the relationships connecting the two-body operators in both representations as
\begin{align}\label{wq}
\mathcal{W}_{\mathbf{q}}(\mathbf{r,r'}) &= \sum_{m} \mathcal{W}(\mathbf{r,r'+R}_m)  e^{i \mathbf{q} \cdot (\mathbf{r}'+\mathbf{R}_m -\mathbf{r})} \\
\mathcal{W}(\mathbf{r,r'}) &= \frac{\Omega}{(2\pi)^3} \int_{\rm BZ} d^3 q \, \mathcal{W}_{\mathbf{q}}(\mathbf{r,r'})e^{i\mathbf{q} \cdot (\mathbf{r-r'})}. \label{transw}
\end{align}

For 2D crystals which are periodic in plane but finite in the out-of-plane direction $z$, we use a mixed representation of the cell-periodic
functions and operators via the following basis,
\begin{equation}
\langle \mathbf{r}| \mathbf{G} z\rangle = \frac{1}{\sqrt{S}} e^{i\mathbf{G \cdot r}} \delta(z-z'),
\end{equation}
where the reciprocal space wave vectors $\mathbf{G}$ and $\mathbf{q}$ are only allowed to have in-plane components.
Then, the relations between the all-space and cell-periodic representations of the scalar quantities and operators must be revised as follows,
\begin{align}
\mathcal{F}^{\tau_{\kappa \alpha}}(\mathbf{r-R}_l) = &\frac{S}{(2\pi)^2}\int_{\rm BZ} d^2 q \, \mathcal{F}^{\tau_{\kappa \alpha}}_{\mathbf{q}}(\mathbf{r})
  e^{i \mathbf{q}\cdot (\mathbf{r-R}_l)},  \\
\mathcal{W}(\mathbf{r,r'}) = & \frac{S}{(2\pi)^2} \int_{\rm BZ} d^2 q \, e^{i{\bf q} \cdot ({\bf r-r'})}
 \mathcal{W}_{\mathbf{q}}(\mathbf{r,r'}),
\end{align}
where $S$ is the surface of the primitive 2D cell.
The converse relations remain unaltered with respect to the 3D case, with the {\em caveat} that the Bravais lattice vectors $\mathbf{R}_l$ now span a 2D plane rather than the 3D space.

\section{Long-range interatomic force constants in 2D}\label{app:longrangeIFC}

We first split the dynamical matrix (we have removed the mass factor for clarity) into a short-range and long range at long wavelength part in 2D~\cite{Royo2021}:
\begin{equation}
D_{\kappa\alpha \kappa'\beta}(\mathbf{q}) = D_{\kappa\alpha \kappa'\beta}^{\mathcal{S}}(\mathbf{q}) + D_{\kappa\alpha \kappa'\beta}^{\mathcal{L}}(\mathbf{q})
\end{equation}
where the nonanalytical, direction-dependent term, is given by the contribution of dipole-dipole, dipole-quadrupole, and quadrupole-quadrupole terms as~\cite{Royo2021}:
\begin{multline}\label{eq:totalDmatrixlong}
D_{\kappa\alpha \kappa'\beta}^{\mathcal{L}}(\mathbf{q}) =   \sum_{\mathbf{G \neq q}}\bigg[
D_{\kappa\alpha \kappa'\beta}^{\mathcal{L}, \text{DD}}(\mathbf{G+q})\\
 + D_{\kappa\alpha \kappa'\beta}^{\mathcal{L}, \text{DQ}}(\mathbf{G+q})  + D_{\kappa\alpha \kappa'\beta}^{\mathcal{L}, \text{QQ}}(\mathbf{G+q})\bigg] - \delta_{\kappa\kappa'}  \! \sum_{\kappa'',\mathbf{G\neq 0}}    \\
 \times \bigg[D_{\kappa\alpha \kappa''\beta}^{\mathcal{L},  \text{DD}}(\mathbf{G}) +  D_{\kappa\alpha \kappa''\beta}^{\mathcal{L}, \text{DQ}}(\mathbf{G})  + D_{\kappa\alpha \kappa''\beta}^{\mathcal{L},  \text{QQ}}(\mathbf{G})\bigg],
\end{multline}
where the dipole-dipole contribution is
\begin{multline}\label{eq:dynmatlong-dip}
D_{\kappa\alpha \kappa'\beta}^{\mathcal{L}, \text{DD}}(\mathbf{q})= \frac{2\pi e^2}{S}\frac{f(|\mathbf{q}|)}{|\mathbf{q}|}
\Bigg[ \frac{(\mathbf{q}\cdot \mathbf{Z}_{\kappa\alpha})^*  (\mathbf{q}\cdot \mathbf{Z}_{\kappa'\beta})}{1+\frac{2\pi f(|\mathbf{q}|)}{|\mathbf{q}|} \mathbf{q}\cdot \boldsymbol{\alpha}^{\parallel} \cdot \mathbf{q}} \\
- \frac{(|\mathbf{q}|Z_{\kappa\alpha z})^* (|\mathbf{q}|Z_{\kappa'\beta z})}{1-2\pi |\mathbf{q}| f(|\mathbf{q}|) \alpha^{\perp}} \Bigg] e^{-i \mathbf{q} \cdot (\tau_{\kappa'\beta}-\tau_{\kappa\alpha})},
\end{multline}
and the dipole-quadrupole one is:
\begin{multline}\label{eq:dynmatlong-dipquad}
D_{\kappa\alpha \kappa'\beta}^{\mathcal{L}, \text{DQ}}(\mathbf{q})= \frac{-2\pi e^2 i}{S}\frac{ f(|\mathbf{q}|)}{|\mathbf{q}|} e^{-i \mathbf{q} \cdot (\tau_{\kappa'\beta}-\tau_{\kappa\alpha})} \\
\times \Bigg[ \frac{(\mathbf{q}\cdot \mathbf{Z}_{\kappa\alpha})^*  (\mathbf{q}\cdot \mathbf{q} \cdot \mathbf{Q}_{\kappa'\beta}) + (\mathbf{q}\cdot \mathbf{Z}_{\kappa\alpha})^* (|\mathbf{q}|^2Q_{\kappa\alpha zz}) }{2\big[1+\frac{2\pi f(|\mathbf{q}|)}{|\mathbf{q}|} \mathbf{q} \cdot \boldsymbol{\alpha}^{\parallel} \cdot \mathbf{q})\big]} \\
- \frac{(|\mathbf{q}|Z_{\kappa\alpha z})^* (|\mathbf{q}|\mathbf{q}\cdot Q_{\kappa'\beta z})}{1-2\pi |\mathbf{q}| f(|\mathbf{q}|) \alpha^{\perp}} + (\kappa\alpha) \leftrightarrow (\kappa'\beta) \Bigg] ,
\end{multline}
and the quadrupole-quadrupole term is
\begin{multline}\label{eq:dynmatlong-quadquad}
D_{\kappa\alpha \kappa'\beta}^{\mathcal{L}, \text{QQ}}(\mathbf{q})=  \frac{-2\pi e^2}{S}\frac{f(|\mathbf{q}|)}{|\mathbf{q}|} e^{-i \mathbf{q} \cdot (\tau_{\kappa'\beta}-\tau_{\kappa\alpha})} \\
\times \Bigg[ \frac{(\mathbf{q}\cdot \mathbf{q}\cdot \mathbf{Q}_{\kappa\alpha})^*  (\mathbf{q}\cdot \mathbf{q} \cdot \mathbf{Q}_{\kappa'\beta}) + (\mathbf{q}\cdot \mathbf{q} \cdot \mathbf{Q}_{\kappa\alpha})^* (|\mathbf{q}|^2 Q_{\kappa'\beta zz}) }{4\big[1+\frac{2\pi f(|\mathbf{q}|)}{|\mathbf{q}|} \mathbf{q} \cdot \boldsymbol{\alpha}^{\parallel} \cdot \mathbf{q} \big]} \\
+\frac{(|\mathbf{q}|^2 Q_{\kappa\alpha zz})^* (\mathbf{q}\cdot \mathbf{q} \cdot \mathbf{Q}_{\kappa'\beta}) + (|\mathbf{q}|^2 Q_{\kappa\alpha zz})^* (|\mathbf{q}|^2Q_{\kappa'\beta zz}) }{ 4\big[1+\frac{2\pi f(|\mathbf{q}|)}{|\mathbf{q}|} \mathbf{q} \cdot \boldsymbol{\alpha}^{\parallel} \cdot \mathbf{q} \big]} \\
- \frac{(|\mathbf{q}|\mathbf{q}\cdot \mathbf{Q}_{\kappa\alpha z})^* (|\mathbf{q}|\mathbf{q}\cdot \mathbf{Q}_{\kappa'\beta z})}{1-2\pi |\mathbf{q}| f(|\mathbf{q}|) \alpha^{\perp}} \Bigg].
\end{multline}

The approximated formulation for the dipole-dipole contribution of Eq.~\eqref{eq:dynmatlong-dip} from Refs.~\onlinecite{Sohier2017a} is given as:
\begin{multline}\label{eq:dynmatlong2D-sohier}
D_{\kappa\alpha\kappa'\beta}^{\mathcal{L}, \text{DDS}}(\mathbf{q})= e^{-\frac{|\mathbf{q}|^2}{4\Lambda^2}}e^{-i\mathbf{q}\cdot (\tau_{\kappa'\beta}-\tau_{\kappa\alpha})}\\
\times \frac{ (\mathbf{q} \cdot \mathbf{Z}_{\kappa\alpha})^* \mathbf{q} \cdot \mathbf{Z}_{\kappa'\beta}}{|\mathbf{q}|\varepsilon^{\text{2D}}(|\mathbf{q}|)}.
\end{multline}

\section{Long-range electron-phonon matrix elements in 3D}\label{app:longrange3D}

As mentioned in Sec.~\ref{sec:range_sep} and \ref{sec:2dcoul}, there is some arbitrariness in the separation between a short and a long range kernel. A simple choice available for 3D materials~\cite{Royo2021} is to set
\begin{align}
v^{\mathcal S} (\textbf{r},\textbf{r}') &= \frac{\textrm{erfc}(|{\bf r} - {\bf r}'|/L)}{|{\bf r} - {\bf r}'|} \\
v^{\mathcal L} (\textbf{r},\textbf{r}') &= \frac{\textrm{erf}(|{\bf r} - {\bf r}'|/L)}{|{\bf r} - {\bf r}'|}
\end{align}
where $L$($=\Lambda^{-1}$ in Ref.~\onlinecite{Royo2021} ) plays the role of range separation parameter.
The long range kernel can be thus represented in separable form using plane waves basis functions  with wave vector $\mathbf{q}$ within the first Brillouin zone and a reciprocal lattice vector $\mathbf{G}$ such that in 3D Eq.~\eqref{eq:2dbasis} is replaced by:
\begin{equation}\label{eq:basisfunctions3D}
\varphi_{\mathbf{qG}}(\mathbf{r}) \equiv \frac{e^{i(\mathbf{G+q})\cdot \mathbf{r}}}{\sqrt{\Omega}},
\end{equation}
where $\Omega$ is the unit cell volume.
The cell periodic part of the bare long-range Coulomb potential can then be written following Eq.~\eqref{eq:separable} as:
\begin{align}\label{eq:barepot3D}
v_{\mathbf{q}}^{\mathcal{L}}(\mathbf{r,r'})  =& \sideset{}{'}\sum_{\mathbf{GG'}} \varphi_{\mathbf{qG}}(\mathbf{r}) \tilde{v}_{\mathbf{q}}^{\mathcal{L}}(\mathbf{G,G'}) \varphi_{\mathbf{qG'}}^{\dagger}(\mathbf{r'}),
\end{align}
where:
\begin{equation}\label{eq:bareCoulomb3DGaussian}
\tilde{v}_{\mathbf{q}}^{\mathcal{L}}(\mathbf{G,G'}) =  \delta_{\mathbf{G,G'}} \frac{4\pi e^2}{|\mathbf{q+G}|^2} e^{-\frac{|\mathbf{q+G}|^2L^2}{4}}.
\end{equation}
The representation \eqref{eq:barepot3D} can thus be restricted to the small space spanned by the functions $\varphi_{\mathbf{qG}}(\mathbf{r})$ with $|\mathbf{q+G}|$ not larger than $1/L$ and we can interpret the factor $f(|\mathbf{q+G}|) = e^{-\frac{|\mathbf{q+G}|^2L^2}{4}}$ as the equivalent in 3D of the range separation function Eq.~\eqref{eq:ffunction} in 2D. 
Interestingly, in 3D the range separation function needs to approach one \textit{quadratically} in $ K = |\mathbf{q+G}|$, and not linearly, in order to have that $[f(K)-1]/K^2$ be analytic.

\begin{figure*}[ht]
  \centering
  \includegraphics[width=0.6\linewidth]{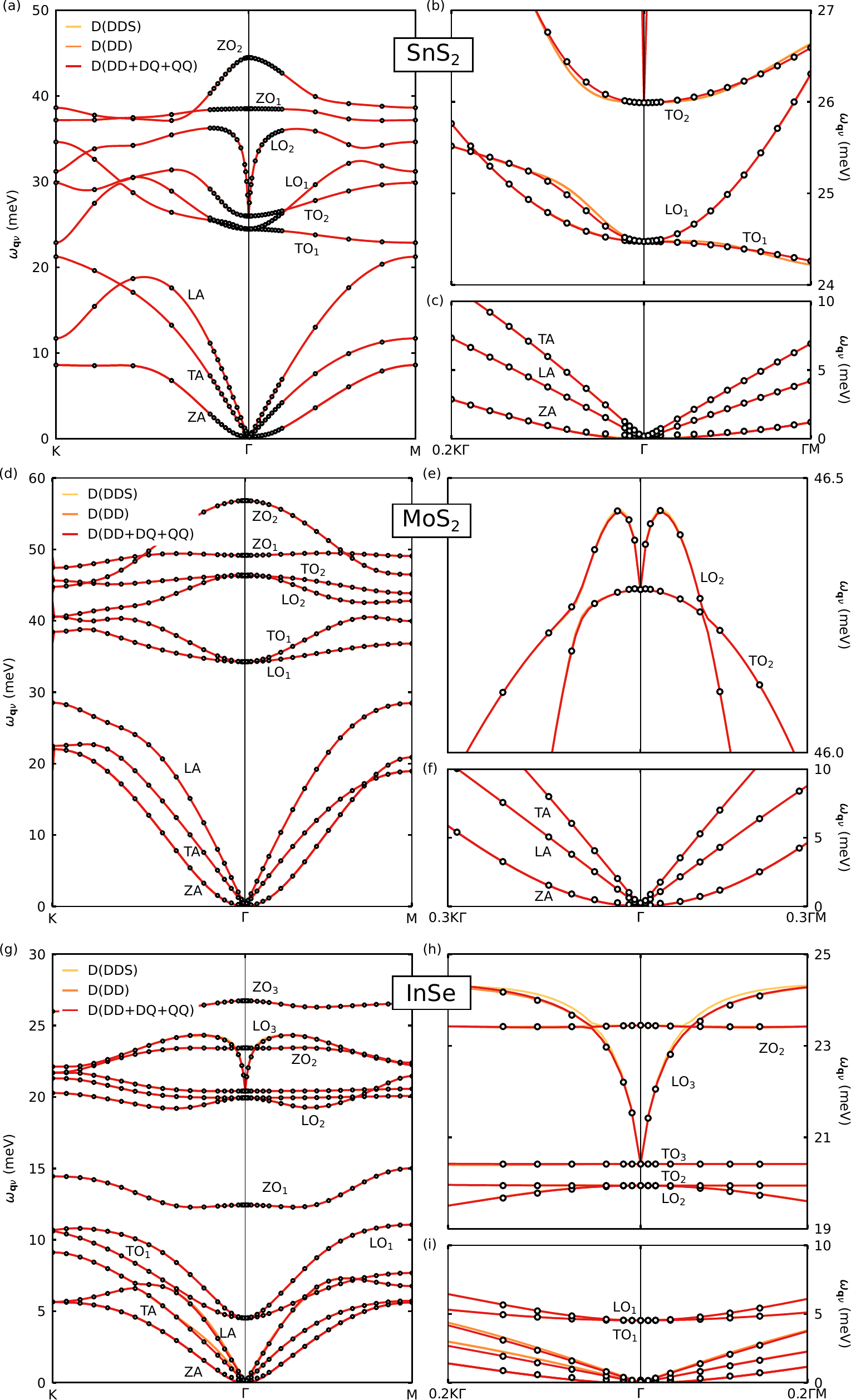}
  \caption{\label{fig:Sns2-phonon}
 Phonon dispersion of (a-c) SnS$_2$, (d-f) MoS$_2$, and (g-j) InSe monolayer calculated with direct density functional perturbation theory calculations (black empty circles) compared with Fourier interpolation where the long-range part of the dynamical matrix (D) includes dipole-dipole (DD), dipole-quadrupole (DQ) and quadrupole-quadrupole (QQ) and where the yellow line refers to the scheme of Ref.~\onlinecite{Sohier2017a}.
  }
\end{figure*}

For sufficiently large $L$, we can limit the sum to $\mathbf{G=G}'=\mathbf{0}$, so that the cell-periodic part of the long-range scattering potential becomes:
\begin{equation}
V_{\mathbf{q}\kappa\alpha}^{\mathcal{L}}(\mathbf{r}) =  -e \varphi_{\mathbf{q0}}^{\mathcal{S}}(\mathbf{r})  \tilde{W}_{\mathbf{q}}^{\mathcal{L}}(\mathbf{0,0})\tilde{\rho}_{\mathbf{q}\kappa\alpha}^{\mathcal{S}}(\mathbf{0}),
\end{equation}
where $\varphi_{\mathbf{qG}}^{\mathcal{S}}(\mathbf{r})$ is given by Eq.~\eqref{eq:srbf},
$\tilde{W}_{\mathbf{q}}^{\mathcal{L}}(\mathbf{G,G}')$ by Eq.~\eqref{eq:tildew}, and
$\tilde{\rho}_{\mathbf{q}\kappa\alpha}^{\mathcal{S}}(\mathbf{G}')$ is the 3D charge-response function which takes the following form in the long-wavelength limit~\cite{Royo2019}:
\begin{multline}
\lim_{\mathbf{q} \to \mathbf{0}} \tilde{\rho}_{\mathbf{q}\kappa\alpha}^{\mathcal{S}}(\mathbf{G'=0}) = \\
\frac{e^{-i\mathbf{q}\cdot \boldsymbol{\tau}_\kappa}}{\sqrt{\Omega}}\Big[ \sum_{\beta} iq_{\beta} Z_{\kappa\alpha\beta} + \sum_{\gamma}\frac{q_\beta q_{\gamma}}{2} Q_{\kappa\alpha\beta\gamma} + \cdots \Big],
\end{multline}
while the screened kernel is:
\begin{equation}
\tilde{W}_{\mathbf{q}}^{\mathcal{L}}(\mathbf{0,0}) 
= \frac{4\pi }{|\mathbf{q}|^2} f(\mathbf{q}) \left[ 1-  \frac{4\pi e^2}{|\mathbf{q}|^2} f(\mathbf{q}) \tilde\chi^{\mathcal{S}}_\mathbf{q}\right]^{-1}
\end{equation}
where the short range polarizability is analytic in $\mathbf{q}$ and can be related to the electronic dielectric tensor $\varepsilon_{\alpha\beta}$ in the long-wavelength limit through:
\begin{align}\label{eq:polarizability3D}
\lim_{\mathbf{q} \to \mathbf{0}}
4\pi e^2 \tilde{\chi}_{\mathbf{q}}^{\mathcal{S}} &= - \sum_{\alpha,\beta}{q}_\alpha  \left(\varepsilon_{\alpha\beta} - \delta_{\alpha\beta}\right) {q}_\beta + \mathcal{O}(\mathbf{q}^4)
\end{align}

Following similar steps as in the 2D case, we then get the final expression:
\begin{multline}\label{eq:glong-final3D}
g_{mn\nu}^{\mathcal{L}}(\mathbf{k},\mathbf{q}) = \sum_{\kappa\alpha} \bigg[ \frac{\hbar}{2M_\kappa \omega_{\nu}(\mathbf{q}) \Omega^2}\bigg]^{\frac{1}{2}}   \frac{4\pi e f(|\mathbf{q}|)}{|\mathbf{q}|^2 \tilde\epsilon(\mathbf{q})}  \\
\times   e^{-i\mathbf{q}\cdot \boldsymbol{\tau}_\kappa}\Big[ \sum_{\beta} iq_{\beta} Z_{\kappa\alpha\beta} + \sum_{\gamma}\frac{q_\beta q_{\gamma}}{2} Q_{\kappa\alpha\beta\gamma} \Big] e_{\kappa\alpha\nu}(\mathbf{q}) \\
\bra{\Psi_{m\mathbf{k+q}}} e^{i\mathbf{q}\cdot \mathbf{r}} [1+i q_{\alpha} V^{{\rm Hxc}, {\cal E}_\alpha}(\mathbf{r})] \ket{\Psi_{n\mathbf{k}}},
\end{multline}
where the effective macroscopic dielectric function reads:
\begin{equation}
\tilde\epsilon(\mathbf{q}) =  \frac{ \mathbf{q}\cdot \boldsymbol{\varepsilon}\cdot\mathbf{q}}{\phantom{^2}|\mathbf{q}|^2}  f(|\mathbf{q}|) +  1 -  f(|\mathbf{q}|)~.
\end{equation}

\begin{figure*}[ht]
  \centering
  \includegraphics[width=0.7\linewidth]{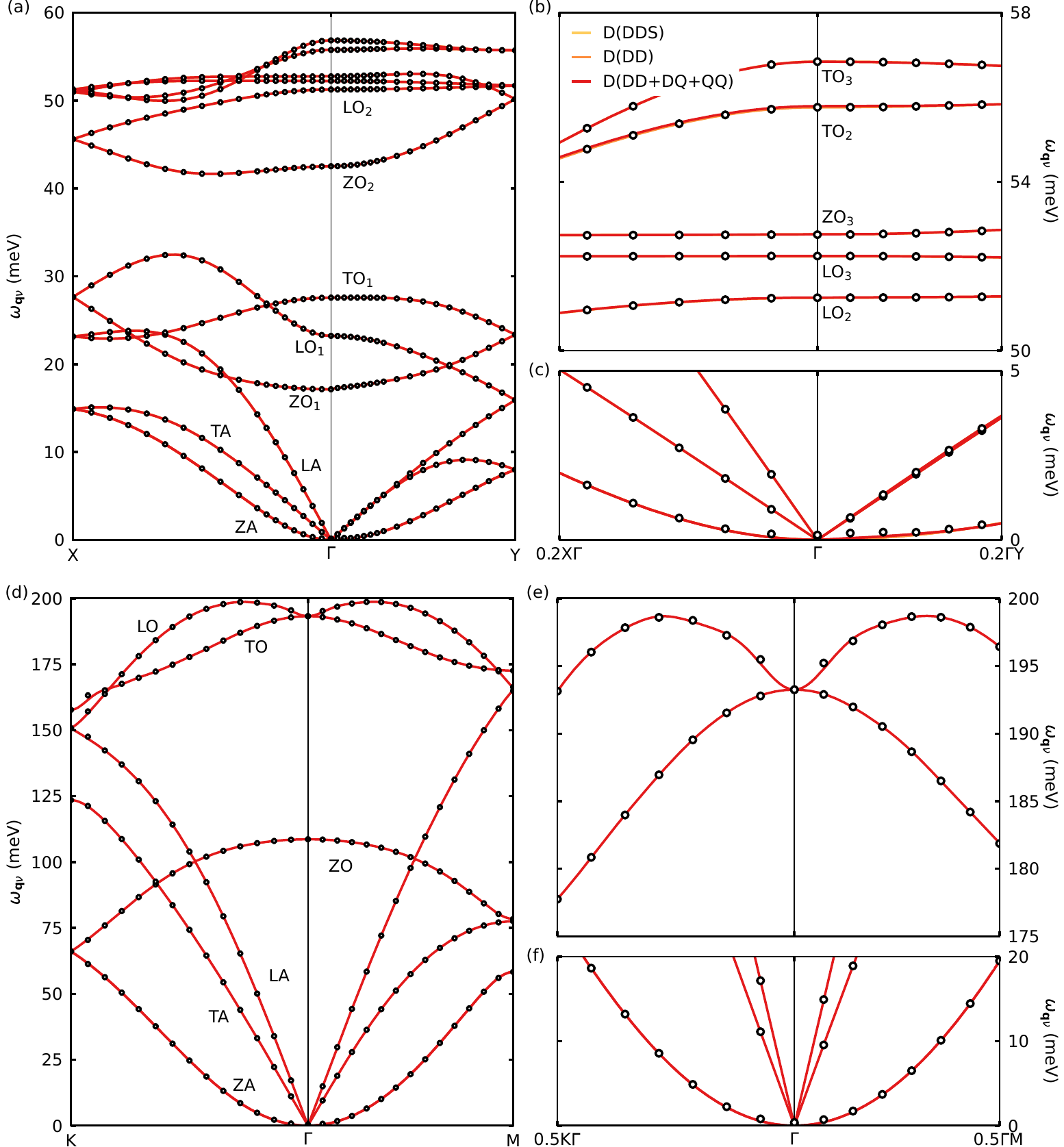}
  \caption{\label{fig:p-phonon}
 Phonon dispersion of phosphorene (a-c) and graphene (d-f) calculated with direct density functional perturbation theory calculations (black empty circles) compared with Fourier interpolation where the long-range part of the dynamical matrix (D) includes dipole-dipole (DD), dipole-quadrupole (DQ) and quadrupole-quadrupole (QQ) and where the yellow line refers to the scheme of Ref.~\onlinecite{Sohier2017a}.
}
\end{figure*}

\section{Wannier interpolation}
\label{sec:wannier}

We now report for each material the choice of initial projections for the Wannier functions as well as interpolated electronic band structures compared with direct DFT calculations.

For SnS$_2$, we choose to focus on the conduction band only since this band is separated in energy.
We perform the interpolation using maximally localized Wannier functions (MLWFs)~\cite{Marzari2012} as implemented in the \textsc{wannier90} software~\cite{Pizzi2020}, starting from an initial projection made on the Sn atom with $s$ character and yielding an atom-centred Wannier function with spread 9.32~\AA$^2$.
The resulting electronic band structure agrees with the DFT one as seen in Fig.~\ref{fig:BS}(a).

In the case of h-BN, two sets of MLWFs are considered separately for the valence (blue) and conduction bands (red), as shown in Fig.~\ref{fig:BS}(d).
For the valance bands, the initial projection is on the N atom with $p$ character and yields two in-plane Wannier functions with spread 1.02~\AA$^2$ and one out-of-plane with spread 1.35~\AA$^2$.
For the conduction band, the initial projections are $p$-orbitals on both B and N atoms, with a resulting average spread of 5.12~\AA$^2$.

For MoS$_2$, we wannierize together the top 2 valence bands and the lowest 8 conduction bands by initially projecting on Mo $d$ orbitals for both spin polarizations, yielding  Wannier functions with spread 4.9~\AA$^2$, 5.0~\AA$^2$, and 5.1~\AA$^2$. The corresponding interpolated band structure is shown in Fig.~\ref{fig:BS}(b).

As far as InSe is concerned, we address separately valence and conduction bands. For the valance bands (blue solid lines in Fig.~\ref{fig:BS}(e)), the initial  projection is on one of the two Se atoms with $p_z$ character and yields two spin-degenerate Wannier functions with spread 15.79~\AA$^2$.
For the conduction bands, the initial projection is an $s$-orbital (for each spin) on one of the two In atoms and gives two Wannier functions with spread of 11.37~\AA$^2$.

For graphene, a single set of 10 MLWFs is created to interpolate both valence and conduction bands, as shown in Fig.~\ref{fig:BS}(c) with blue lines.
The initial Wannier projections consist of hybrid $sp^2$ orbitals on one carbon atom and  $p_z$ orbitals on both carbon atoms, resulting in 6 (including spin degeneracy) MLWFs with spread 0.61~\AA$^2$ and 4 with spread  0.98~\AA$^2$.

Finally, for phosphorene we consider valence and conduction bands together by constructing 32 (including spin) MLWFs that accurately reproduce DFT results around the band gap, as shown in Fig.~\ref{fig:BS}(f).
As  initial projections on each of the four P atoms we take $p$- and $s$-orbitals, giving
8 MLWFs with spread 1.92~\AA$^2$, 16 MLWFs with spread 2.06~\AA$^2$, and 8 MLWFs with spread 2.36~\AA$^2$.

\section{Additional phonon dispersion and mobility results}\label{app:phonon}

For completeness, we report here the figures showing the DFPT phonon dispersions and their interpolations for SnS$_2$,  MoS$_2$ and InSe in Fig.~\ref{fig:Sns2-phonon},  and for phosphorene and graphene in Fig.~\ref{fig:p-phonon}.
Finally, we give explicit room temperature mobility values for various integration fine grids as well as their extrapolated values for all the
materials investigated here in Table~\ref{table2} as well as a comparison between different long-range treatments in Table~\ref{table3}.

%\clearpage

\begin{longtable*}{r r r | r r | r r | r}
\caption[Room temperature drift and Hall mobility]{
Room temperature fine grid convergence of drift and Hall mobility using adaptive smearing and dipole velocity.
The symbol $\infty$ for the fine grids means that the value is obtained by extrapolating the linear dependence on inverse grid size as explained in Ref.~\onlinecite{Ponce2021}.
SERTA mobility refers to the mobility in the self-energy relaxation time approximation.
The ``-e'' or ``-h" suffix after the material's name refers to electron or hole mobility, respectively.
In all cases we report the mobility including quadrupoles, except for graphene.  
For phosphorene, results along the two principal directions ($xx$ and $yy$) are reported separately.} \label{table2} \\
\hline
\endfirsthead
\multicolumn{8}{c}%
{{ \tablename\ \thetable{} -- continued from previous page}} \\
\hline Material   &  Window & Coarse grids  & Fine grids   &  SERTA mobility & BTE mobility & BTE Hall & Hall mobility \\
	        &    eV   & \textbf{k/q}  & \textbf{k/q} &       cm$^2$/Vs &   cm$^2$/Vs  & factor   &     cm$^2$/Vs   \\
 \hline
\endhead

\hline \multicolumn{8}{|r|}{{Continued on next page}} \\ \hline
\endfoot

\hline \hline
\endlastfoot

 Material   &  Window & Coarse grids  & Fine grids   &  SERTA mobility & BTE mobility & BTE Hall & Hall mobility   \\
	        &    eV   & \textbf{k/q}  & \textbf{k/q} &       cm$^2$/Vs &   cm$^2$/Vs  & factor   &     cm$^2$/Vs   \\
\hline
SnS$_2$-e   &   0.3   &        16$^2$ &      300$^2$ &           16.79 &        23.52 &    0.979 &   23.03 \\
            &         &               &      400$^2$ &           16.78 &        23.58 &    0.992 &   23.39 \\
            &         &               &      500$^2$ &           16.68 &        23.45 &    0.999 &   23.43 \\
            &         &               &      600$^2$ &           16.69 &        23.49 &    1.000 &   23.49 \\
            &         &               &      700$^2$ &           16.62 &        23.42 &    1.004 &   23.51 \\
            &         &               &     $\infty$ &           16.53 &        23.38 &    1.023 &   23.92 \\
MoS$_2$-e   &   0.2   &       18$^2$  &      300$^2$ &          116.66 &       134.39 &    1.066 &  143.26 \\
            &         &               &      400$^2$ &          114.81 &       131.09 &    1.069 &  140.14 \\
            &         &               &      500$^2$ &          115.95 &       132.37 &    1.069 &  141.50 \\
            &         &               &      600$^2$ &          117.39 &       133.19 &    1.072 &  142.78 \\
            &         &               &      800$^2$ &          116.84 &       132.75 &    1.074 &  142.57 \\
            &         &               &     $\infty$ &          117.20 &       131.78 &    1.078 &  142.06 \\
MoS$_2$-h   &   0.25  &       18$^2$  &      300$^2$ &           60.04 &        72.70 &    1.576 &  114.58 \\
            &         &               &      400$^2$ &           60.35 &        73.18 &    1.588 &  116.21 \\
            &         &               &      500$^2$ &           60.55 &        73.33 &    1.594 &  116.89 \\
            &         &               &      600$^2$ &           61.16 &        73.41 &    1.604 &  117.75 \\
            &         &               &     $\infty$ &           61.97 &        74.14 &    1.629 &  120.77 \\
     BN-e   &   0.4   &       16$^2$  &      200$^2$ &          131.87 &       141.18 &    0.912 &  128.76 \\
            &         &               &      300$^2$ &          130.64 &       143.28 &    0.903 &  129.38 \\
            &         &               &      400$^2$ &          128.06 &       139.26 &    0.912 &  127.01 \\
            &         &               &      500$^2$ &          127.27 &       138.87 &    0.909 &  126.23 \\
            &         &               &      600$^2$ &          126.42 &       138.10 &    0.911 &  125.81 \\
            &         &               &     $\infty$ &          124.06 &       137.08 &    0.909 &  124.61 \\
     BN-h   &   0.4   &       16$^2$  &      200$^2$ &          539.45 &       665.66 &    0.990 &  659.00 \\
            &         &               &      300$^2$ &          501.15 &       641.54 &    1.016 &  651.80 \\
            &         &               &      400$^2$ &          495.52 &       650.36 &    1.016 &  660.77 \\
            &         &               &      500$^2$ &          485.26 &       640.82 &    1.015 &  650.43 \\
            &         &               &      600$^2$ &          480.64 &       638.16 &    1.017 &  649.01 \\
            &         &               &     $\infty$ &          447.65 &       626.50 &    1.017 &  637.15 \\
   InSe-e   &   0.3   &       16$^2$  &      200$^2$ &           68.90 &       122.20 &    1.085 &  132.59 \\
            &         &               &      300$^2$ &           64.03 &       115.28 &    1.127 &  129.92 \\
            &         &               &      400$^2$ &           63.05 &       113.07 &    1.132 &  128.00 \\
            &         &               &      500$^2$ &           62.21 &       111.18 &    1.136 &  126.30 \\
            &         &               &      600$^2$ &           61.67 &       110.26 &    1.139 &  125.59 \\
            &         &               &      800$^2$ &           61.38 &       110.09 &    1.144 &  125.94 \\
            &         &               &     $\infty$ &           59.70 &       106.34 &    1.153 &  122.61 \\
   InSe-h   &   0.3   &       16$^2$  &      100$^2$ &            0.42 &         0.56 &    1.534 &    0.86 \\
            &         &               &      200$^2$ &            0.42 &         0.56 &    1.432 &    0.80 \\
            &         &               &      300$^2$ &            0.42 &         0.56 &    1.420 &    0.80 \\
            &         &               &      400$^2$ &            0.42 &         0.56 &    1.417 &    0.79 \\
            &         &               &     $\infty$ &            0.42 &         0.56 &    1.400 &    0.78 \\
 P-e ($xx$) &   0.2   &       16$^2$  &      200$^2$ &           31.58 &        31.44 &    1.172 &   36.85 \\
            &         &               &      300$^2$ &           31.63 &        32.26 &    1.187 &   38.29 \\
            &         &               &      400$^2$ &           31.69 &        32.25 &    1.193 &   38.47 \\
            &         &               &      500$^2$ &           31.57 &        32.23 &    1.194 &   38.48 \\
            &         &               &      600$^2$ &           31.62 &        32.22 &    1.197 &   38.57 \\
            &         &               &     $\infty$ &           31.64 &        32.18 &    1.206 &   38.81 \\
 P-e ($yy$) &   0.2   &       16$^2$  &      200$^2$ &          251.93 &       252.68 &    1.172 &  296.14 \\
            &         &               &      300$^2$ &          249.18 &       254.97 &    1.187 &  302.65 \\
            &         &               &      400$^2$ &          249.58 &       254.28 &    1.193 &  303.36 \\
            &         &               &      500$^2$ &          247.70 &       253.73 &    1.194 &  302.95 \\
            &         &               &      600$^2$ &          248.59 &       253.80 &    1.197 &  303.80 \\
            &         &               &     $\infty$ &          247.23 &       252.41 &    1.206 &  304.41 \\
 P-h ($xx$) &   0.3   &       16$^2$  &      200$^2$ &           21.36 &        26.57 &    1.107 &   29.41 \\
            &         &               &      300$^2$ &           21.40 &        26.39 &    1.108 &   29.24 \\
            &         &               &      400$^2$ &           21.34 &        26.40 &    1.127 &   29.75 \\
            &         &               &      500$^2$ &           21.21 &        26.27 &    1.147 &   30.13 \\
            &         &               &      600$^2$ &           21.15 &        26.20 &    1.148 &   30.08 \\
            &         &               &     $\infty$ &           20.92 &        26.05 &    1.194 &   31.10 \\
 P-h ($yy$) &   0.3   &       16$^2$  &      200$^2$ &          553.77 &       504.52 &    1.107 &  558.50 \\
            &         &               &      300$^2$ &          525.16 &       491.12 &    1.108 &  544.16 \\
            &         &               &      400$^2$ &          511.78 &       480.22 &    1.127 &  541.21 \\
            &         &               &      500$^2$ &          494.05 &       465.65 &    1.147 &  534.10 \\
            &         &               &      600$^2$ &          494.57 &       464.75 &    1.148 &  533.53 \\
            &         &               &      700$^2$ &          490.12 &       460.06 &    1.156 &  531.83 \\
            &         &               &     $\infty$ &          461.40 &       435.64 &    1.193 &  519.72 \\
 Graphene-e+h &  0.5  &       12$^2$  &     1000$^2$ &         1437330 &      1224730 &    3.019 & 3697460 \\
            &         &               &     1600$^2$ &         1371750 &      1160470 &    3.423 & 3972289 \\
            &         &               &     2000$^2$ &         1366710 &      1174670 &    4.005 & 4704553 \\
            &         &               &     2400$^2$ &         1308210 &      1091250 &    3.820 & 4168575 \\
            &         &               &     $\infty$ &         1249560 &      1051630 &    4.549 & 4783865 \\
\end{longtable*}

\begin{longtable*}{ l  l  r  r  r  r |  r  r  r  r}
\caption[Room temperature drift and Hall mobility]{
Room temperature drift and Hall mobility using adaptive smearing and dipole velocity with and without spin-orbit coupling (SOC) included.
SERTA mobility refers to the mobility in the self-energy relaxation time approximation, while BTE refers to the mobility calculated using the iterative solution of the Boltzmann transport equation.
The ``-e'' or ``-h" suffix after the material's name refers to electron or hole mobility, respectively.
The \textbf{k/q}-point grids reported in parenthesis refers to the coarse and fine electron and phonon momentum grids, respectively.
We compare various long-range treatments where the long-range part of the dynamical matrix (D) and electron-phonon matrix elements (G) includes dipole-dipole (DDS) and monopole-dipole (eDS) with the scheme of Refs.~\onlinecite{Sohier2016,Sohier2017a}, dipole-dipole (DD), dipole-quadrupole (DQ), quadrupole-quadrupole (QQ), monopole-dipole (eD), and monopole-quadrupole (eQ).  
We also report if the pseudopotential includes non-linear core corrections (NLCC) or not.} \label{table3} \\
\hline
\endfirsthead
\multicolumn{10}{c}%
{{ \tablename\ \thetable{} -- continued from previous page}} \\
\hline      &                 & \multicolumn{4}{c}{No SOC} & \multicolumn{4}{|c}{SOC} \\
            &                 & \multicolumn{2}{c}{Drift mobility} & \multicolumn{2}{c}{Hall mobility} & \multicolumn{2}{|c}{Drift mobility} & \multicolumn{2}{c}{Hall mobility} \\
            & Long-range      & SERTA & BTE & SERTA & BTE  & SERTA & BTE & SERTA & BTE \\
Material    &   treatment     &      cm$^2$/Vs &   cm$^2$/Vs  &   cm$^2$/Vs  &  cm$^2$/Vs &   cm$^2$/Vs  &   cm$^2$/Vs  &   cm$^2$/Vs  &   cm$^2$/Vs \\
 \hline
\endhead

\hline \multicolumn{10}{|r|}{{Continued on next page}} \\ \hline
\endfoot

\hline \hline
\endlastfoot
            &                 & \multicolumn{4}{c}{No SOC} & \multicolumn{4}{|c}{SOC} \\
            &                 & \multicolumn{2}{c}{Drift mobility} & \multicolumn{2}{c}{Hall mobility} & \multicolumn{2}{|c}{Drift mobility} & \multicolumn{2}{c}{Hall mobility} \\
            & Long-range      & SERTA & BTE & SERTA & BTE  & SERTA & BTE & SERTA & BTE \\
Material    &   treatment     &      cm$^2$/Vs &   cm$^2$/Vs  &   cm$^2$/Vs  &  cm$^2$/Vs &   cm$^2$/Vs  &   cm$^2$/Vs  &   cm$^2$/Vs &   cm$^2$/Vs \\
\hline
\hline 
SnS$_2$-e              & No long-range          &  31.97 &  29.74 &  33.18 &  31.00 &  - &  -  &  -   &  -   \\
(16$^2$ \textbf{k/q})  & D(DDS) + G(eDS)        &  16.35 &  23.28 &  17.59 &  23.38 &   - &    - &   - &   -  \\
(600$^2$ \textbf{k/q}) & D(DD) + G(eD)          &  16.70 &  23.49 &  17.98 &  23.56 &   - &    - &   - &   -  \\
without NLCC           & D(DD+DQ+QQ) + G(eD)    &  16.70 &  23.49 &  17.98 &  23.57 &  -  &  - &  -  & -  \\
                       & D(DD) + G(eD+eQ)       &  16.68 &  23.49 &  17.95 &  23.51 &  -  & -  &  -  &  - \\
                       & D(DD+DQ+QQ) + G(eD+eQ) &  16.69 &  23.49 &  17.95 &  23.51 &   - &    - &   - &   -   \\
\hline
MoS$_2$-e              & No long-range          &  178.18 &  159.67 & 200.41 & 175.30 & 183.13 & 163.82 & 212.73 & 185.26  \\
(18$^2$ \textbf{k/q})  & D(DDS) + G(eDS)        &  157.58 &  152.71 & 178.53 & 166.51 & 160.76 & 163.55 & 187.41 & 189.14  \\
(500$^2$ \textbf{k/q}) & D(DD) + G(eD)          &  158.52 &  153.03 & 179.49 & 166.88 & 161.75 & 156.35 & 188.48 & 175.08  \\
with NLCC              & D(DD+DQ+QQ) + G(eD)    &  158.70 &  153.19 & 179.70 & 167.09 & 160.73 & 155.27 & 186.42  & 173.70  \\
                       & D(DD) + G(eD+eQ)       &  113.78 &  129.90 & 119.90 & 136.51 & 115.98 & 132.40 & 124.33 & 141.58  \\
                       & D(DD+DQ+QQ) + G(eD+eQ) &  114.24 &  130.22 & 120.44 & 136.90 & 115.95 & 132.37 & 124.37 & 141.57  \\ [3pt]
MoS$_2$-h              & No long-range          &   14.38 &   14.18 &  34.52 &  32.65 & 119.03 & 108.08 & 431.59 & 345.76  \\
(18$^2$ \textbf{k/q})  & D(DDS) + G(eDS)        &   13.90 &   13.98 &  31.97 &  30.51 &  97.80 &  96.22 & 229.36 & 203.17  \\
(500$^2$ \textbf{k/q}) & D(DD) + G(eD)          &   13.92 &   13.99 &  32.08 &  30.60 &  98.54 &  96.60 & 233.18 & 205.61  \\
with NLCC              & D(DD+DQ+QQ) + G(eD)    &   13.95 &   14.02 &  32.07 &  30.60 &  98.90 &  96.48 & 231.06 & 203.24  \\
                       & D(DD) + G(eD+eQ)       &   11.63 &   13.07 &  22.10 &  22.54 &  59.84 &  72.45 &  94.22 & 94.437  \\
                       & D(DD+DQ+QQ) + G(eD+eQ) &   11.62 &   13.06 &  22.19 &  22.64 &  60.55 &  73.33 &  96.49 & 117.42  \\
\hline
     BN-e              &                        &        &         &        &        &    &     &    &  \\
(16$^2$ \textbf{k/q})  & D(DDS) + G(eDS)        &  96.94 &   61.86 & 166.26 & 175.55 & -  &  -  & -  & - \\
(500$^2$ \textbf{k/q}) & D(DD) + G(eD)          &  98.32 &   62.88 & 168.55 & 178.98 & -  &  -  & -  & - \\
without NLCC           & D(DD+DQ+QQ) + G(eD)    & 127.27 &  138.87 & 100.87 & 126.20 & -  &  -  & -  & - \\[3pt]
     BN-h              &                        &        &         &        &        &    &     &    &  \\
(16$^2$ \textbf{k/q})  & D(DDS) + G(eDS)        & 303.17 &  861.28 & 299.22 & 969.31 & -  &  -  & -  & - \\
(500$^2$ \textbf{k/q}) & D(DD) + G(eD)          & 308.31 &  878.68 & 303.40 & 984.70 & -  &  -  & -  & - \\
 without NLCC          & D(DD+DQ+QQ) + G(eD+eQ) & 485.26 &  640.82 & 472.67 & 650.22 & -  &  -  & -  & - \\
\hline
     BN-e              & No long-range          &  148.07 &  168.58 &  313.96 &  405.16 &  148.06 &  168.57 &  313.94 &  405.14 \\
(16$^2$ \textbf{k/q})  & D(DDS) + G(eDS)        &   49.48 &   96.73 &   61.10 &  132.94 &   49.48 &   96.73 &   61.10 &  132.94 \\
(500$^2$ \textbf{k/q}) & D(DD) + G(eD)          &   50.64 &   98.49 &   62.52 &  137.16 &   50.64 &   98.49 &   62.52 &  137.15 \\
with NLCC              & D(DD+DQ+QQ) + G(eD)    &   53.25 &   97.08 &   65.66 &  133.33 &   53.25 &   65.66 &   97.08 &  133.32 \\
                       & D(DD) + G(eD+eQ)       &  100.50 &  151.81 &  120.67 &  206.72 &  100.50 &  151.80 &  120.66 &  206.70 \\ 
                       & D(DD+DQ+QQ) + G(eD+eQ) &  101.38 &  128.89 &  153.62 &  235.15 &  101.38 &  128.88 &  153.62 &  235.13 \\ [3pt]
     BN-h              & No long-range          & 1619.14 & 1369.60 & 1777.93 & 1469.98 & 1619.14 & 1369.60 & 1777.93 & 1469.98 \\
(16$^2$ \textbf{k/q})  & D(DDS) + G(eDS)        &  255.05 &  866.39 &  254.10 &  968.47 &  255.05 &  866.39 &  254.10 &  968.47 \\
(500$^2$ \textbf{k/q}) & D(DD) + G(eD)          &  262.21 &  893.49 &  260.09 &  997.86 &  262.21 &  893.48 &  260.09 &  997.85 \\
 with NLCC             & D(DD+DQ+QQ) + G(eD)    &  276.68 &  903.90 &  273.30 & 1020.49 &  276.68 &  903.90 &  273.30 & 1020.49 \\
                       & D(DD) + G(eD+eQ)       &  613.67 &  596.56 &  738.28 &  745.53 &  613.67 &  596.56 &  738.28 &  745.53 \\  
                       & D(DD+DQ+QQ) + G(eD+eQ) &  636.38 &  837.63 &  617.34 &  847.13 &  636.38 &  837.62 &  617.34 &  847.12 \\                      
\hline                           
   InSe-e              &                        &        &        &       &        &   &    &  &  \\
(16$^2$ \textbf{k/q})  & D(DDS) + G(eDS)        &  56.44 & 100.58 & 71.02 & 112.47 &  57.25 &  101.95 &  72.25 & 114.28 \\
(600$^2$ \textbf{k/q}) & D(DD) + G(eD)          &  58.64 & 103.50 & 73.89 & 115.69 &  59.49 &  104.94 &  75.10 & 117.52 \\
 with NLCC             & D(DD+DQ+QQ) + G(eD+eQ) &  60.88 & 108.95 & 78.85 & 123.86 &  61.67 &  110.28 &  79.96 & 125.53 \\ [3pt]
   InSe-h              &                        &        &        &       &        &   &    &  &  \\
(16$^2$ \textbf{k/q})  & D(DDS) + G(eDS)        &   0.27 &   0.16 &  0.28 &   0.14 &   0.41 &    0.55 &   0.43 &   0.79 \\
(400$^2$ \textbf{k/q}) & D(DD) + G(eD)          &   0.45 &   0.36 &  0.59 &   0.69 &   0.42 &    0.55 &   0.44 &   0.80 \\
 with NLCC             & D(DD+DQ+QQ) + G(eD+eQ) &   0.46 &   0.36 &  0.59 &   0.69 &   0.42 &    0.56 &   0.44  &  0.80 \\ 
\end{longtable*}

\begin{acknowledgments}
The authors would like to thank J.-M. Lihm, G. Brunin and X. Gonze for useful discussions;
S.P. acknowledges support from the F.R.S.-FNRS as well as from the European Unions Horizon 2020 Research and Innovation Program, under the Marie
Skłodowska-Curie Grant Agreement (SELPH2D, No. 839217);
N.M. acknowledges support from the Swiss National Science Foundation and the NCCR MARVEL;
M.S. and M.R. acknowledge support from Ministerio de Ciencia y Innovaci\'on (MICINN-Spain) through Grant No. PID2019-108573GB-C22; from Severo Ochoa FUNFUTURE center of excellence (CEX2019-000917-S); from Generalitat de Catalunya (Grant No. 2017 SGR1506); and from the European Research Council (ERC) under the European Union's
 Horizon 2020 research and innovation program (Grant Agreement No. 724529).
M.G. acknowledges support from the Italian Ministry for University and Research through the Levi-Montalcini program.
Computational resources have been provided by the PRACE-21 resources MareNostrum at BSC-CNS, the Supercomputing Center of Galicia (CESGA), and by the Consortium des \'Equipements de Calcul Intensif (C\'ECI), funded by the Fonds de la Recherche Scientifique de Belgique (F.R.S.-FNRS) under Grant No. 2.5020.11 and by the Walloon Region
as well as computational resources awarded on the Belgian share of the EuroHPC LUMI supercomputer.
\end{acknowledgments}

\end{document}